\documentclass[a4paper,11pt]{article}
\pdfoutput=1 

\usepackage{jheppub}
\usepackage{float}
\usepackage{graphicx}
\usepackage{amsmath} 
\usepackage{ulem}

\usepackage{amsmath,graphicx,amssymb,subfigure,bbm}

\usepackage[utf8]{inputenc}
\usepackage[T1]{fontenc}

\usepackage{bbold}

\usepackage[all]{xy}

\usepackage{verbatim}

\newcommand{\be}{\begin{equation}}
\newcommand{\ee}{\end{equation}}
\newcommand{\bea}{\begin{eqnarray}}
\newcommand{\eea}{\end{eqnarray}}

\def\beq{\begin{equation}}
\def\eeq{\end{equation}}
\def\beqa{\begin{eqnarray}}
\def\eeqa{\end{eqnarray}}

\begin{document}

\title{\boldmath Spontaneous chiral symmetry breaking in  holographic soft wall models}

\author[a]{Alfonso Ballon-Bayona,}
\author[b]{Luis A. H. Mamani,}
\author[c]{and Diego M. Rodrigues}
\affiliation[a]{Instituto de F\'{i}sica, Universidade
Federal do Rio de Janeiro, \\
Caixa Postal 68528, RJ 21941-972, Brazil.}
\affiliation[b]{Centro de Ci\^encias Exatas Naturais e Tecnológicas,\\ Universidade Estadual da Regi\~ao Tocantina do Maranh\~ao,\\ Rua Godofredo Viana 1300, 65901- 480, Imperatriz, MA, Brazil}
\affiliation[c]{Instituto de Física Teórica, UNESP-Universidade Estadual Paulista,\\ R. Dr. Bento T. Ferraz 271, Bl. II, Sao Paulo 01140-070, SP, Brazil}

\emailAdd{aballonb@if.ufrj.br}
\emailAdd{luis.mamani@uemasul.edu.br}
\emailAdd{diego.m.rodrigues@unesp.br}

\abstract{
We investigate non-linear extensions of the holographic soft wall model proposed by Karch, Katz, Son and Stephanov \cite{Karch:2006pv} including non-minimal couplings in the five-dimensional action. 
The non-minimal couplings bring a new parameter $a_0$ which controls the transition between spontaneous and explicit symmetry breaking near the limit of massless quarks (the chiral limit). In the physical region (positive quark mass), we show that above a critical value of the parameter $a_0$ the chiral condensate $\langle \bar{q} q \rangle$ is finite in the chiral limit, signifying spontaneous chiral symmetry breaking. This result is supported by the lightest states arising in the spectrum of the pseudoscalar mesons, which become massless in the chiral limit and are therefore intrepreted as Nambu-Goldstone bosons. Moreover, the decay constants of the pseudoscalar mesons also support this conclusion, as well as the Gell-Mann-Oakes-Renner (GOR) relation satisfied by the lightest states. We also calculate the spectrum of scalar, vector, and axial-vector mesons with their corresponding decay constants. We describe the evolution of masses and decay constants with the increasing of the quark mass and for the physical mass we compare our results against available experimental data. Finally,  we do not find instabilities in our model for the physical region (positive quark mass). 
}

\maketitle
\flushbottom

\section{Introduction}

The investigation of spontaneous symmetry breaking (SSB) is a very interesting subject, essential in condensed matter and particle physics contexts. It is related with the symmetry of the quantum field theory (QFT) action and the vacuum solution of the theory. It is believed that SSB is realized only in systems with infinite degrees of freedom. An example of the realization of SSB in QFT is the emergence of Nambu-Goldstone bosons in the theory, which are related to the breaking of a global symmetry breaking via the Goldstone theorem \cite{PhysRev.127.965}. Moreover, in such systems, the vacuum expectation value does not have the same symmetry as the action. In particle physics, the pions are an example of pseudo Nambu-Goldstone bosons. In turn, quantum chromodynamics (QCD) has an approximate $U(2)_{L}\times U(2)_R$ symmetry in the limit of light quark masses, considering just up and down quarks. The symmetry breaking leads to $U(2)_{L}\times U(2)_R\to U(2)_V$ and it gives rise to pseudo-Nambu-Goldstone bosons, see for instance \cite{Peskin:1995ev,Maggiore:2005qv}. This symmetry breaking occurs in the low energy regime of QCD, or strong coupling regime, where perturbative techniques are not reliable anymore. However, there are effective theories that one may use in this regime, like QCD sum rules, chiral effective theories, Dyson-Schwinger equations or even numerical methods introduced by Lattice QCD. In the past years Lattice QCD became a powerful computational tool for extracting information of QCD at low energies. Nevertheless, there are still difficulties and limitations to be overcome for this fascinating subject; for an investigation of chiral symmetry breaking in Lattice QCD with two light flavors, see for instance Ref.~\cite{Engel:2014cka}.

An alternative theoretical framework for investigating strongly coupled systems was provided by the Anti-de Sitter/Conformal Field Theory (AdS/CFT) correspondence, proposed at the end of the 90s \cite{Maldacena:1997re}. Using as a guide the holographic dictionary \cite{Gubser:1998bc, Witten:1998qj}, one may map any operator of a strongly coupled conformal field theory into its dual classical field in the gravitational side of the duality living in an AdS space with one dimension greater than the space where the conformal field theory lives. In the case of four dimensions, one may break the conformal symmetry and introduce symmetries known to be present in QCD so that the original conformal field theory becomes more similar to QCD. This is the motivation behind a bottom-up approach that led to the construction of phenomenological AdS/QCD models known as the hard wall model \cite{Erlich:2005qh,DaRold:2005mxj} and the soft wall model \cite{Karch:2006pv}. There have been a lot of progress on  this approach, see for instance Refs.~\cite{Ghoroku:2005vt, Colangelo:2008us, Gherghetta:2009ac, Abidin:2009aj, Grigoryan:2007wn, Vega:2008af, Kwee:2007dd, Kwee:2007nq, Chelabi:2015gpc, Cui:2013xva, Sui:2009xe, Fang:2016nfj, Ballon-Bayona:2014oma, Ballon-Bayona:2017bwk, Contreras:2021onc, Cao:2021tcr, FolcoCapossoli:2019imm} and references therein. However, it is worth pointing out that the gravitational background is not obtained solving the Einstein equations, this means that the metric is fixed being AdS. Nevertheless, one may consider this naive approximation as a toy model to gain intuition. A more consistent version of this approach takes into consideration the gravitational background obtained solving the Einstein equations, see for instance Refs.~\cite{Kiritsis:2006ua, Gursoy:2007er, Gubser:2008yx, Li:2013oda, Ballon-Bayona:2017sxa, Ballon-Bayona:2018ddm, Ballon-Bayona:2021tzw} and the references therein. 
The bottom-up approach is complementary to the  top-down approach, whose goal is to find string theory duals to strongly coupled QCD-like theories, see for instance Refs.~\cite{Klebanov:2000hb,Maldacena:2000yy,Karch:2002sh, Sakai:2004cn,Bigazzi:2014qsa}. For a review of the top-down approach see for instance \cite{Edelstein:2006kw,Erdmenger:2007cm,Nunez:2010sf}.

One of the aims of this work is the investigation of spontaneous chiral symmetry breaking within holographic models for QCD. The first attempts to describe chiral symmetry breaking in the bottom-up approach were proposed in Refs.~\cite{Erlich:2005qh, DaRold:2005mxj} where the quark mass operator is mapped to a 5d tachyonic field. 
These bottom-up models as well as their top-down counterparts, see e.g. \cite{Evans:2004ia,Sakai:2004cn,Sakai:2005yt}, have provided very useful insights for the construction of more realistic holographic models for QCD. There has been recent progress on the description of spontaneous chiral symmetry breaking in bottom-up holographic QCD models inspired by string theory \cite{Casero:2007ae,Iatrakis:2010jb,Iatrakis:2010zf,Jarvinen:2011qe} and its relation to the violation of the Breitenlohner-Freedman bound \cite{Alho:2013dka,BitaghsirFadafan:2018efw}. 

In this paper we restrict ourselves to the holographic soft wall model proposed in Ref.~\cite{Karch:2006pv} where the dilaton is quadratic in the radial coordinate leading to meson masses organized in linear trajectories in the excitation number, i.e. $m_n^2 \sim n$. It is important to point out that the original version of the linear soft wall model does not describe spontaneous chiral symmetry breaking \cite{Colangelo:2008us}. It was suggested in \cite{Karch:2006pv} that the addition of non-linear terms in the action for the tachyonic field would allow for a non-linear description of chiral symmetry breaking that may be compatible with spontaneous symmetry breaking. Following this suggestion the authors of Ref.~\cite{Gherghetta:2009ac} built a phenomenological model compatible with spontaneous symmetry breaking in the sense that the quark condensate is finite in the chiral limit. In our opinion, this model provides good results, but there are a few caveats that we point out: i) the chiral condensate is given as an input instead of arising dynamically from solving the tachyon differential equation, 
ii) the dilaton is non-monotonic and negative near the AdS boundary, 
iii) an instability arises in the scalar sector first observed in Ref.~\cite{Sui:2009xe}, see also \cite{Li:2013oda}. Following the ideas of Refs.~\cite{Casero:2007ae, Iatrakis:2010zf, Iatrakis:2010jb} the authors of Ref.~\cite{Ballon-Bayona:2020qpq} showed that the non-linear extension of the original soft wall model \cite{Karch:2006pv} based on a quadratic dilaton and a  Higgs-like potential for the tachyonic field indeed allows for a non-linear realisation of chiral symmetry breaking. However, it was shown in \cite{Ballon-Bayona:2020qpq} that the chiral condensate still vanishes in the massless limit and therefore  chiral symmetry breaking is explicit. 

An attempt to describe spontaneous chiral symmetry breaking in the soft wall model was motivated by the discussion on the sign of an effective dilaton field, responsible for conformal symmetry breaking and the mass generation in the infrared. A negative dilaton profile was proposed in Refs.~\cite{deTeramond:2009xk, Zuo:2009dz} as an alternative to describe the hadronic spectrum and spontaneous chiral symmetry breaking. However, there has been a debate about the phenomenological consequences of the negative dilaton profile. It has been shown in  Ref.~\cite{Karch:2010eg} that the negative dilaton drives to the emergence of an unphysical massless scalar state in the vectorial sector. Besides that, the negative dilaton 
leads to unphysical masses for higher spin mesons \cite{Karch:2010eg}. Motivated by this very interesting discussion, the authors of Ref.~\cite{Chelabi:2015gpc} suggested that an interpolation function between an effective negative dilaton in the ultraviolet (UV) region and positive in the infrared (IR) region might provide a description of spontaneous chiral symmetry breaking. Nevertheless, the authors of that paper did not calculate the spectrum of the pseudoscalar mesons at zero temperature and proved the emergence of the pions as Nambu-Goldstone bosons. In this paper we work on the ideas of Ref.~\cite{Chelabi:2015gpc} and reinterpret them in terms of non-minimal dilaton couplings in the 5d action. Following the techniques used in Ref.~\cite{Ballon-Bayona:2020qpq} we solve the tachyon differential equation and find that the new parameter $a_0$
controlling the non-minimal couplings allows us to describe
 spontaneous chiral symmetry breaking in the chiral limit. We show explicitly that, above a critical value for $a_0$, the chiral condensate is nonzero in the chiral limit signalling spontaneous chiral symmetry breaking, and that this result is consistent with the emergence of massless modes in the pseudoscalar sector, i.e., the pions as Nambu-Goldstone bosons. As a check of consistency we also calculate the decay constants and show that the lightest states in the pseudoscalar sector satisfy the Gell-Mann-Oakes-Renner (GOR) relation.

This paper is organized as follows. Sec.~\ref{Sec:SoftWall} is a short review of the holographic soft wall model and its non-linear extension. In Sec.~\ref{Sec:non-linearSoftWall} we present the non-linear extension of the soft wall model including the non-minimal couplings. We consider two interpolating functions for those couplings, models of type I and models of type II. We solve the tachyon differential equation in its asymptotic regions and then we solve it numerically. Near  the chiral limit we describe the transition between spontaneous and explicit chiral symmetry breaking driven by the parameter $a_0$ characterizing the non-minimal couplings. The free parameters of our model are fixed at the end of that section. In Sec.~\ref{Sec:Spectrum} we expand the five-dimensional action up to second order in the fields in order to get the equation of motions describing the vector, scalar, axial-vector, and pseudoscalar mesons. The corresponding spectrum is calculated and compared against results available in the literature. In Sec.~\ref{Sec:DecayConstants} we calculate the decay constants corresponding to vector, scalar, axial-vector, and pseudoscalar mesons, showing how they are related to the normalization constants of the wave functions. At the end of the section we derive the GOR relation. We present our conclusions in Sec.~\ref{Sec:Conclusions}. Additional details about the numerical procedure and results are displayed in Appendix \ref{Sec:NumericalAnalysis}. Meanwhile, in Appendix \ref{app:KKexp} we present the Kaluza-Klein expansion. Details about the derivation of the decay constants are presented in Appendix \ref{Sec:Decay}.
Details of the masses and decay constants in the linear soft wall model are reviewed in Appendix \ref{MesonsSW}.
A short discussion about the linear soft wall model with negative dilaton is presented in Appendix \ref{Sec:LinearDilNeg}, while the non-linear extension is investigated in Appendix \ref{Se:non-linearNegativeDil}. Finally, we derive the GOR relation using a toy model in Appendix \ref{Sec:GORToyModel}.

\section{Soft wall models with minimal dilaton couplings}
\label{Sec:SoftWall}

\subsection{The original soft wall model}

One of the seminal papers considering a bottom-up approach in holographic QCD is the soft wall model \cite{Karch:2006pv}. In \cite{Karch:2006pv}, the authors introduced a scalar field, the dilaton, in a 5d Anti-de-Sitter space in order to incorporate effects of  conformal symmetry breaking and generate an infrared mass gap. They showed that a dilaton growing quadratically in the radial coordinate far from the boundary leads to an approximate linear behavior of the mass spectrum, i.e., $m^2\propto n$. It is worth pointing out that the dilaton proposed in \cite{Karch:2006pv} was not obtained solving the Einstein equations. Then, one might think that the confinement properties introduced by the dilaton field are ``artificial''. However, it has been shown that the quadratic behaviour proposed in \cite{Karch:2006pv} is in fact the right asymptotic behaviour for the dilaton in confining holographic QCD backgrounds found by solving the Einstein-dilaton equations \cite{Gursoy:2007er}, see also \cite{Ballon-Bayona:2017sxa}.

Thus, we consider the soft wall model as a first step for a realistic description is good enough, at least to extract general properties of the dual field theory. Thus, having identified the fields in the gravitational side of the duality one can write the five-dimensional action, which is given by
\noindent
\begin{equation}
S = - \int d^5 x \sqrt{-g} \, e^{-\Phi} \, {\rm Tr} \Big [| D_m X |^2 + m_X^2 |X|^2 + \frac{1}{4g_5^2} {F_{mn}^{(L)}}^2 + \frac{1}{4 g_5^2} {F_{mn}^{(R)}}^2 \Big ] \, , \label{SoftWallAction}
\end{equation}
where $\Phi=\Phi(z)$ is the dilaton field, $X$ is the bifundamental scalar field (or tachyonic field) dual to the quark mass operator $\bar q q$, $m_X^2=-3$ the mass of the tachyonic field, and the field strength $F_{nm}^{(L/R)}$ are defined by 
 \begin{align}
F_{mn}^{(L/R)} &= \partial_m A_n^{(L/R)} - \partial_n A_m^{(L/R)} - i [A_m^{(L/R)},A_n^{(L/R)}] \, , \nonumber \\
D_m X &= \partial_m X - i A_m^{(L)} X + i X A_m^{(R)} \, , \label{fdstrength}
\end{align} 
\noindent
where $A_n^{(L/R)}$ represent the gauge fields, and $D_{m}X$ the covariant derivative. As usual in the soft wall model approach, the 5d metric $g_{mn}$ is the AdS spacetime in Poincar\'e coordinates. The 5d coupling in the vectorial sector is fixed as $g_5^2=12\pi^2/N_{c}$, with $N_c$ the number of colors, in order to reproduce the perturbative QCD result for the current correlators at small distances \cite{Erlich:2005qh}. The action \eqref{SoftWallAction} describes the breaking of the $SU(N_f)_L \times SU(N_f)_R$ gauge symmetry due to the presence of a nonzero tachyonic field $X$. The generators of the $su(N_f)$ algebra are normalized as ${\rm Tr}(T^a T^b) = \delta^{ab}/2$ and we will focus on the two-flavor case $N_f=2$.

A very nice feature of the soft wall model is the possibility of describing approximate linear trajectories  for the meson spectrum, i.e. $m_n^2 \sim n$, by simply imposing an asymptotic quadratic behaviour for the dilaton field, i.e. $\Phi(z) = \phi_{\infty} z^2 $  at large $z$ (far from the boundary).  As realised in \cite{Karch:2006pv}, if the dilaton quadratic ansatz is taken for all $z$ the effective Schr\"odinger potential of the field perturbations dual to the mesons takes the form of an harmonic oscillator with orbital momentum and the linear behaviour $m_n^2 \sim n$ becomes exact. Appendix \ref{MesonsSW} provides a short review of the results for vector and scalar mesons in the (linear) soft wall model.

The (linear) soft wall model \eqref{SoftWallAction} leads to a linear differential equation for the tachyonic field $X$. Regularity of the action in the infrared region (large $z$) fixes the source and VEV coefficients in the UV and it turns out that the chiral condensate is proportional to the quark mass and vanishes in the chiral limit.

\subsection{The non-linear soft wall model}

Non-linear extensions of the soft wall model have been investigated in the literature, see for instance \cite{Gherghetta:2009ac,Chelabi:2015gpc, Ballon-Bayona:2020qpq}. The non-linearity is carried out by the tachyon potential. Then, the action is a modified version of \eqref{SoftWallAction}
\noindent
\begin{equation}
S = - \int d^5 x \sqrt{-g} \, e^{-\Phi} \, {\rm Tr} \Big [| D_m X |^2 + V(|X|) + \frac{1}{4 g_5^2} {F_{mn}^{(L)}}^2 + \frac{1}{4 g_5^2} {F_{mn}^{(R)}}^2 \Big ] \, . \label{HiggsAction}
\end{equation}
\noindent
where the non-Abelian field strengths $F_{mn}^{(L/R)}$ and covariant derivative $D_m X$ given by \eqref{fdstrength}, while the tachyon potential has the Higgs form 
\noindent
\begin{equation}
V(|X|) = m_X^2 |X|^2 + \lambda |X|^4 \,,
\label{Eq:Higgspot}
\end{equation}
\noindent
where $\lambda$ is a free parameter.  The 5d metric is again the AdS spacetime in Poincar\'e coordinates. 
The form of the Higgs potential \eqref{Eq:Higgspot} is motivated by the fact that it provides a minimal description of spontaneous breaking of a local symmetry in the standard model and may be possibly extended to AdS space in order to describe the gravity dual of the spontaneous breaking of a global symmetry. The potential in \eqref{Eq:Higgspot} seems also a good ansatz if we want to build a holographic realization of the Coleman-Witten theorem \cite{Coleman:1980mx}.

The action in \eqref{HiggsAction} was used in the phenomenological model of Ref.~\cite{Gherghetta:2009ac} with $\lambda<0$ for the Higgs potential \eqref{Eq:Higgspot}. Instead of solving for the tachyonic field for a given dilaton field an inverse method was proposed: given a tachyonic field compatible with spontaneous chiral symmetry breaking the dilaton field was reconstructed. The ansatz for the tachyonic field in \cite{Gherghetta:2009ac} is singular at large $z$ and the effective dilaton field found in \cite{Gherghetta:2009ac} is quadratic at large $z$ (far from the AdS boundary) but becomes non-monotonic and negative as $z$ approaches zero (near the AdS boundary).

In \cite{Ballon-Bayona:2020qpq}, the same action in \eqref{HiggsAction} was considered for a quadratic dilaton field and the chiral condensate was found dynamically by solving the tachyonic field differential equation. The main result in \cite{Ballon-Bayona:2020qpq} was that the only non-trivial solution for the tachyonic field is regular far from the boundary and leads to a chiral condensate growing non-linearly with the quark mass. However, it turns out that this regular solution vanishes in the chiral limit and therefore spontaneous chiral symmetry breaking is absent. It was found in \cite{Ballon-Bayona:2020qpq} that for the case of $\lambda>0$ the model provides very interesting results in the limit of heavy quarks. The meson masses and decay constants obtained in the model for the axial and pseudo-scalar sector display a behaviour qualitatively similar to the one found in perturbative QCD.

Note that the tachyonic field in \eqref{HiggsAction} couples minimally with the dilaton, i.e., the factor $e^{-\Phi}$ in front of the trace, depends on the dilaton field only. In the next section we shall investigate how to generalize the form of this dilaton coupling. We will see that keeping the  minimal dilaton coupling only at large $z$ (where $\Phi$ is large) but deviating from it at small $z$ (where $\Phi$ is small) it is possible to describe spontaneous chiral symmetry breaking in the chiral limit. 

\section{Non-linear soft wall models with non-minimal dilaton couplings}
\label{Sec:non-linearSoftWall}

The gravitational action \eqref{HiggsAction} can be generalized  considering two non-minimal couplings in the form
\begin{align}
S &= - \int d^5 x \sqrt{-g}\,{\rm Tr}\Big \{ e^{-a(\Phi)} \Big[  |D_m X|^2 + V(|X|)   \Big]  
+ \frac{e^{- b(\Phi)}}{4 g_5^2} \,
\Big [{F_{mn}^{(L)}}^2 + {F_{mn}^{(R)}}^2 \Big ] \Big \}  \,.  \label{5dmodel}
\end{align}
\noindent
The fields of this action are the same as presented previously, the difference in relation to the previous action is the addition of the functions: $e^{-a(\Phi)}$, and $e^{-b(\Phi)}$, which are in principle non trivial functions of the dilaton field.

Having specified the gravitational action describing the quarks of the dual field theory, we now consider the gravitational background, which is characterised by a Poincar\'e invariant metric
\noindent
\begin{equation}
ds^2 = e^{2 A_s(z)} \Big ( - dt^2 + d \vec{x}^2 + dz^2 \Big ) \, ,
\end{equation}
where $A_s(z)$ is the warp factor in the string frame. As usual in soft wall models we fix the warp factor and dilaton as
\noindent
\begin{equation}
A_s(z) = - \ln{\left(\frac{z}{\ell}\right)}  \quad, \quad 
\Phi(z) = \phi_{\infty} z^2 \,, 
\end{equation}
That is, the 5d metric is AdS in Poincar\'e coordinates and the dilaton is quadratic in the radial coordinate $z$. 
The parameter $\ell$ denotes the AdS radius and from here on we set $\ell=1$. Note that $\phi_{\infty}$ is a parameter with units of energy squared. The original soft wall model corresponds to the particular case $a(\Phi)=b(\Phi)=\Phi$. For that particular case, as shown in \cite{Karch:2006pv}, a quadratic dilaton is sufficient to guarantee a linear behaviour for the meson Regge trajectories.  In this work we
consider more general dilaton couplings $a(\Phi)$ and $b(\Phi)$. We  consider for simplicity the case $N_f=2$ where the model describes the gauge symmetry breaking $SU(2) \times SU(2) \to SU(2)$ which is the gravity dual of chiral symmetry breaking in QCD with 2 flavours.

\subsection{Background field equations}

For the background fields we take a Lorentz invariant ansatz 
\begin{equation}
A_{m}^{(L/R)} = 0 \quad, \quad 
2 X(z) =  v(z) I_{2 \times 2} \,.
\end{equation}
Under this ansatz the action takes the 1d form
\begin{equation}
S_0= - V_4 \int dz \, e^{3A_s - a} \Big [ \frac12 (\partial_z v)^2 + e^{2A_s} U(v)  \Big ] \, ,
\end{equation}
where
\begin{equation}
U(v) = \frac{m_X^2}{2} v^2 + \frac{\lambda}{8} v^4 \, , \label{Upot}
\end{equation}
is the potential for the 1d function $v(z)$ and $V_4 = \int d^4 x$. We remind the reader that $A_s$ and $a$ depend only on $z$. The critical points of the potential are obtained solving the equation $\frac{dU(v)}{dv}=0$, as $m_X^2=-3$, we get three solutions, two of them corresponding to minimum and the other to the trivial solution (the local maximum)
\noindent
\begin{equation}
v=\pm\sqrt{\frac{6}{\lambda}},\qquad v=0.
\end{equation}

The Euler-Lagrange equation for $v(z)$ can be written as
\begin{equation}
\Big [ \partial_z + (3 A_s' - a') \Big ] \partial_z v - e^{2A_s} \frac{dU}{dv} = 0 \,, \label{veq}
\end{equation}
\noindent
where $'$ denotes $d/dz$. Note that this equation simplifies when evaluated at the critical points where the derivative of the potential is zero,
\noindent
\begin{equation}\label{Eq:TachyonCriticalPoint}
\left(\ln{v'(z)}\right)' + (3 A_s - a)'=0,\quad\to\quad v=C_0+D_0 \int_{0}^{z}e^{-(3A_s(x)-a(x))}dx.
\end{equation}
\noindent
This means that the tachyon has at least a constant as a solution and the other solution is divergent. Plugging this solution in \eqref{veq}, considering the regular solution, we get an expression for $C_0$ whose solutions are: $C_0=\pm\sqrt{6/\lambda}$ and $C_0=0$. If we look for solutions in the IR, the corrections to $C_0$ can be calculated considering corrections in the form $\mathcal{O}(z^{-1}, z^{-2}, \cdots)$.

In turn, considering $A_s = - \ln z$ and the potential $U(v)$ given by \eqref{Upot} the tachyon field equation \eqref{veq} takes the form
\begin{equation}\label{Eq:TachyonDE}
\Big [z^2 \partial_z^2 - \left ( 3 + z \, a'  \right ) z \partial_z - m_X^2 \Big ] v - \frac{\lambda}{2} v^3 = 0 \,. 
\end{equation}
\noindent
Another interesting property of this differential equation is the scaling symmetry in $\sqrt{\lambda}$, this means that the equation is the same under the transformation $\sqrt{\lambda}\,v\to \tilde{v}$. One may take advantage of this property. For example, we may solve the differential equation for $\tilde{v}$, then, we get the solution for $v$ scaling $\tilde{v}$, i.e., using the relation $v=\tilde{v}/\sqrt{\lambda}$. We will see the applicability of this symmetry in the next section.

\subsection{Interpolations for the dilaton coupling}

So far, we have the differential equation for the tachyon, it is lacking the explicit form of the non-minimal coupling $a(\Phi)$, and $b(\Phi)$. In the following, we motivate our choices for the functions $a(\Phi)$ and $b(\Phi)$.

\begin{figure}[ht!]
\centering
\includegraphics[width=7cm]{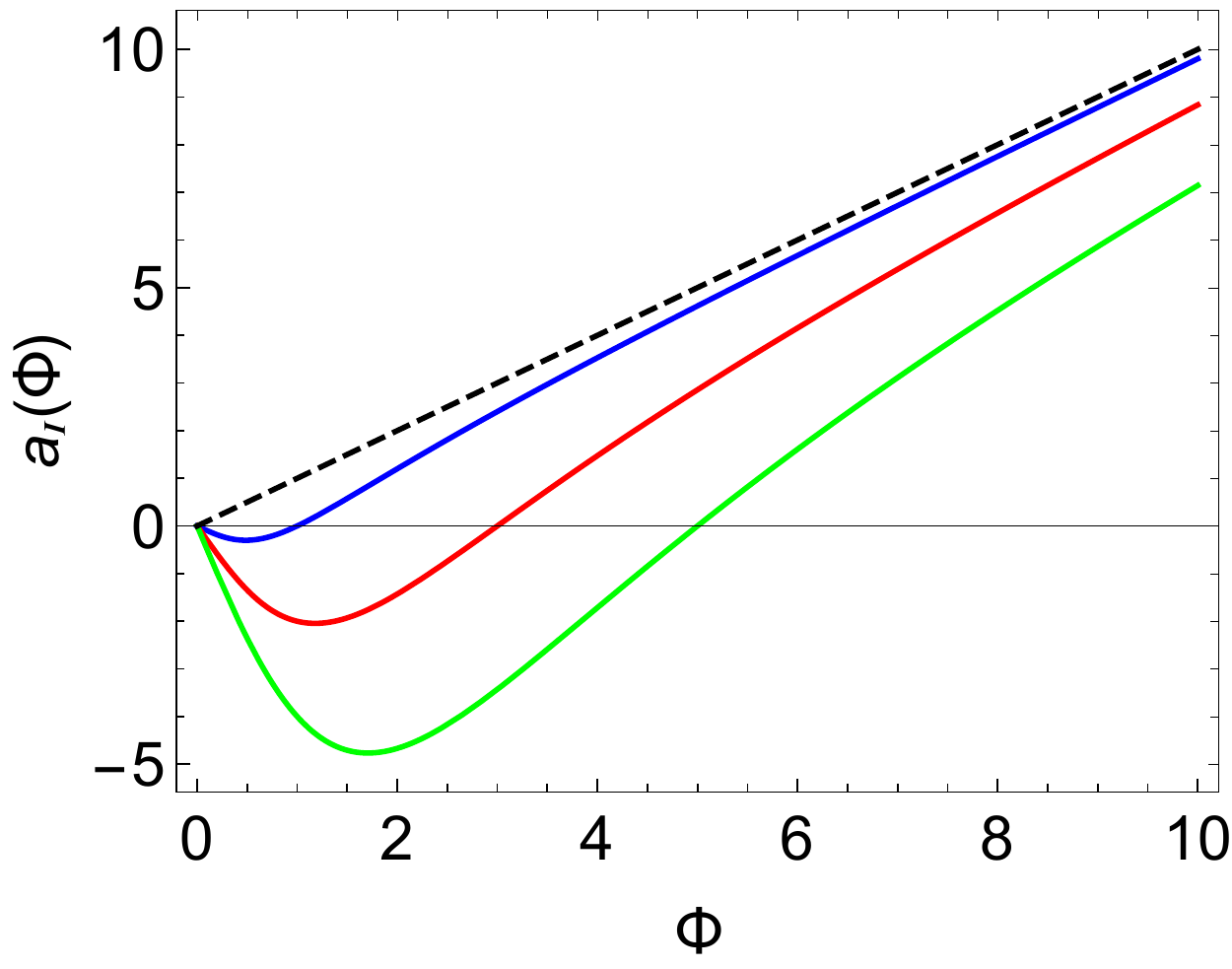}\hfill 
\includegraphics[width=7cm]{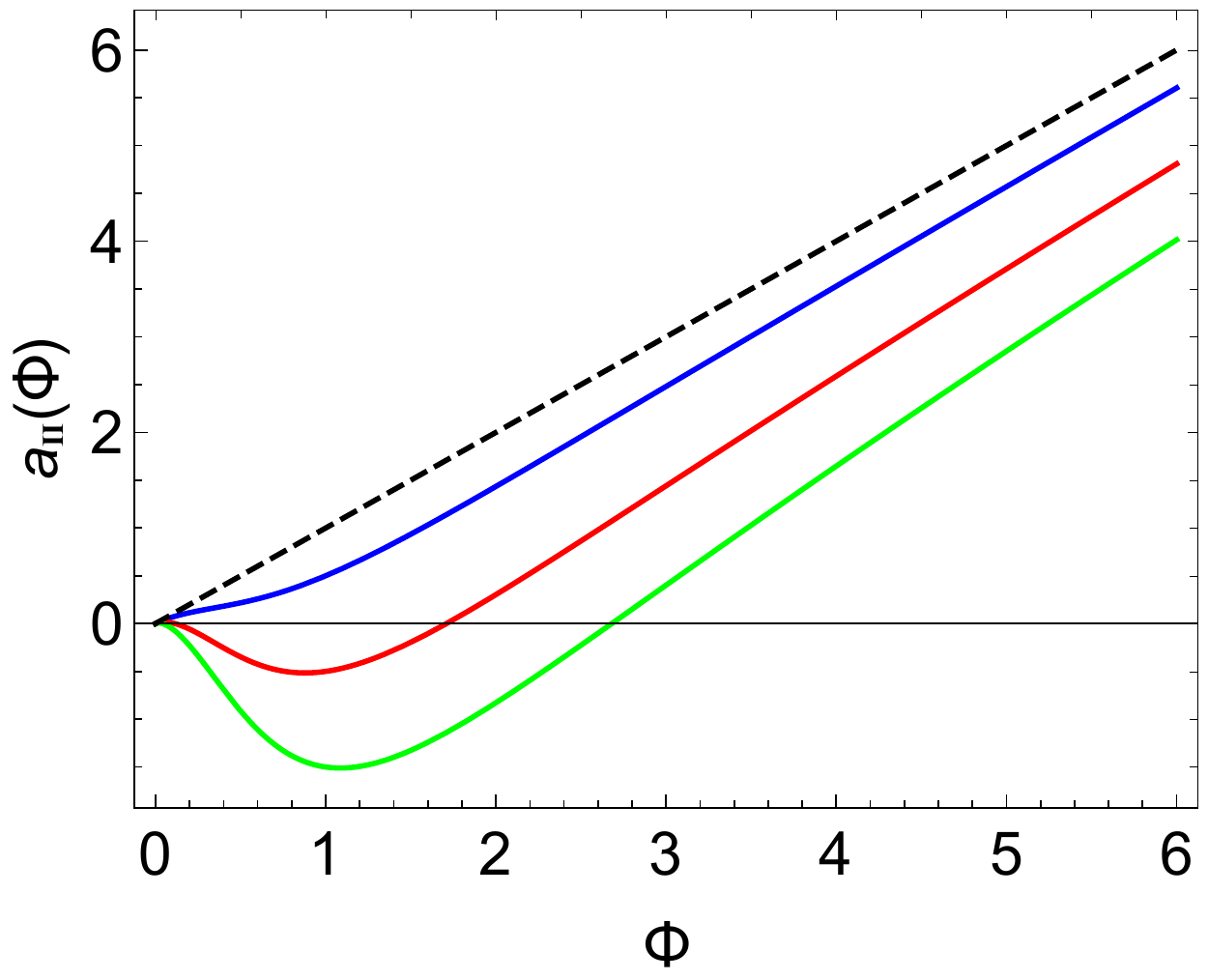}
\caption{
The non-minimal dilaton couplings $a_{I}(\Phi)$ (left panel) and $a_{II}(\Phi)$ (right panel) for the selected values $a_0=1$ (blue), $a_0=3$ (red) and $a_0=5$ (green). The black dashed lines in both panels  correspond to the limit $a_0 \to 0$ where the dilaton couplings reduce to the minimal form $a(\Phi)=\Phi$.  
}
\label{Fig:Interpolations}
\end{figure}

The  non-minimal couplings $a(\Phi)$ and $b(\Phi)$ were  originally considered to be equal and both interpreted as an effective dilaton field $\tilde \Phi = a(\Phi)=b(\Phi)$ so that the effective dilaton field $\tilde \Phi$  couples minimally to the tachyon and gauge fields. An effective negative dilaton field $\tilde \Phi$ quadratic in the radial coordinate allows for a description of spontaneous chiral symmetry breaking, see Appendix \ref{Sec:LinearDilNeg} for a short review. However, the authors of Ref.~\cite{Karch:2010eg} showed that a non-physical state arises in the vectorial sector when the  dilaton remains negative in the IR. This means that the dilaton must be positive in the IR in order to avoid non-physical states. We believe this was the motivation of the authors of Ref.~\cite{Chelabi:2015gpc} for building an interpolation for the effective dilaton field, whose asymptotic form reduces to a negative $\tilde \Phi$ (quadratic in the radial coordinate) in the UV and a positive $\tilde \Phi$ in the IR (also quadratic in the radial coordinate).
Following these considerations we impose that the non-minimal coupling reduces to the minimal coupling in the IR regime, namely
\noindent
\begin{equation}\label{Eq:aIR}
a( \Phi \to \infty) = \Phi \, .
\end{equation}
We remind the reader that in our framework the dilaton $\Phi$ is always quadratic in the radial coordinate, i.e. $\Phi(z) = \phi_{\infty} z^2$. 
It turns out that spontaneous chiral symmetry breaking can be realised as long as we impose the following UV asymptotic behaviour \cite{Chelabi:2015gpc} 
\noindent
\begin{equation}\label{Eq:aUV}
a_{I}(\Phi \to 0) = - a_0 \Phi \,. 
\end{equation} 
\noindent
 Having fixed the asymptotic form of the function $a(\Phi)$, one may build an interpolating function considering these asymptotic constraints.  We consider first the following simple interpolation\,\footnote{The choice of this function was motivated by the interpolation presented in Ref.~\cite{Chelabi:2015gpc}, in our notation it would take the form $a(\Phi) = \Phi \Big [ - a_0  + (1 + a_0) \tanh \left (\frac{\Phi}{a_0} \right) \Big ]$. Note that the interpolation using powers of the dilaton field is smoothest than the interpolation using hyperbolic function.}
\begin{enumerate}
\item[] Models of type I: 
\begin{equation}\label{Eq:Dilaton1}
a_I(\Phi) = \Phi \left [ \frac{ - a_0^2 + \Phi^2 }{ a_0 + \Phi^2 }\right ]  \,.
\end{equation}
\end{enumerate}
 However, as long as the coupling $a(\Phi)$ becomes minimal in the IR, the asymptotic form in the UV should not be necessarily negative. As we will see, it will be sufficient to impose that $a(\Phi)$ is negative for intermediate values of $\Phi$ (or equivalently $z$). This choice will provide improved results compared to models of type I. Its asymptotic behavior in the UV is such that
\noindent
\begin{equation}
a_{II}(\Phi\to 0)=\Phi.
\end{equation}
\noindent
Considering this case we build the following interpolation function
\noindent
\begin{enumerate}
\item[] Models of type II: 
\begin{equation}\label{Eq:Dilaton2}
a_{II}(\Phi) = \Phi-\frac{a_0\,\Phi^{3/2}}{1+\Phi^2}
\,.
\end{equation}
\end{enumerate}
\noindent
The non-minimal dilaton couplings \eqref{Eq:Dilaton1} and \eqref{Eq:Dilaton2} are displayed in Fig. \ref{Fig:Interpolations} for selected values of $a_0$. 
The non-minimal couplings can be thought as deformations of the original soft wall model characterized by a new free parameter $a_0$; we will show later that we need this additional parameter to induce a transition between explicit and spontaneous chiral symmetry breaking.

\noindent

As regards the dilaton coupling $b(\Phi)$ to the 5d gauge fields we will consider the following possibilities:
\begin{itemize}
\item[] Models of type A:  
\begin{equation}
b(\Phi)=a(\Phi)    
\end{equation}
\end{itemize}
That is, in models of type A the gauge fields couple non-minimally to the dilaton in exactly the same way as the tachyonic field. 
\begin{itemize}
\item []Models of type B:  
\begin{equation}
b(\Phi)=\Phi    
\end{equation}
In this case only the tachyonic field couples non-minimally to the dilaton.         
\end{itemize}

Table \ref{tab:Models} summarizes the four different models considered in this work. 
\begin{table}[]
    \centering
        \begin{tabular}{c|c|c|c|c}
    \hline
    \hline
Model & IA  & IB  & IIA & IIB  \\
         \hline
Coupling $a(\Phi)$ & $a_I(\Phi)$ &  $a_I(\Phi)$  & $a_{II}(\Phi)$ &  $a_{II}(\Phi)$  \\
    \hline 
Coupling $b(\Phi)$ & $a_I(\Phi)$ & $\Phi$ &  $a_{II}(\Phi)$ & $\Phi$  \\
    \hline 
    \hline
    \end{tabular}
    \caption{Non-minimal couplings $a(\Phi)$ and $b(\Phi)$ in models of type IA, IB, IIA and IIB.}
    \label{tab:Models}
\end{table}

\subsection{Violation of the BF bound}

\begin{figure}[ht!]
\centering
\includegraphics[width=7cm]{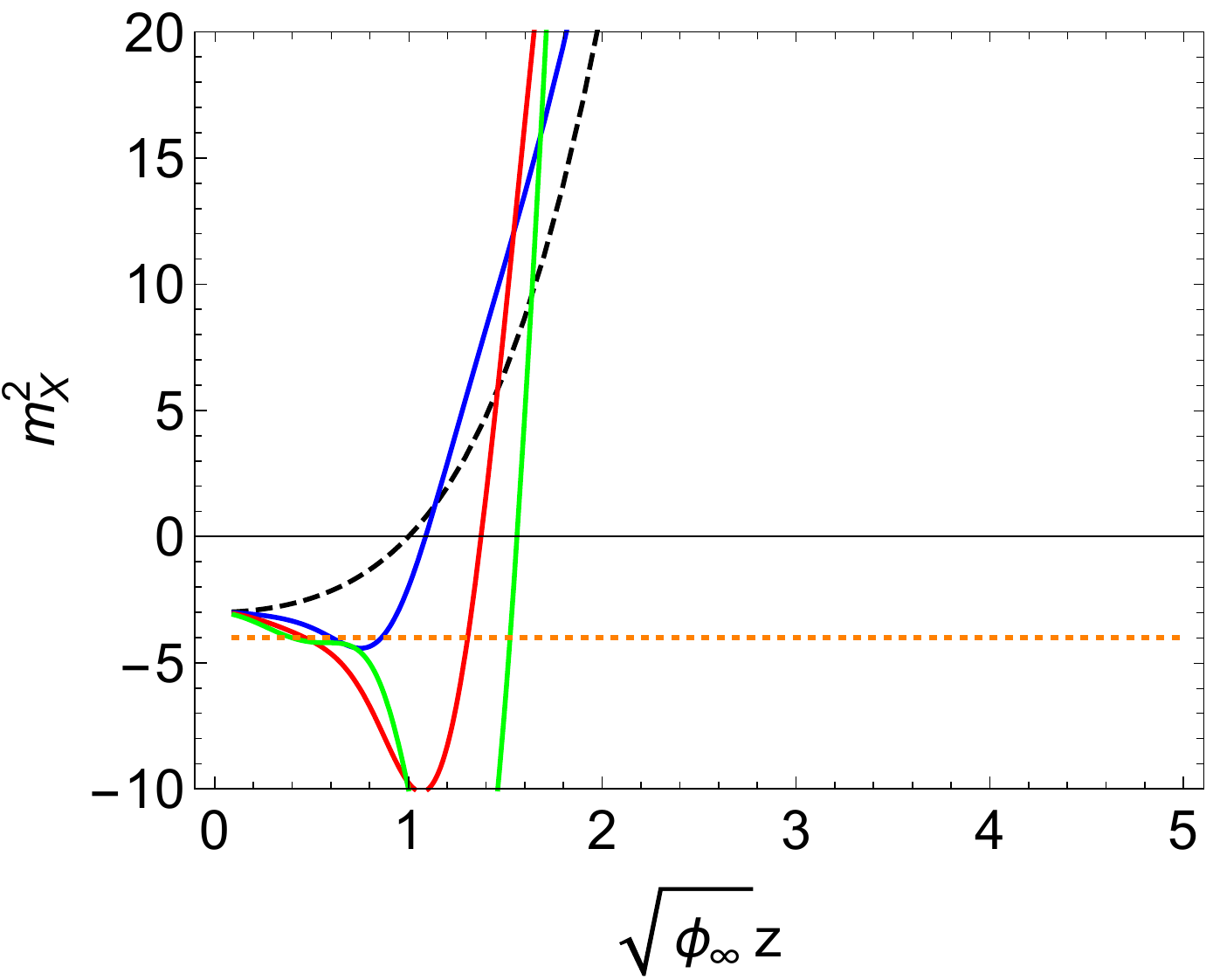}\hfill 
\includegraphics[width=7cm]{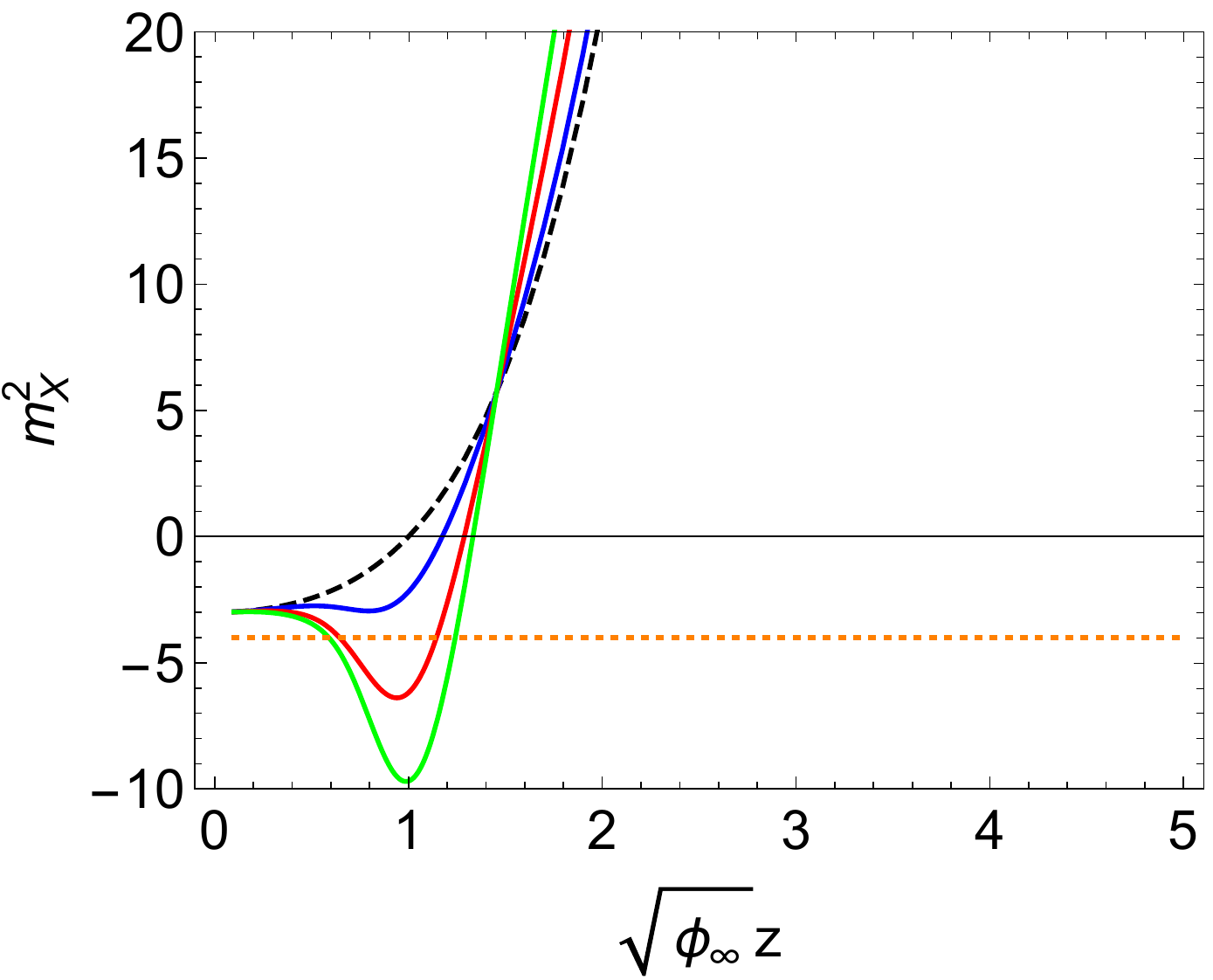}
\caption{
The evolution of the 5d effective mass $m_X^2$ with the dimensionless radial coordinate $\sqrt{\phi_{\infty}} z$ for models of type I (left panel) and models of type II (right panel). In both panels the blue, red and green solid lines correspond to $a_0=1$, $a_0=3$ and $a_0=5$ respectively. The black dashed line in both panels correspond to the limit $a_0 \to 0$ where the dilaton couplings become minimal (original soft wall model). The horizontal line (orange dotted) represents the BF bound  $m_{BF}^2 = -4$.}
\label{Fig:RunningMass}
\end{figure}

In this subsection we investigate the violation of the Breitenlohner-Freedman lower bound $m_{BF}^2 = -4$ \cite{Breitenlohner:1982bm} for the 5d effective mass of a scalar perturbation $S(x,z)$ around the trivial solution $v(z)=0$. This scalar perturbation satisfies the linear differential equation
\begin{equation}
\Big [ \partial_z + 3 A_s' - a' \Big ] \partial_z S 
+ \Box S - e^{2 A_s} m_X^2 S = 0 \,.
\end{equation}
Redefining the scalar field as 
\begin{equation}
S = e^{\frac{a}{2}} \bar S \, ,
\end{equation}
the tachyonic differential equation takes the $AdS_5$ form 
\begin{equation}
\Big [ \partial_z + 3 A_s'  \Big ] \partial_z \bar S
 + \Box \bar S - e^{2A_s}  
\bar m_X^2(z) \bar S  = 0 \, ,
\end{equation}
where
\begin{equation}
\bar m_X^2(z) = m_X^2 - e^{2A_s} \Big [ \frac{a''}{2} + \frac{a'}{2} (3 A_s' - \frac{a'}{2}) \Big ] \, ,
\end{equation}
is an effective 5d running mass. Since $A_s = - \ln z$ is the AdS warp factor the evolution of the effective 5d mass $m_X^2$ with the radial direction $z$ may imply the violation of the BF found for the scalar field $S(x,z)$; namely there maybe some regions in $z$ where $m_X^2(z) < -4$. The violation of the BF found indicates the instability of the trivial solution $v(z)=0$ which is the gravity dual of a chirally symmetric vacuum. This is what we expect if we want to  describe spontaneous chiral symmetry breaking because if we consider the limit of massless quarks (chiral limit),  the instability of the trivial solution $v(z)=0$ indicates the presence of a non-trivial solution $v(z) \neq 0$ that breaks the chiral symmetry spontaneously. In Fig. \ref{Fig:RunningMass} we plot the 5d effective mass $m_X^2$ as a function of the dimensionless radial coordinate $\sqrt{\phi_{\infty}} z$. The violation of the BF bound at intermediate values of $z$ indicates the instability of the trivial vacuum.  

In the next subsection we solve the non-linear differential equation \eqref{Eq:TachyonDE} for the tachyonic field $v(z)$ and find a non-trivial solution that is consistent with the spontaneous breaking of chiral symmetry in the chiral limit. 

\noindent

\subsection{The background solution for the tachyonic field}

Once the non-minimal coupling was fixed one may solve the differential equation \eqref{Eq:TachyonDE} in the asymptotic regions. Below we describe this procedure for the UV region (near the boundary) and the IR region (far from the boundary). 

\medskip

{\bf UV asymptotics}

\medskip

In the UV, or close to the boundary, we consider the power ansatz for the rescaled tachyon field $\widetilde{v}(z) = \sqrt{\lambda} v(z)$. The asymptotic solution
to \eqref{Eq:TachyonDE} takes the form
\noindent
\begin{equation}\label{Eq:AnsatzTachyonUV}
\widetilde{v}=\widetilde{c}_1\, z+\widetilde{d}_3\,z^3\ln{z}+\widetilde{c}_3\, z^3+\widetilde{c}_4\,z^4+\widetilde{d}_5\,z^5\ln{z}+\widetilde{c}_5\,z^5+\cdots
\end{equation}
\noindent
It is worth mentioning that the parameter $\widetilde{c}_1$ is related to the quark mass $m_q$ and chiral condensate $\Sigma=\langle \bar q q \rangle$ in the dual field theory through the relations $\widetilde{c}_1=\zeta\, \widetilde{m}_q$ and $\widetilde{c}_3=\widetilde{\Sigma}/(2\zeta)$ \cite{Ballon-Bayona:2020qpq}, where we have defined $\widetilde{m}_q = \sqrt{\lambda} m_q$ and $\widetilde{\Sigma} = \sqrt{\lambda} \Sigma$. The normalization constant $\zeta$ is fixed as $\zeta=\sqrt{N_c}/(2\pi)$ by large $N_c$ counting rules for the scalar correlation function \cite{Cherman:2008eh}. For models of type I, the coefficients: $\widetilde{d}_3$, $\widetilde{c}_4$, $\cdots$ in Eq.~\eqref{Eq:AnsatzTachyonUV} are given by
\noindent
\begin{equation}
\begin{split}
\widetilde{d}_3=&\frac{\widetilde{c}_1\,\phi_{\infty}}{4}\left(\frac{\widetilde{c}_1^{\,\,2}}{\phi_{\infty}}-4\,a_{0}\right),\qquad \widetilde{c}_4=0,\qquad \widetilde{d}_5=\frac{3\, \widetilde{c}_1\,\phi_{\infty}^2}{64}\left(\frac{\widetilde{c}_1^{\,\,2}}{\phi_{\infty}}-4\,a_{0}\right)^2,\\
\widetilde{c}_5=&\frac{\widetilde{c}_1\,\phi_{\infty}^2}{256}\left(\frac{\widetilde{c}_1^{\,\,2}}{\phi_{\infty}}-4\,a_{0}\right)\left(\frac{48\,\widetilde{c}_3}{\widetilde{c}_1\,\phi_{\infty}}-\frac{9\,\widetilde{c}_1^{\,\,3}}{\widetilde{c}_1\,\phi_{\infty}}+20\,a_{0}\right).
\end{split}
\end{equation}
\noindent
while for models of type II are:
\noindent
\begin{equation}
\begin{split}
\widetilde{d}_3=&\frac{\widetilde{c}_1\,\phi_{\infty}}{4}\left(\frac{\widetilde{c}_1^{\,\,2}}{\phi_{\infty}}+4\right),\qquad \widetilde{c}_4=-a_0\,\widetilde{c}_1\phi_{\infty}^{3/2},\qquad \widetilde{d}_5=\frac{3\, \widetilde{c}_1\,\phi_{\infty}^2}{64}\left(\frac{\widetilde{c}_1^{\,\,2}}{\phi_{\infty}}+4\right)^2,\\
\widetilde{c}_5=&\frac{\widetilde{c}_1\,\phi_{\infty}^2}{256}\left(\frac{\widetilde{c}_1^{\,\,2}}{\phi_{\infty}}+4\right)\left(\frac{48\,\widetilde{c}_3}{\widetilde{c}_1\phi_{\infty}}-\frac{9\,\widetilde{c}_1^{\,\,2}}{\phi_{\infty}}-20\right).
\end{split}
\end{equation}
\noindent

\medskip

{\bf IR asymptotics}

\medskip

In the IR region there are two independent solutions, see Eq.~\eqref{Eq:TachyonCriticalPoint}. In order to avoid singularities in the action \eqref{5dmodel}, we choose the regular solution. The subleading terms may be obtained considering the power ansatz
\noindent
\begin{equation}\label{Eq:TachyonPosDilIR}
\widetilde{v}=\widetilde{C}_0+\frac{\widetilde{C}_2}{z^2}+\frac{\widetilde{C}_4}{z^4}+\frac{\widetilde{C}_5}{z^5}+\cdots.
\end{equation}
\noindent
Thus, plugging them into the differential equation \eqref{Eq:TachyonDE} and solving order-by-order we get the coefficients for models of type I
\noindent
\begin{equation}
\begin{split}
\widetilde{C}_2=&\,\frac{\widetilde{C}_0}{8\phi_{\infty}}\left(\widetilde{C}_0^2-6\right),\\ 
\widetilde{C}_4=&\,\frac{3\widetilde{C}_0}{128\phi_{\infty}^2}\left(\widetilde{C}_0^2-10\right)\left(\widetilde{C}_0^2-6\right),\\
\widetilde{C}_5=&\,0.
\end{split}
\end{equation}
\noindent
While the coefficients for models of type II are:
\noindent
\begin{equation}
\begin{split}
\widetilde{C}_2=&\,\frac{\widetilde{C}_0}{8\phi_{\infty}}\left(\widetilde{C}_0^2-6\right),\\ 
\widetilde{C}_4=&\,\frac{3\widetilde{C}_0}{128\phi_{\infty}^2}\left(\widetilde{C}_0^2-10\right)\left(\widetilde{C}_0^2-6\right),\\
\widetilde{C}_5=&\,\frac{a_0\,\widetilde{C}_0}{40\phi_{\infty}^{5/2}}\left(\widetilde{C}_0^2-6\right).
\end{split}
\end{equation}
\noindent

It is worth pointing out that the contribution of the parameter $a_0$ is relevant in the UV regime, while in the IR it contributes in terms greater than $\widetilde{C}_4$. Note also that the leading term in the IR is constant, while it becomes an exact solution when $\widetilde{C}_0=6$, which coincides with the minimum of the tachyon potential. 

\medskip

{\bf Numerical results}

\medskip

\begin{figure}[ht!]
\centering
\includegraphics[width=7cm]{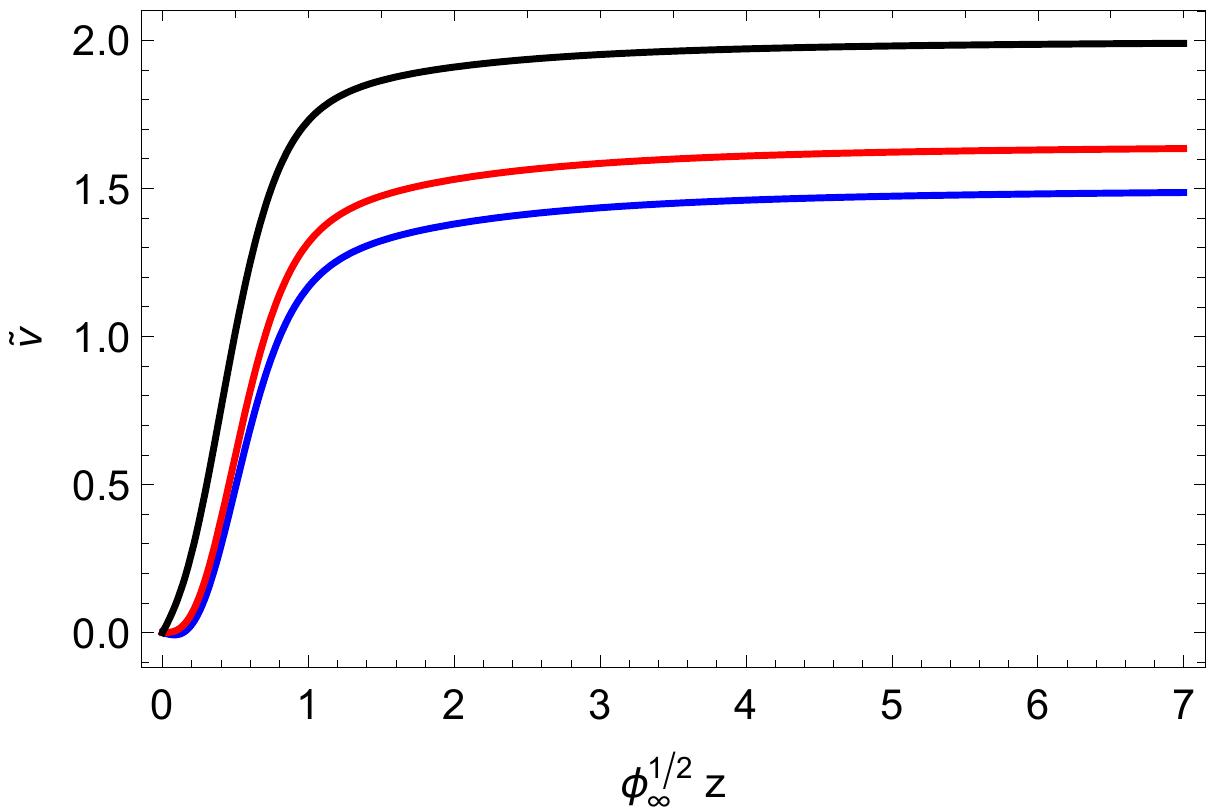}\hfill 
\includegraphics[width=7cm]{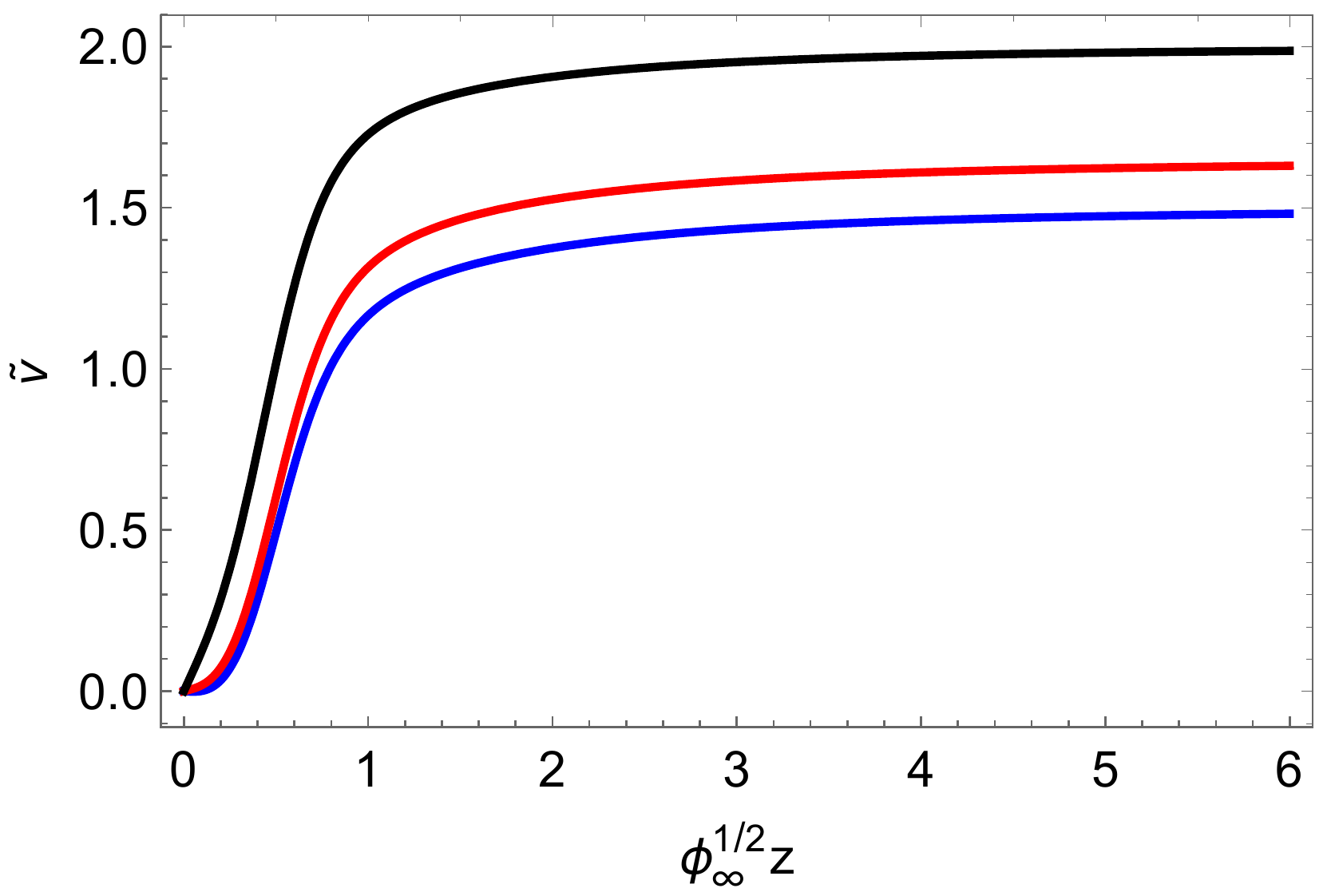}
\caption{
Left: The tachyon $\widetilde{v}$ as a function of $\phi_{\infty}^{1/2}\,z$ for models of type I. Right: The tachyon $\widetilde{v}$ as a function of $\phi_{\infty}^{1/2}\,z$ for models of type II.
}
\label{Fig:Tachyon}
\end{figure}

In the sequence, we present and discuss the numerical results obtained solving the differential equation \eqref{Eq:TachyonDE}. To solve the problem numerically we use as ``initial condition'' the asymptotic solution obtained in the IR. Then, we integrate numerically from the IR to the UV varying the independent parameter $\widetilde{C}_0$ once $a_0$ was fixed. Finally, we read off the values for $\widetilde{c}_1$ and $\widetilde{c}_3$ by matching the numerical solution against the asymptotic solution in the UV. To finish this section we plot the tachyon profile for selected values of the parameter $\widetilde{C}_0$. In the left panel of Fig.~\ref{Fig:Tachyon} we display the tachyon obtained in models of type I for $\widetilde{C}_0=1.5$ (Blue), $\widetilde{C}_0=1.648$ (Red), and $\widetilde{C}_0=2.0$ (Black). In the right panel of this figure we display the tachyon obtained in models of type II for selected values of $\widetilde{C}_0$

\noindent

\subsection{Spontaneous chiral symmetry breaking/Chiral limit}
\label{Subsec:SpontXSB}

\begin{figure}[ht!]
\centering
\includegraphics[width=7cm]{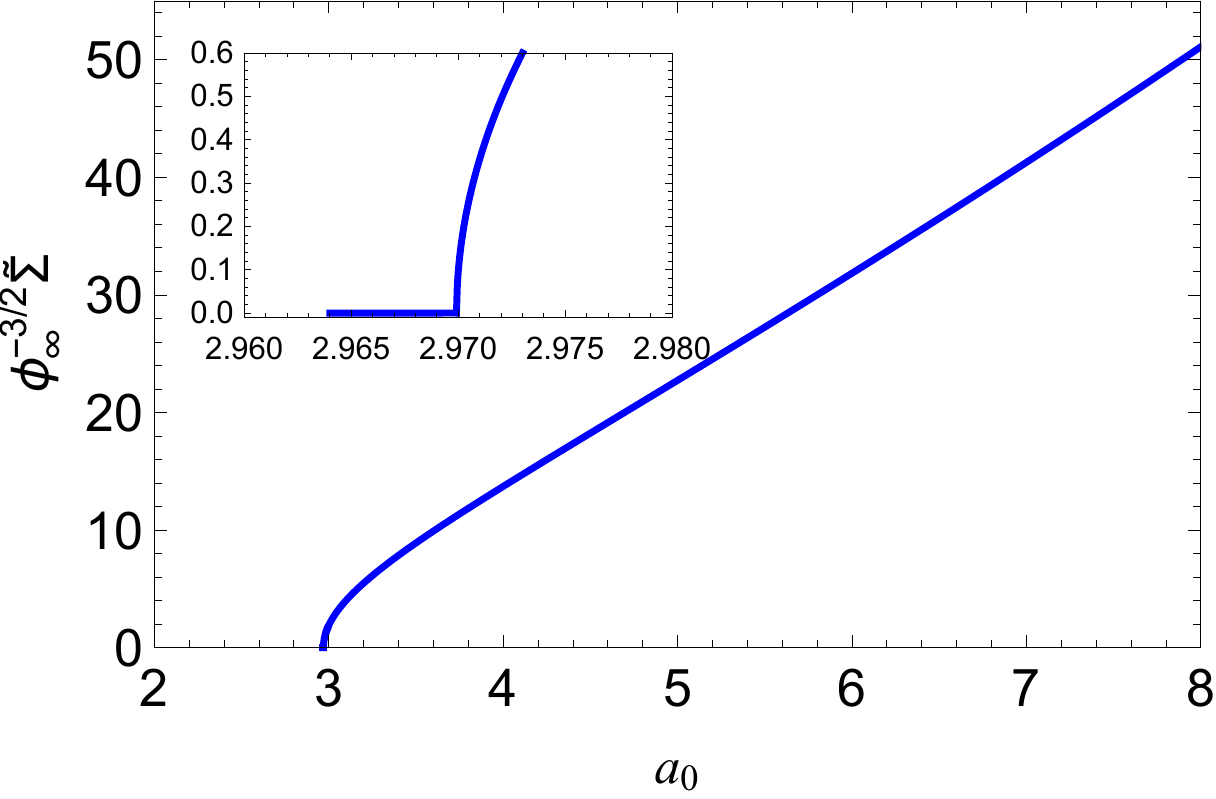}\hfill 
\includegraphics[width=7cm]{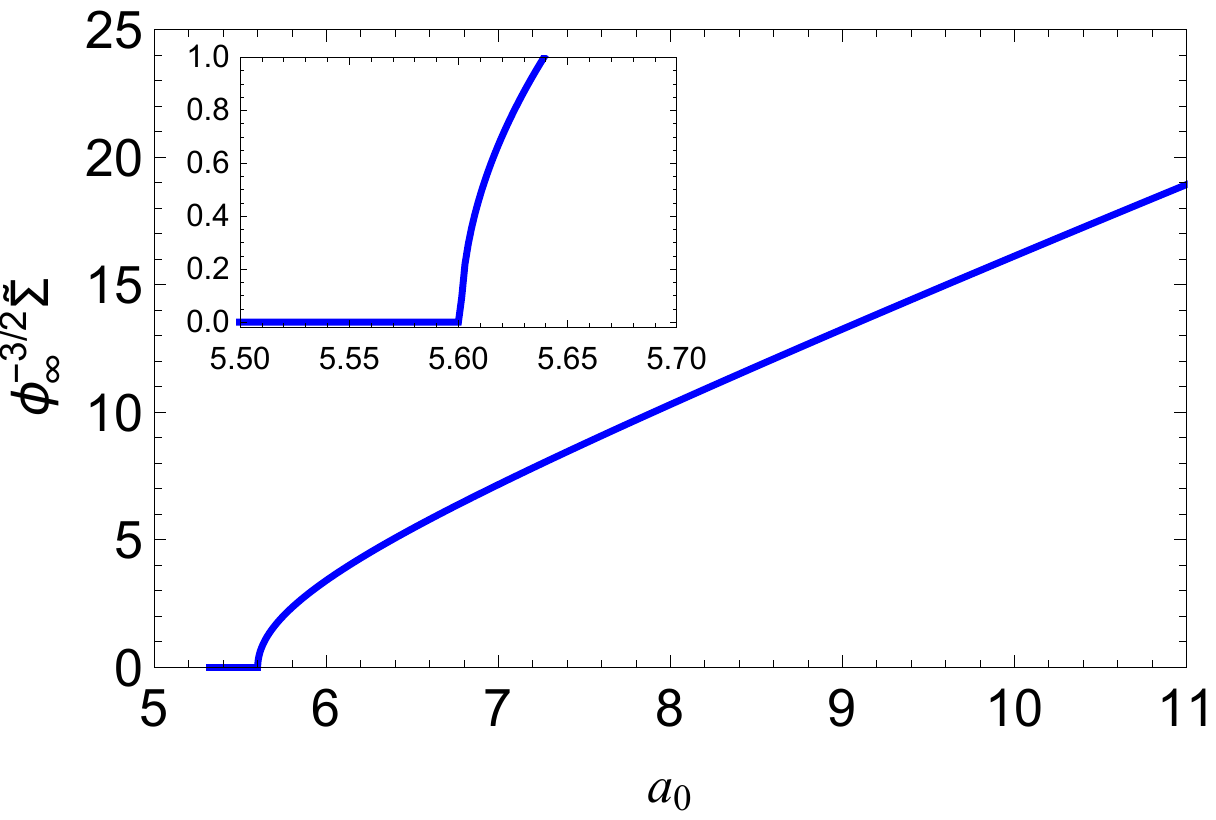}
\caption{
Left: The dimensionless condensate $\phi_{\infty}^{-3/2} \widetilde{\Sigma}$ as a function of $a_0$ in the chiral limit, i.e., $c_1=0$, for models of type I. Right: The dimensionless condensate $\phi_{\infty}^{-3/2}\widetilde{\Sigma}$ as a function of $a_0$ in the chiral limit for models of type II.
}
\label{Fig:PhiC3}
\end{figure}

Here we investigate the limit of massless quarks, which is reached in the limit of zero $\widetilde{c}_1$ since $\widetilde{c}_1\propto \widetilde{m}_q$, where $\widetilde{m}_q$ is the rescaled quark mass. In the left panel of Fig.~\ref{Fig:PhiC3} we display the numerical results of $\phi_{\infty}^{-3/2}\widetilde{c}_3$ as a function of the parameter $a_0$ obtained in models of type I. As can be seen, there is a critical value at $a_0=a_{0_{c}}\approx 2.970$ where $\phi_{\infty}^{-3/2}\widetilde{c}_3$ starts to increase, see the inset for details. This figure is showing us that there is a nonzero condensate, $\phi_{\infty}^{-3/2}\widetilde{c}_3\neq 0$, for $a_0\geq a_{0_{c}}$. It is also interesting to point out that the condensate is zero in the region $a_0<a_{0_{c}}$, where the symmetry breaking is explicit. Therefore, there is a transition between explicit and spontaneous symmetry breaking at $a_{0_c}$. In turn, the right panel of Fig.~\ref{Fig:PhiC3} shows the same results for models of type II. As can be seen, for this Model we have $a_{0_c}\approx 5.60$. 

\noindent

\begin{figure}[ht!]
\centering
\includegraphics[width=7cm]{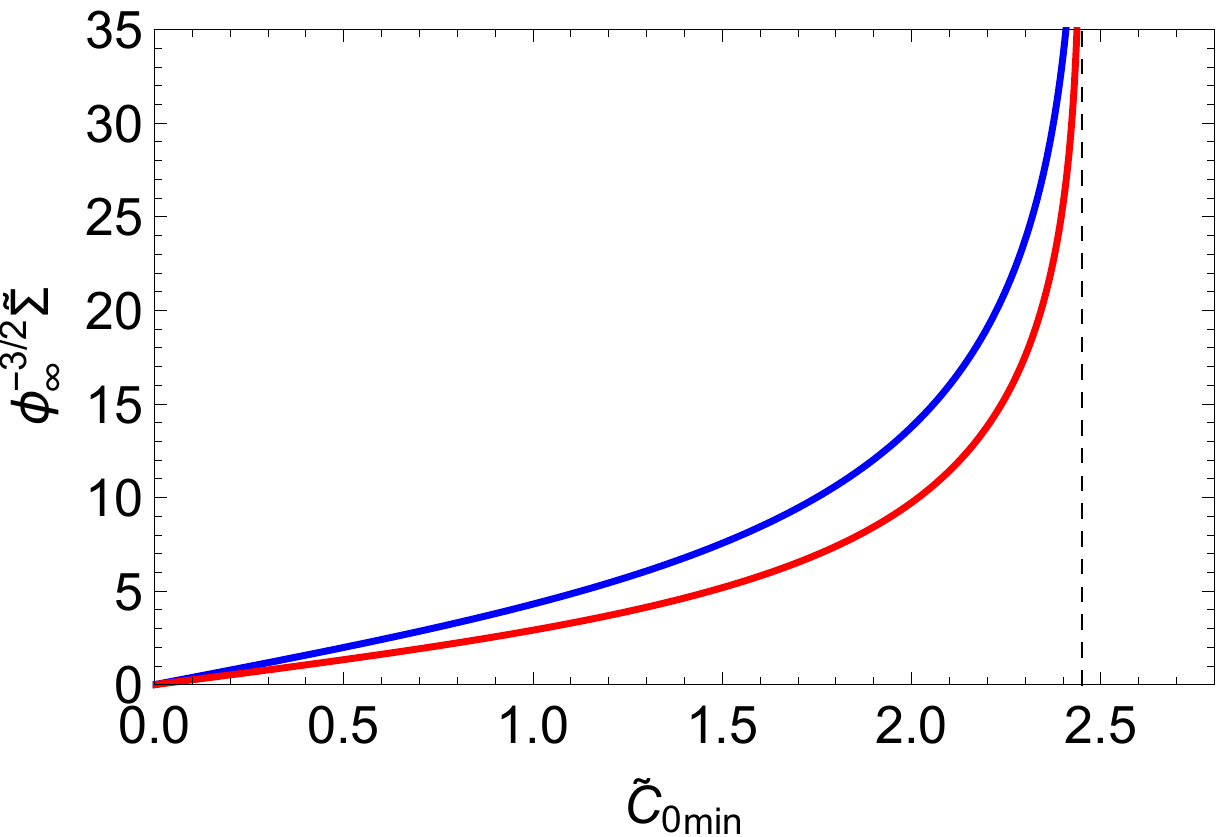}
\hfill
\includegraphics[width=7cm]{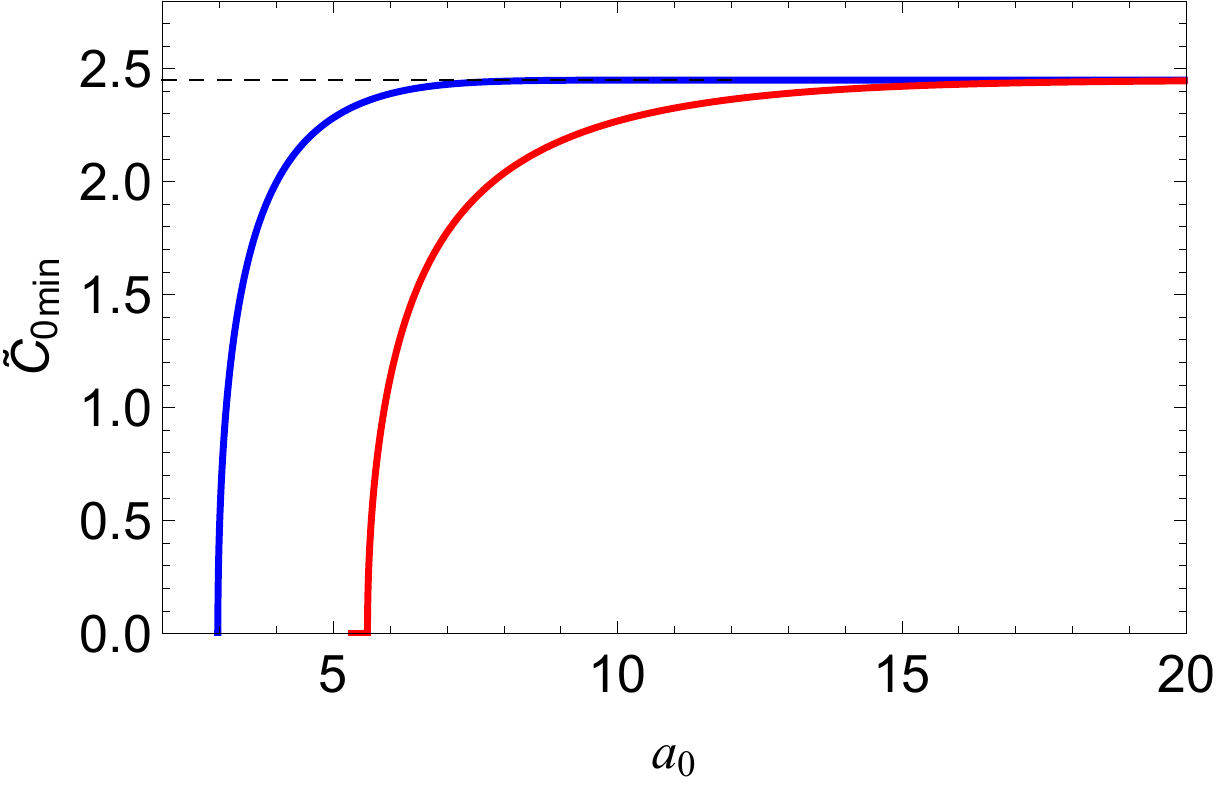}
\caption{
Left: The dimensionless condensate $\phi_{\infty}^{-3/2}\widetilde{\Sigma}$ as a function of $\widetilde{C}_{0_{\text{min}}}$ in the chiral limit, blue solid line are the results for models of type I, while red solid line for models of type II.
Right: $\widetilde{C}_{0_{\text{min}}}$ as a function of $a_0$. In both figures, dashed lines represent the minimum of the Higgs potential $\widetilde{C}_{0_{\text{max}}}=\sqrt{6}$.
}
\label{Fig:C0MinC3PhiC0Min}
\end{figure}

Furthermore, we display the corresponding results of the dimensionless chiral condensate $\phi_{\infty}^{-3/2}\widetilde{\Sigma}$ as a function of $\widetilde{C}_{0_{\text{min}}}$
in the left panel of Fig.~\ref{Fig:C0MinC3PhiC0Min} with solid blue line for models of type I and solid red line for models of type II. Note that $\widetilde{C}_{0_{\text{min}}}$ corresponds to each $\phi_{\infty}^{-3/2} \widetilde{\Sigma}$ calculated in the chiral limit, this means that each $\widetilde{C}_{0_{\text{min}}}$ corresponds to each $a_0$, as shown in the right panel of this figure. As can be seen, the value of $\phi_{\infty}^{-3/2}\widetilde{\Sigma}$ goes to zero when $\widetilde{C}_{0_{\text{min}}}$ goes to zero. However, it diverges when $\widetilde{C}_{0_{\text{min}}}$ approaches to $\widetilde{C}_{0_{\text{max}}}=\sqrt{6}$ represented by vertical dashed line. In turn, in the right panel of this figure we display the results of $\widetilde{C}_{0_{\text{min}}}$ as a function of $a_0$ for both Models I and II. In this figure we may identify the physical region where $\widetilde{c}_1>0$. Therefore, fixing the value of $a_0$, the physical region belongs to the interval $\widetilde{C}_{0_{\text{min}}}< \widetilde{C}_0 < \sqrt{6}$. It is also worth mentioning that for values of $a_0$ in the interval $a_0\leq 6.954$, one can go continuously from one minimum of the tachyon potential, $-\sqrt{6}$, to the other, $\sqrt{6}$, passing through the trivial solution. However, for values larger than $a_0\approx 6.954$, $\widetilde{C}_0$ is constrained to take values close to the minimum of the tachyon potential, and the trivial solution. We explain details on the numerical results in Appendix \ref{Sec:NumericalAnalysis}. 

Finally, in order to correctly describe spontaneous symmetry breaking, these results must be consistent with the emergence of pseudo-Nambu-Goldstone bosons in the spectrum. We will calculate the spectrum in the following section and check if these states are present in the holographic model.

\subsection{Chiral condensate in the physical region}

\begin{figure}[ht!]
\centering
\includegraphics[width=7cm]{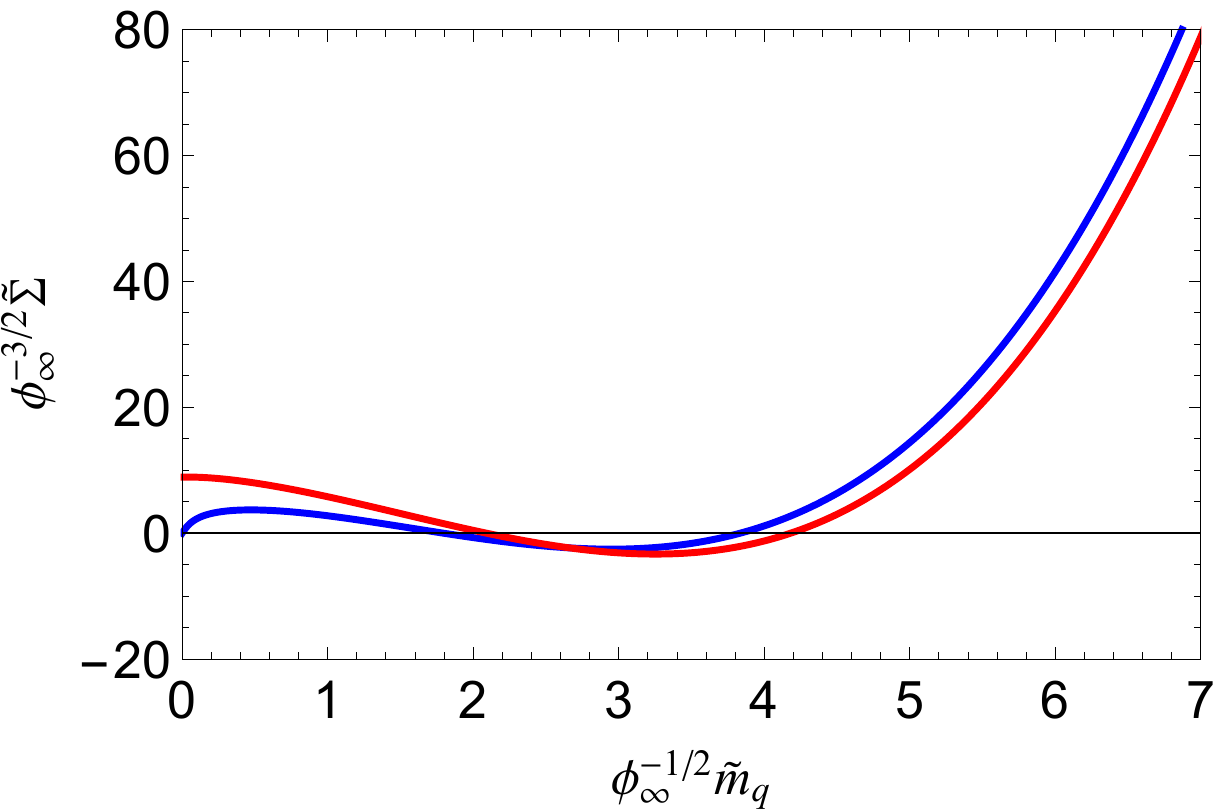}
\hfill
\includegraphics[width=7cm]{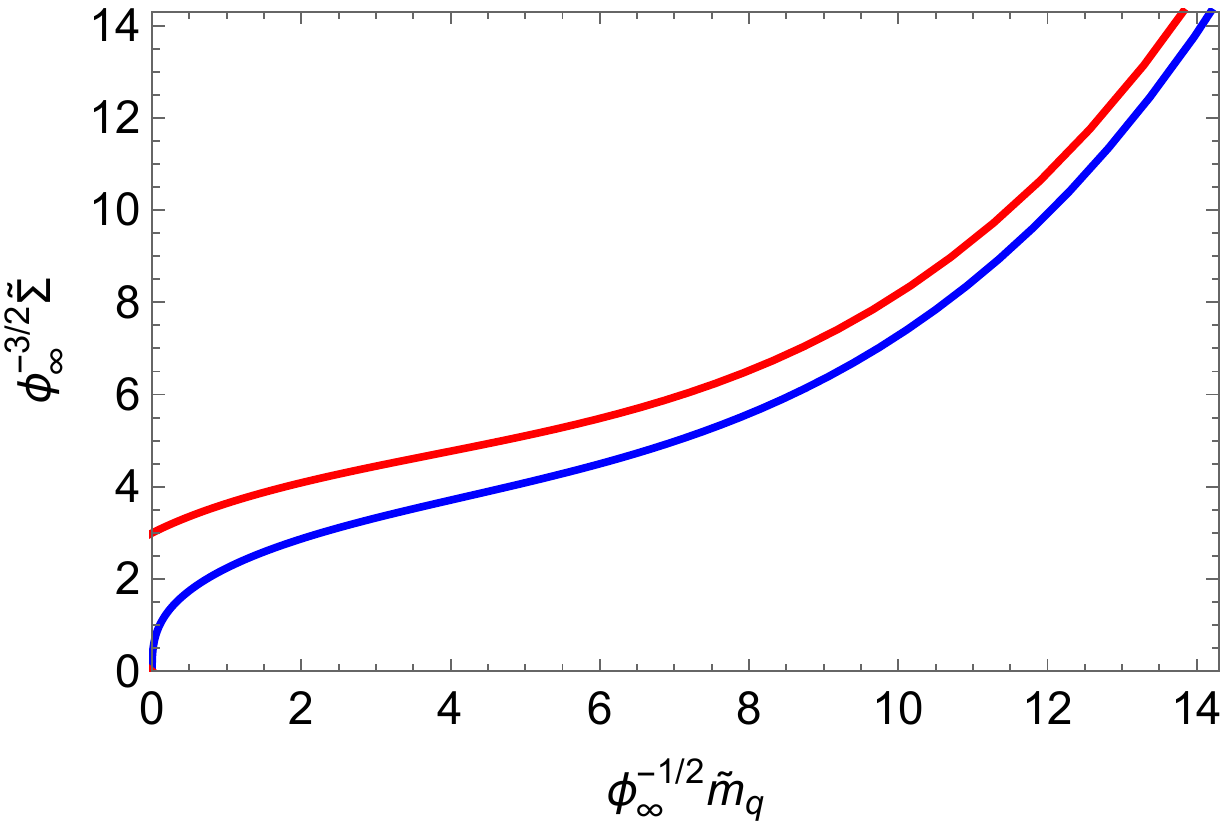}
\caption{
Left: The dimensionless condensate $\phi_{\infty}^{-3/2} \widetilde{\Sigma}$ as a function of the dimensionless quark mass $\phi_{\infty}^{-1/2}\widetilde{m}_q$ for models of type I, blue line are the results for $a_0=2.8$, while red line for $a_0=3.5$.
Right: The dimensionless condensate $\phi_{\infty}^{-3/2} \widetilde{\Sigma}$ as a function of the dimensionless quark mass $\phi_{\infty}^{-1/2}\widetilde{m}_q$ for models of type II, blue line are the results for $a_0= 5.6$, while red line for $a_0= 6.5$.
}
\label{Fig:mqc3}
\end{figure}

Here we show results for the rescaled chiral condensate $\widetilde{\Sigma}$ as a function of the rescaled quark mass $\widetilde{m}_q$,
obtained from the relations $\widetilde{c}_3= \widetilde{\Sigma}/(2\zeta)$ and $\widetilde{c}_1=\widetilde{m}_q$, in units of $\phi_{\infty}$ in the physical region $\widetilde{m}_q>0$. Our numerical results for models of type I are displayed in the left panel of Fig.~\ref{Fig:mqc3}, where the blue line represents the results for $a_0=2.8$, and the red line the results for $a_0=3.5$. As can be seen, the chiral condensate is nonzero in the chiral limit for $a_0>a_{0_c}\approx 2.970$ (red line). It is worth pointing out that the chiral condensate becomes negative in the intermediate region. In turn, our numerical results for models of type II are displayed in the right panel of Fig.~\ref{Fig:mqc3}. As can be seen, the chiral condensate becomes a monotonic increasing function of the quark mass. Analogous to models of type I, we have nonzero chiral condensate in the chiral limit for $a_0>a_{0_c}\approx 5.60$ (red line).

For a detailed discussion of the numerical procedure and to see the symmetric behavior of the tachyon see Appendix \ref{Sec:NumericalAnalysis}. 

\subsection{Fixing the model parameters}
\label{Subsec:Parameters}

To finish this section we are going to fix the parameters of the model. The dilaton parameter $\phi_{\infty}$ is fixed to  $\phi_{\infty}=(0.388\,\text{GeV})^2$, as in the soft wall model \cite{Herzog:2006ra}. This leads to a good description of the meson spectrum in the vectorial sector. As described in subsection \ref{Subsec:SpontXSB}, models of type I (type II) lead to spontaneous chiral symmetry breaking when $a_0 \geq 2.97$ ($a_0 \geq 5.6$). For models of type I (type II) we choose $a_0=3.5$ ($a_0=6.5$),  corresponding to $\phi_{\infty}^{-3/2}\widetilde{\Sigma}/(2\zeta)=8.8988$ ($\phi_{\infty}^{-3/2}\widetilde{\Sigma}/(2\zeta)=5.4367$), see Fig.~\ref{Fig:PhiC3}.  In the next section we will see that these parameter choices allows us to obtain good results for the meson spectrum in the scalar sector. 

Considering the relation between the parameter $\widetilde{c}_3$ and the rescaled chiral condensate, $\widetilde{c}_3=\widetilde{\Sigma}/(2\zeta)$ \cite{Ballon-Bayona:2020qpq}, where $\zeta=\sqrt{N_c}/(2\pi)$ \cite{Cherman:2008eh}, and recalling the definition $\widetilde{\Sigma}= \sqrt{\lambda} \langle \bar q q \rangle$, we obtain the relation
\noindent
\begin{equation}
\phi_{\infty}^{-3/2}\frac{\sqrt{\lambda} \langle \bar q q \rangle}{2\zeta}=\phi_{\infty}^{-3/2}\tilde{c}_3.
\end{equation}
\noindent
For fixed $\phi_{\infty}$ and $a_0$, different choices of the parameter $\lambda$ lead to different values for the chiral condensate $\langle \bar q q \rangle$. We fix the parameter $\lambda$ separately for each type of model in order to obtain good results for the meson spectrum in the axial-vector and pseudo-scalar sector; namely in order to avoid crossings between the fundamental and excited states in the axial and pseudo-scalar sector $\lambda$ has to be above some critical value that depends of each model type. 

For the models of type IA, IB, IIA and IIB we choose $\lambda=160$, $\lambda=380$, $\lambda=60$ and $\lambda=413$ respectively. These choices of parameters lead to $\langle \bar q q \rangle =(0.245\,\text{GeV})^3$ for model IA, $\langle \bar q q \rangle =(0.283\,\text{GeV})^3$ for models IB and IIA and  $\langle \bar q q \rangle =(0.205\,\text{GeV})^3$ for model IIB.  These results are of the same order as those obtained in lattice QCD, see for instance \cite{Shuryak:1988ck, Giusti:1998wy, Fukaya:2009fh,McNeile:2012xh}.

Table \ref{tab:Parameters} summarizes our choice of parameters for the four different type of models. Finally, the only remaining parameter in our models is the (current) quark mass $m_q$. In the next section we will initially treat $m_q$ as a free parameter so that we can investigate the evolution from the chiral limit $m_q=0$ to the heavy quark regime (large $m_q$). For the scalar, axial-vector and pseudo-scalar mesons it will be necessary to fix $m_q$ close to the physical quark mass in order to compare our results for the masses and decay constants against experimental data and other results available in the literature.

\begin{table}[ht]
\centering
\begin{tabular}{l |c|c|c|c}
\hline 
\hline
 Parameter & Model IA & Model IB & Model IIA & Model IIB \\
\hline 
 $a_0$ & $3.5$  & $3.5$  & $6.5$ & $6.5$ \\
 $\phi_{\infty}$ & $(0.388\,\text{GeV})^2$  &  $(0.388\,\text{GeV})^2$  & $(0.388\,\text{GeV})^2$ & $(0.388\,\text{GeV})^2$ \\
$\lambda$ & $160$  & $380$  & $60$ & $413$ \\
\hline
\hline
\end{tabular}
\caption{
Parameters in the four different models fixed in order to reproduce the meson spectrum.
}
\label{tab:Parameters}
\end{table}

\section{Meson spectrum}
\label{Sec:Spectrum}

In this section we calculate the masses of mesons in the vectorial, scalar, axial and pseudo-scalar sector. First we obtain the 5d differential equations for the  tachyonic and gauge field perturbations. Decomposing these perturbations into irreducible representations of the Lorentz group we identify the  normalizable solutions with the wave functions representing mesons in the 4d field theory. In the vectorial, scalar and axial sector we recast the differential equations in a Schr\"odinger form and find the squared masses as the energy eigenvalues in the Schr\"odinger problem. For the pseudo-scalar mesons the spectrum wil be obtained by solving directly the coupled differential equations satisfied by the normalizable modes. 

\subsection{Expanding the action} 
\label{SubSec:ExpAction}

Since the background gauge fields are zero we can expand trivially around it and consider the fluctuations $A_{m}^{(L/R)}$. It is convenient to rewrite those fluctuations in terms of vectorial and axial gauge fields 
\begin{equation}
A_{m}^{(L/R)} = V_m \pm A_m  \,. \label{gaugefluct}
\end{equation}
On the other hand, the tachyon field has a non-trivial background solution and we can expand around it as 
\begin{equation}
2 X = e^{2 i \pi} (v + S)  \, , \label{Tachyonfluct}     
\end{equation}
where $\pi = \pi^c T^c$  and $S$ is a real scalar field. Note that the fields $\pi^c$ transform as a triplet in the adjoint representation of $SU(2)$. They will give rise to a description of 4d pions whereas the field $S$ will correspond to scalar mesons. 
We expand the action \eqref{5dmodel} using \eqref{gaugefluct} and \eqref{Tachyonfluct}. The expansion can be written as 
\begin{equation}
S = S_0 + S_1 + S_2 + \dots \, ,
\end{equation}
where
\begin{align}
S_0 &= - \int d^4 x \int dz \,   e^{3 A_s -a}  \Big [  \frac12 (\partial_z v )^2 + e^{2A_s} U(v) \Big ]  \, , \nonumber \\
S_1&= - \int d^4 x \int dz  \,   e^{3 A_s - a}  \Big [ 
(\partial_z v) (\partial_z S)  + e^{2A_s} \frac{dU}{dv} S \Big ] \, ,\nonumber \\
S_2 &= - \int d^4 x \int dz \, \Big \{   e^{3 A_s -a}  \Big [ \frac12 (\partial_{\hat m} S )^2   + \frac12 e^{2 A_s} \frac{d^2 U}{dv^2}S^2 
 \Big ] \nonumber \\
&+ \frac{1}{g_5^2} e^{A_s - b} \,
\Big [ \frac14 {v_{\hat m \hat n}^c}^2 + \frac14 {a_{\hat m \hat n}^c}^2 + \frac{\beta}{2} (\partial_{\hat m} \pi^c - A_{\hat m}^c)^2 \Big ] \Big \} .\label{5dmodelv2}
\end{align}
where 
\begin{equation}\label{Eq:Beta}
\beta \equiv      e^{2A_s +b-a} g_5^2 v^2 \,. 
\end{equation}
and $v_{\hat m \hat n}^c = \partial_{\hat m} V_{\hat n}^c - \partial_{\hat n} V_{\hat m}^c$ (similarly for $a_{\hat m \hat n}^c$). The indices $(\hat m, \hat n)$ correspond to coordinates in the 5d flat metric $\eta_{\hat m \hat n}$ and the index $c=(1,2,3)$ corresponds to the adjoint representation of the $SU(2)$ isospin group. The actions $S_0$ and $S_1$ become surface terms after replacing the tachyon field equation in \eqref{veq}. 

The action $S_2$ is responsible for the dynamics of the field fluctuations $S$, $V^{\hat m,c}$, $A^{\hat m,c}$ and $\pi^c$. The variation of $S_2$ can be written as 
\begin{equation}
\delta S_2 = \delta S_2^{Bulk} + \delta S_2^{Bdy} \,.
\end{equation}
The bulk term $\delta S_2^{Bulk}$ can be written as
\begin{align}
\delta S_2^{Bulk} &= \int d^4 x \int dz \Big \{  e^{3 A_s -a} \delta S \Big [ e^{a - 3 A_s} \partial_{\hat m} ( e^{3A_s -a} \partial^{\hat m} S) - e^{2A_s} \frac{d^2 U}{dv^2} S \Big ] \nonumber \\
&+ \frac{1}{g_5^2} \delta V_{\hat n}^c  \partial_{\hat m} ( e^{A_s -b} v^{\hat m \hat n}_c ) + \frac{1}{g_5^2} e^{A_s -b} \delta A_{\hat n}^c \Big [ e^{b - A_s} \partial_{\hat m} (e^{A_s -b} a^{\hat m \hat n}_c )  \nonumber \\
&+ \beta (\partial^{\hat m} \pi^c - A^{\hat m,c}) \Big ] 
+ \frac{1}{g_5^2} \delta \pi^c \partial_{\hat m} \Big [ e^{A_s - b} 
\beta (\partial^{\hat m} \pi^c - A^{\hat m,c}) \Big ] \Big \} \,.
\end{align}

The boundary term $\delta S_2^{Bdy}$ can be written as
\begin{align}
\delta S_2^{Bdy} &= - \int d^4 x \int dz \, \partial_{\hat m} \Big \{ 
e^{3 A_s - a} (\partial^{\hat m} S) \delta S 
+ \frac{1}{g_5^2} e^{A_s - b} v^{\hat m \hat n}_c \delta V_{\hat n}^c
 \nonumber \\
&+ \frac{1}{g_5^2} e^{A_s - b} a^{\hat m \hat n}_c \delta A_{\hat n}^c 
+ \frac{1}{g_5^2} \beta e^{A_s -b} (\partial^{\hat m} \pi^c -  A^{\hat m,c}) \delta \pi^c \Big \} \, .
\end{align}
Imposing periodic boundary conditions in the $x^{\mu}$ coordinates we end up with a boundary term in the $z$ coordinate: 
\begin{align}
\delta S_2^{Bdy} &= - \int d^4 x \Big \{ e^{3 A_s -a} (\partial_z S) \delta S 
+ \frac{1}{g_5^2} e^{A_s -b} v^{\hat z \hat \mu}_c \delta V_{\hat \mu}^c \nonumber \\
&+ \frac{1}{g_5^2} e^{A_s -b} a^{\hat z \hat \mu}_c \delta A_{\hat \mu}^c 
+ \frac{1}{g_5^2} \beta e^{A_s -b} (\partial^{\hat z} \pi^c -  A^{\hat z,c}) \delta \pi^c \Big \}_{z=\epsilon}^{z \to \infty} \,. 
\label{deltaS2bdy}
\end{align}
The quadratic behaviour $\Phi(z) \sim z^2$ at large $z$ implies that the terms with couplings $\exp{(-a)}$ and $\exp{(-b)}$ vanish exponentially in the limit $z \to \infty$. Imposing also Dirichlet boundary conditions at $z= \epsilon$ one guarantees the vanishing of \eqref{deltaS2bdy}. 

The vanishing of $\delta S_2$ then implies the condition $\delta S_2^{Bulk}=0$ for the bulk term. This leads to the field equations
\begin{align}
& e^{a - 3 A_s} \partial_{\hat m} \left ( e^{3 A_s - a} \partial^{\hat m} S \right )
- e^{2 A_s} \frac{d^2 U}{dv^2} S = 0 \, ,\nonumber \\
& \partial_{\hat m} \Big [ e^{A_s - b} \beta (\partial^{\hat m} \pi^a - A^{\hat m,a} ) \Big ] = 0 \, , \nonumber \\
& \partial_{\hat m} \left ( e^{A_s - b} v_a^{\hat m \hat n}   \right ) = 0  \, , \nonumber \\
& e^{b- A_s} \partial_{\hat m} \left ( e^{A_s -b} a_a^{\hat m \hat n} \right )
+ \beta (\partial^{\hat n} \pi - A^{a, \hat n} ) = 0 \, . \label{ELeqsflucts}
\end{align}

Note that the action and field equations are invariant under the vectorial gauge symmetry 
\begin{equation}
V^{\hat m,c} \to V^{\hat m,c} - \partial^{\hat m} \lambda_V^c \, ,
\label{vecsym}
\end{equation}
and the (residual) axial gauge symmetry 
\begin{equation}
A^{\hat m,c} \to A^{\hat m,c} - \partial^{\hat m} \lambda_A^c 
\quad , \quad  \pi^c \to \pi^c - \lambda_A^c \, . \label{axialsym}
\end{equation}
The gauge fields can be decomposed as $V_{\hat m} = (V_z , V_{\hat \mu})$ , $A_{\hat m} = (A_z , A_{\hat \mu})$ and also the derivatives $\partial_{\hat m} = (\partial_z , \partial_{\hat \mu})$.
The vectorial gauge symmetry in \eqref{vecsym} allows us to set $V_z=0$ whilst the residual axial gauge symmetry in \eqref{axialsym} can be used to set $A_z=0$.  The 4d vectors $V_{\hat \mu}$ and $A_{\hat \mu}$ admit also the Lorentz decomposition
\begin{equation}
V_{\hat \mu, c} = V_{\hat \mu,c}^{\perp} + \partial_{\hat \mu} \xi^c \, \quad , \quad 
A_{\hat \mu , c} = A_{\hat \mu,c}^{\perp} + \partial_{\hat \mu} \varphi^c \,. 
\end{equation}
The scalar fields $\xi^c$ are not dynamical and can be set to zero. Using these results the 5d  equations \eqref{ELeqsflucts} become the set of independent field equations 
\begin{align}
& \Big [ \partial_z + 3 A_s' - a' \Big ] \partial_z S 
+ \Box S - e^{2 A_s} \frac{d^2 U}{dv^2} S = 0  \quad 
({\rm scalar \, sector}) \, , \label{Seq} \\ 
& \Big [ \partial_z + A_s' - b' \Big ] \partial_z V^{\hat \mu,c}_{\perp} + \Box   V^{\hat \mu,c}_{\perp}   
 =0 \quad ({\rm vectorial \, sector}) \, , \label{Veq}  \\
& \Big [ \partial_z + A_s' - b' \Big ] \partial_z A^{\hat \mu,c}_{\perp} + \Box A^{\hat \mu,c}_{\perp} -  \beta A^{\hat \mu,c}_{\perp} = 0 \quad ({\rm axial \, sector}) \, , \label{Aeq}  \\
&\Big [ \partial_z + A_s' - b' \Big ] \partial_z  \varphi^c +   \beta (  \pi^c  -  \varphi^c )  = 0 \, , \nonumber \\
& - \partial_z \Box \varphi^c + \beta  \, \partial_z \pi^c  = 0 
\quad (\text{pseudo-scalar sector}) \,, \label{pieq}
\end{align}

Note that the coupled differential equations in \eqref{pieq} for the pseudo-scalar sector can be combined into the single differential equation
\begin{eqnarray}
\partial_z \Big [ \partial_z + A_s' - b' + (\ln \beta)' \Big ] \Pi^c + (\Box  - \beta )\Pi^c = 0 \, ,  \label{Pieq}
\end{eqnarray}
where we have defined $\Pi^c \equiv \partial_z \pi^c$.

\subsection{Vector mesons}
\label{Subsec:VM}

Let us start with the equation of motion describing the vector mesons, i.e., Eq.~\eqref{Veq}. Considering the Fourier transform on $V_{\mu}$, $V_{\mu}(x^{\mu},z)\to V_{\mu}(k^{\mu},z)$, then, setting $\square\to m_V^2$, the resulting equation may be rewritten in the Schrödinger-like form through the transformation $V_{\mu}= \eta_{\mu} e^{-B_{V}} \psi_{v_n}$, where $2B_{V}=A_s-b$ and $\eta_{\mu}$ is a (transverse) polarisation vector. The Schrödinger-like equation reads 
\noindent
\begin{equation}\label{Eq:SchrodingerVector}
-\partial^2_z\,\psi_{v_n} +V_{V}\,\psi_{v_n} =m_{V}^2\, \psi_{v_n},
\end{equation}
\noindent
where the potential is given by
\begin{equation}\label{Eq:VectorPot}
V_{V}=\left(\partial_z B_{V}\right)^2+\partial^2_{z} B_{V}.
\end{equation}
\noindent

\begin{figure}[ht!]
\centering
\includegraphics[width=7cm]{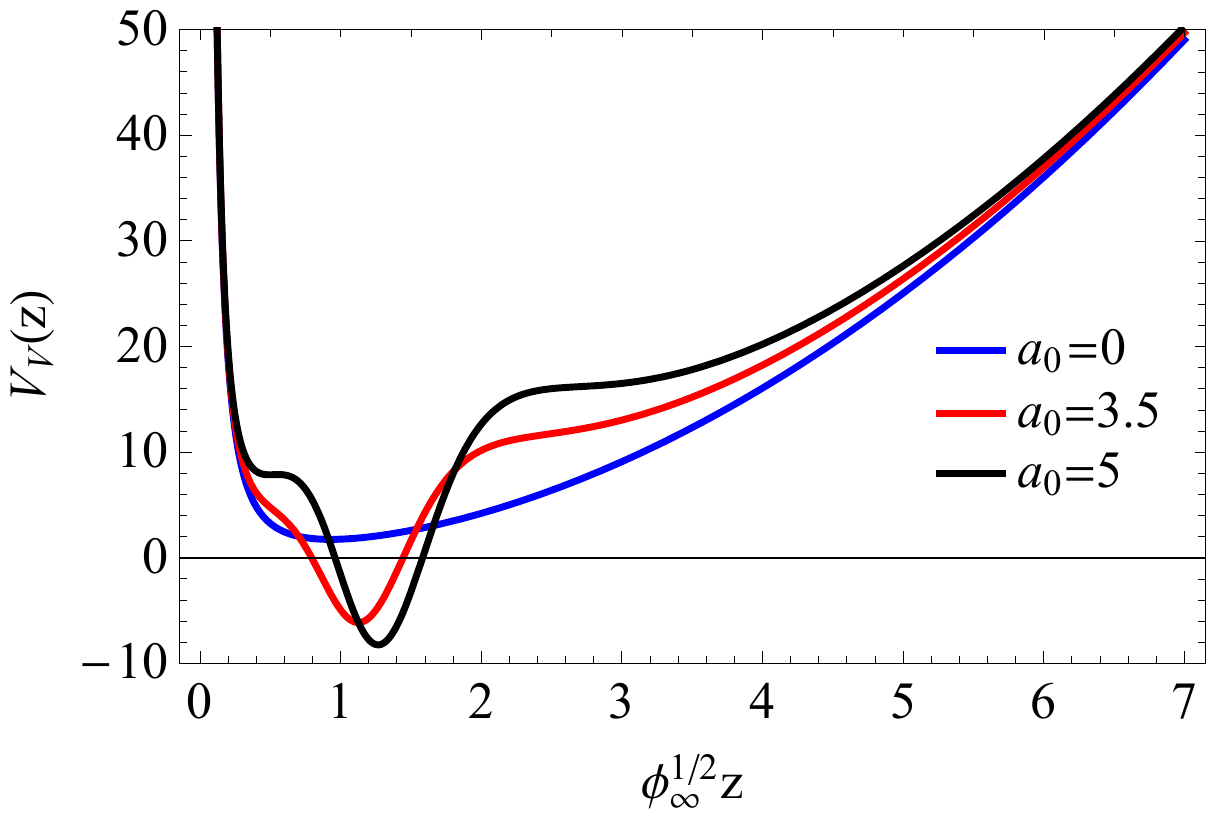}\hfill
\includegraphics[width=7cm]{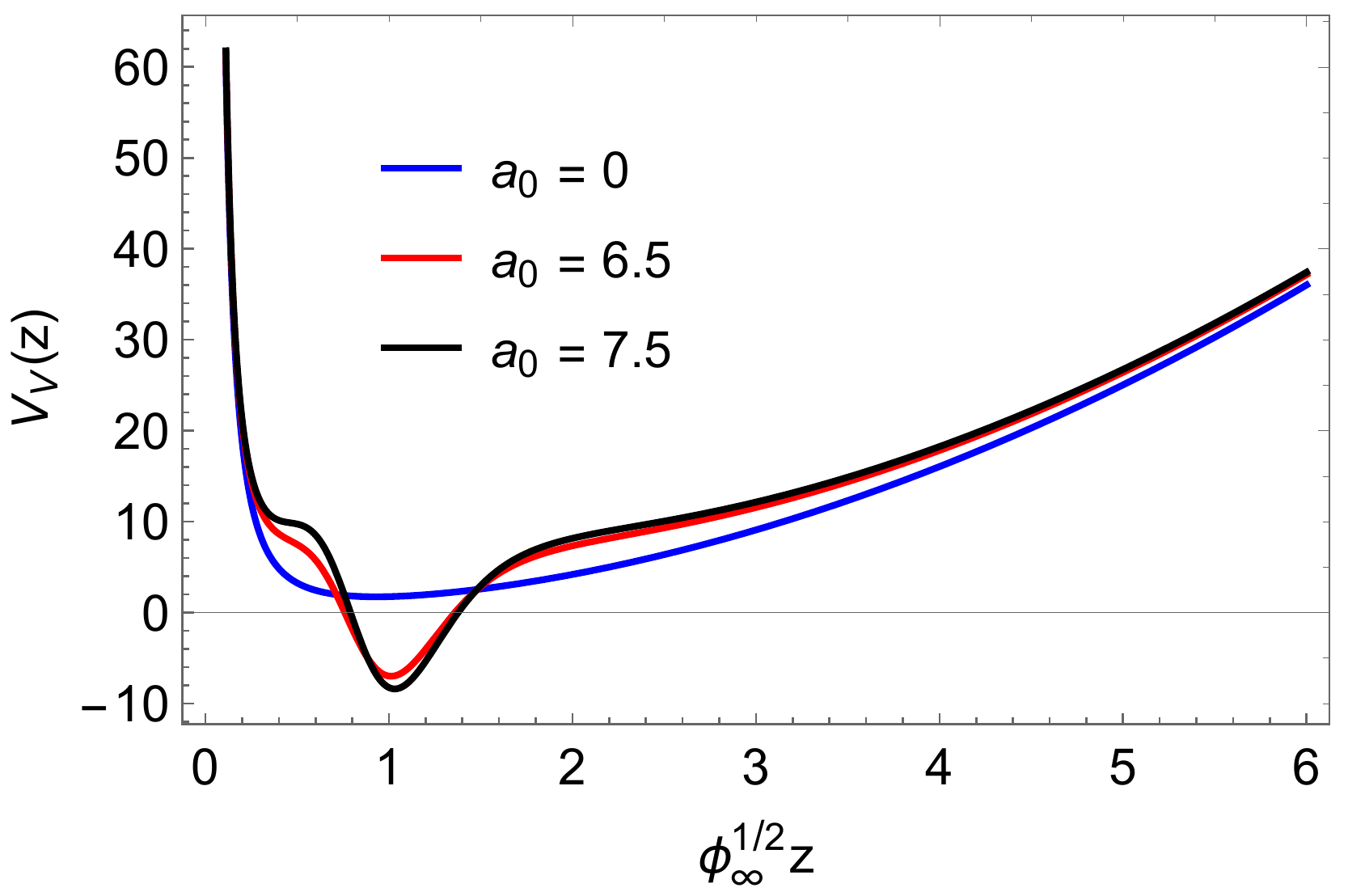}
\caption{
Left: The potential of the Schr\"odinger-like equation for selected values of $a_0$ for models of type I. As can be seen, increasing $a_0$ the potential becomes negative. Right: The same as left panel for models of type II
}
\label{Fig:VectorPotential}
\end{figure}

\begin{figure}[ht!]
\centering
\includegraphics[width=7cm]{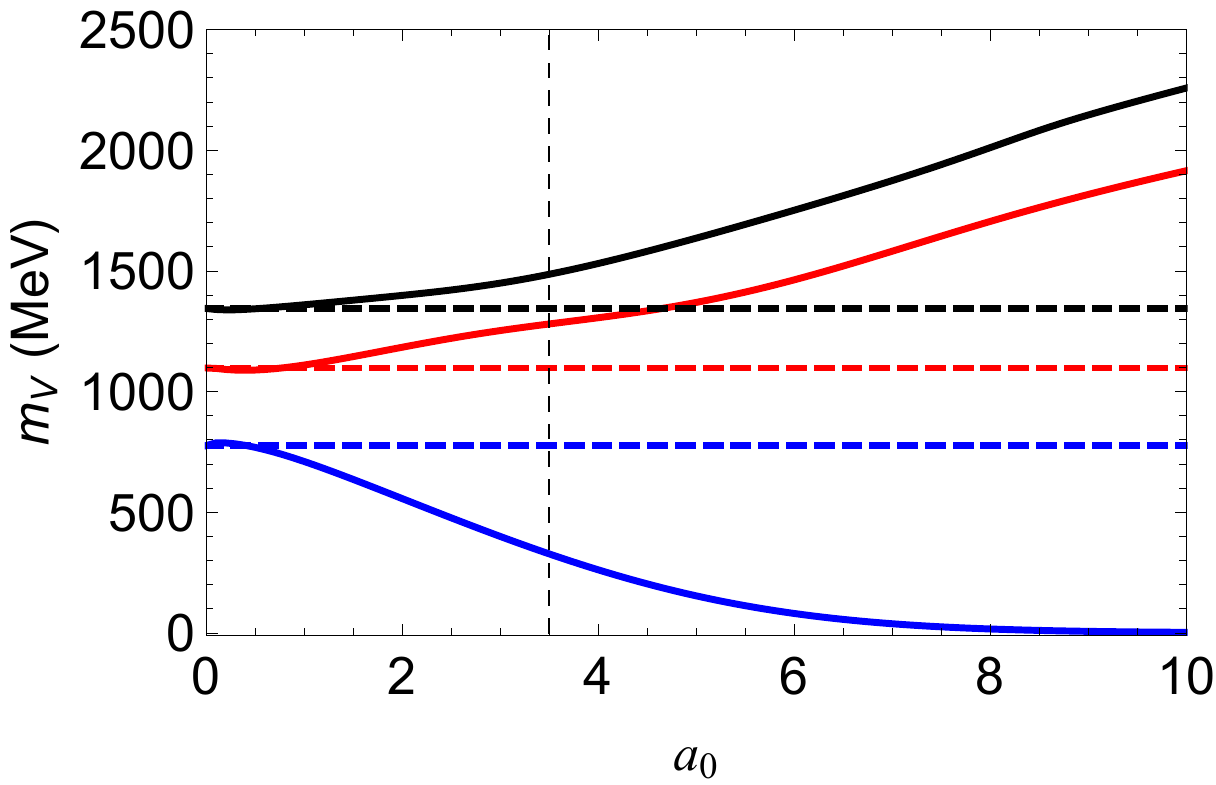}\hfill
\includegraphics[width=7cm]{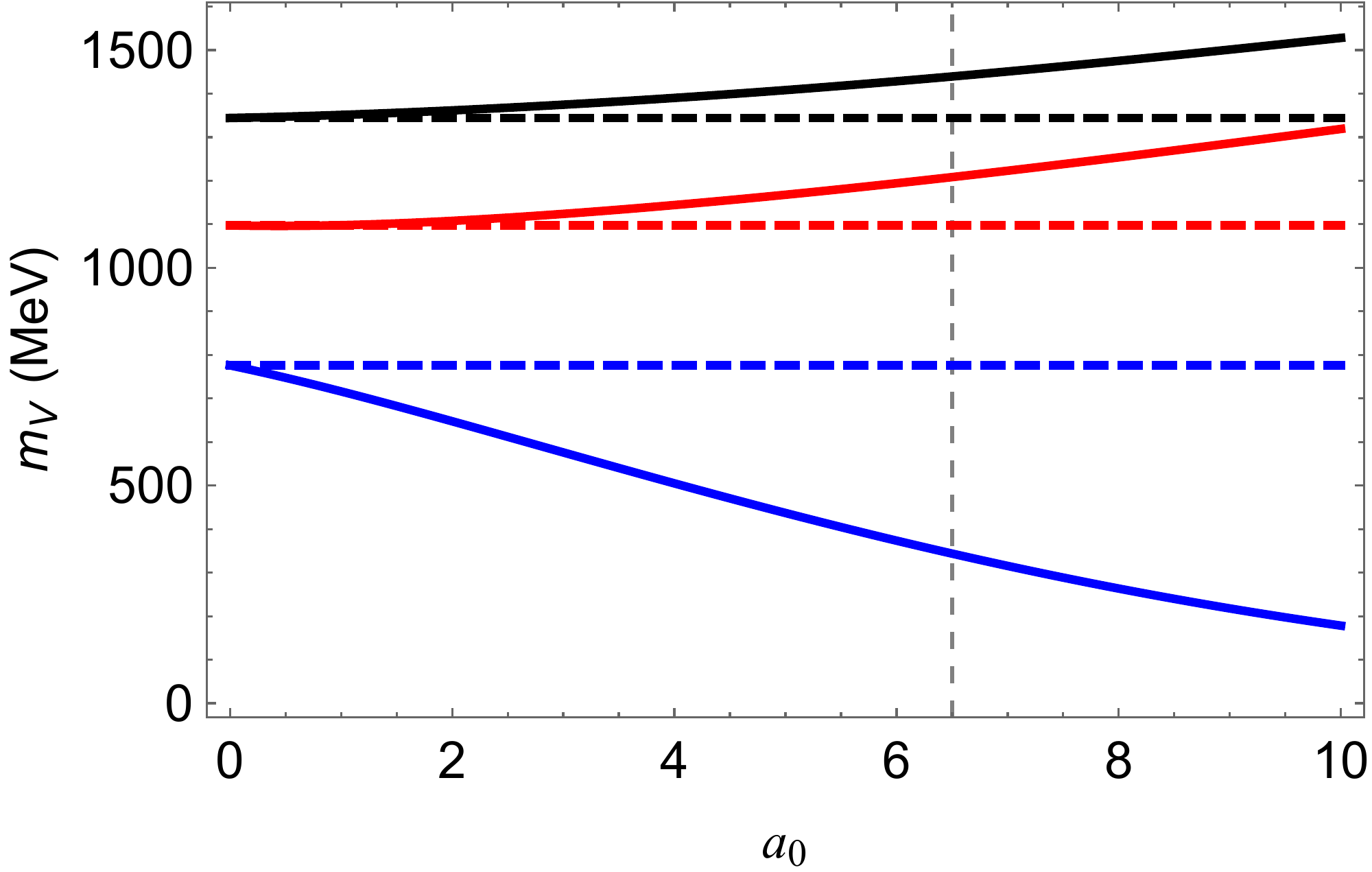}
\caption{
Left: Masses of vector mesons $M_V$ as functions of the parameter $a_0$ for models of type I, solid lines represent the results for $a=b$  (model IA), while dashed lines for $a\neq b$ (model IB) . Right: Masses of vector mesons $M_V$ functions of the parameter $a_0$ for models of type II. Solid lines represent the results for $a=b$ (model IIA), while dashed lines for $a\neq b$ (model IIB).
}
\label{Fig:PhiMvDil2}
\end{figure}

Firstly, let us consider Model IA where $b=a$. In this case the solution of the eigenvalue problem is obtained numerically. A plot of the potential \eqref{Eq:VectorPot} for different values of $a_0$ is displayed on the left panel of Fig.~\ref{Fig:VectorPotential}. In turn, our numerical results for the masses as a function of $a_0$ are displayed on the left panel of Fig.~\ref{Fig:PhiMvDil2} with solid lines. As can be seen, the masses are sensitive to the parameter $a_0$, special attention requires the mass of the ground state which decreases with the increasing of $a_0$, while the masses of the other states increase with the increasing of $a_0$. It is worth mentioning that in the limit of zero $a_0$, the results reduce to those obtained in the linear soft wall model displayed with dashed lines \cite{Karch:2006pv,Gherghetta:2009ac, Li:2013oda, Ballon-Bayona:2020qpq}. 
Note that the coupling $a(\Phi)$ reduces to $a(\Phi) = - a_0 \Phi$ for models of type I  in the limit of large $a_0$. This corresponds to an effetive negative dilaton profile. As explained in Ref.~\cite{Karch:2010eg}, the potential of the Schr\"odinger-like equation is insensitive to the signal of the dilaton field, meaning that the spectrum will not change in relation to the original soft wall model. However, as shown in Ref.~\cite{Karch:2010eg}, the longitudinal fluctuation of the field $V_{\mu}$ will give rise to a massless state. In turn, in the models of type I a massless mode arises naturally by solving the Schr\"odinger-like equation in the region of large $a_0$. The parameter $a_0$ controls the minimum of the potential well, see the left panel of Fig.~\ref{Fig:VectorPotential}, opening the possibility to the emergence of this state as can seen on the left panel of Fig.~\ref{Fig:PhiMvDil2}. Regardless the origin of the massless mode is, it is a non-physical state and must be avoided. In the following analysis we choose $a_0=3.5$, such that the ground state is not so light. 

Now consider models of type IB where $a\neq b$, with $b=\Phi=\phi_{\infty}\,z^2$, and $a$ is given by \eqref{Eq:Dilaton1}. Note that $b$ is the dilaton field of the linear soft wall model. Hence, in this case the problem has the exact solution \cite{Karch:2006pv}
\noindent
\begin{equation}
m^{2}_{V}=4\phi_{\infty}(1+n),\qquad n=0,1,2,\cdots
\end{equation}
\noindent
The free parameter $\phi_{\infty}$ may be fixed by comparing the first state with the corresponding experimental result of the $\rho$ meson \cite{Herzog:2006ra}, providing us with the value $\phi_{\infty}=(388\,\text{MeV})^2$. In Model IB the potential corresponds to the case $a_0=0$ in the left panel of Fig.~\ref{Fig:VectorPotential}, while the masses as functions of $a_0$ are displayed  in the left panel of Fig.~\ref{Fig:PhiMvDil2} with dashed lines.

The results obtained above for models of type I can be extended for models of type II, which is described by Eq.~\eqref{Eq:Dilaton2}. For Model IIA, where $a=b$, the potential is displayed on the right panel of Fig.~\ref{Fig:VectorPotential}, while the masses as  functions of $a_0$ are displayed on the right panel of Fig.~\ref{Fig:PhiMvDil2}. The main difference between models of type IIA with respect to models of type IA is that the mass of the ground states decreases slowly with $a_0$. Note also that the results for Model IIB ($b=\Phi=\phi_{\infty}z^2$) are equivalent to results for Model IB.

Finally, considering $a_0=3.5$ for models of type I, and $a_0=6.5$ for models of type II we calculated the mass of the vector mesons. We display our results  and compare them against some results available in the literature in Table~\ref{Taba:VectorSF}. As can be seen from the table, excluding the ground state, our results for all models are close to the experimental results available in particle data group (PDG) Ref.~\cite{Zyla:2020zbs}. Note also that the results for the ground state in  Models IB and IIB are the same are closer to the experimental result. This is because they are equivalent to the original soft wall model and the parameter $\phi_{\infty}$ was fixed in that model in order to match the experimental result for the $\rho$ meson mass.

\begin{table}[ht]
\centering
\begin{tabular}{l |c|c|c|c|c|c}
\hline 
\hline
 $n$ & Model IA & Model IB & Model IIA & Model IIB & GKK \cite{Gherghetta:2009ac} & $\rho$ experimental \cite{Zyla:2020zbs} \\
  & ($a=b$) & ($a\neq b$) & ($a=b$) & ($a\neq b$) &  &  \\
\hline 
 $0$ & 327  & 776  & 344 &776 & 475  & $776\pm 1$  \\
 $1$ & 1280 & 1097 & 1208 &1097 & 1129 & $1282\pm 37$  \\
 $2$ & 1486 & 1344 & 1439 &1344 & 1429 & $1465\pm 25$ \\
 $3$ & 1662 & 1552 & 1632 &1552 & 1674 & $1720\pm 20$  \\
 $4$ & 1823 & 1735 & 1802 &1735 & 1884 & $1909\pm 30$  \\
 $5$ & 2116 & 1901 & 1958 &1901 & 2072 & $2149\pm 17$ \\
 $6$ & 2250 & 2053 & 2104 &2053 & 2243 & $2265\pm 40$ \\
\hline\hline
\end{tabular}
\caption{
Masses of vector mesons (in MeV) obtained in models of type I for $a_0=3.5$, models of type II for $a_0=6.5$, compared against the holographic models  \cite{Gherghetta:2009ac} and experimental results from PDG \cite{Zyla:2020zbs}.
}
\label{Taba:VectorSF}
\end{table}

\subsection{Scalar mesons}
\label{Sec:ScalarL}

Let us move on to the scalar mesons which are described by Eq.~\eqref{Seq}. Considering the Fourier transform on $S(x^{\mu},z)$, $S(x^{\mu},z)\to S(k^{\mu},z)$, then, replacing $\square\to m_s^2$, and implementing the transformation $S=e^{-B_S}\psi_{s_{n}}(z)$, where $2B_S=3A_s-a$, the equation \eqref{Seq} may be rewritten in the Schr\"odinger-like form
\noindent
\begin{equation}\label{Eq:SchroScalarEq}
-\partial_z^2 \psi_{s_{n}}+V_S\,\psi_{s_{n}}=m_s^2\,\psi_{s_{n}},
\end{equation}
\noindent
where the potential is given by
\noindent
\begin{equation}\label{Eq:Schro/potScalarL}
V_S=(\partial_z B_S)^2+\partial_z^2B_S
+e^{2A_s}\left(m_X^{2}+\frac{3\,\lambda}{2}v^2(z)\right).
\end{equation}
\noindent
Note that the scaling symmetry of the tachyon differential equation arises here due to the product $\lambda\,v^2$ in the potential. This means that the spectrum of scalar mesons depend only on $\widetilde{v}= \sqrt{\lambda} v$ which in turn is independent of $\lambda$ as long as $\lambda>0$. 

The masses of scalar mesons depend on the quark mass due to the presence of the tachyon field in the potential. For $\lambda=0$, the potential \eqref{Eq:Schro/potScalarL} reduces to results available in the literature for the linear soft wall model, see for instance Ref.~\cite{Vega:2008af} (see also \cite{Colangelo:2008us}). We find that, for fixed $a_0$, the masses of scalar mesons are monotonically increasing functions of the parameter $\tilde m_q = \sqrt{\lambda} m_q$. This means that for  fixed $m_q$ ($\lambda$) they are increasing functions of $\lambda$ ($m_q$). This is consistent with the results found in Ref.~\cite{Ballon-Bayona:2020qpq} for the case $a_0=0$ where the dilaton coupling becomes minimal.

\begin{figure}[ht!]
\centering
\includegraphics[width=7cm]{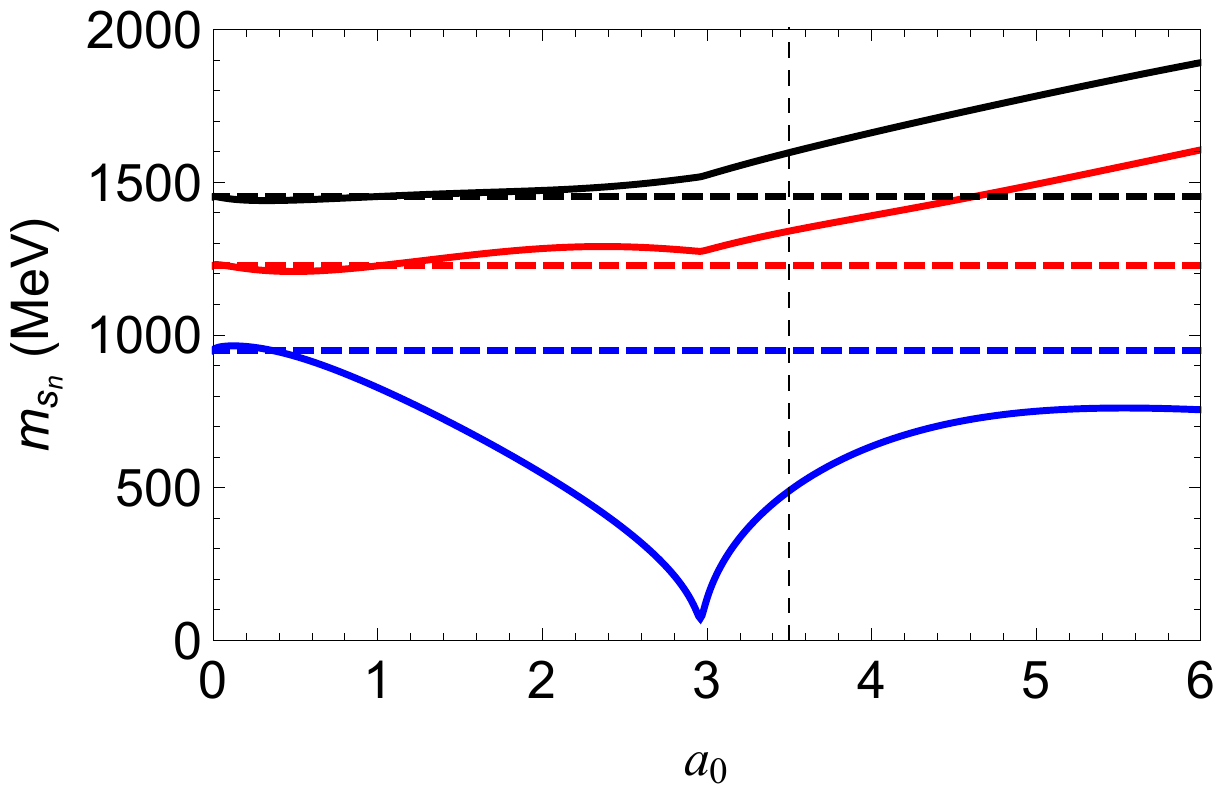}\hfill 
\includegraphics[width=7cm]{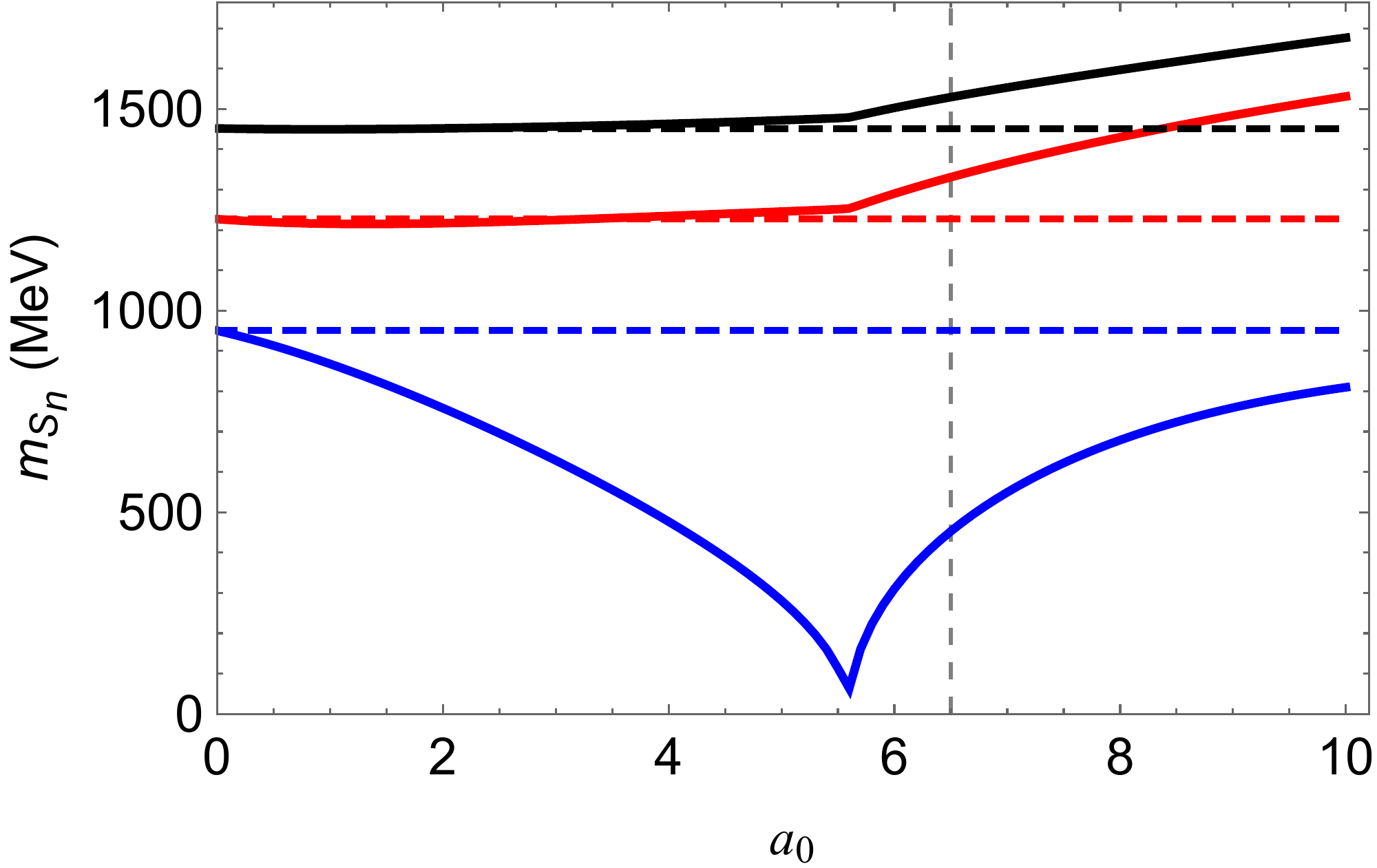}
\caption{
Left: Masses of scalar mesons as functions of the parameter $a_0$. Solid lines represent the results for Models IA and IB while dashed lines represent the results of the linear soft model. Right: Masses of scalar mesons as functions of $a_0$. Solid lines represent the results for Models IIA and IIB ($a\neq b$) while dashed lines represent the results of the linear soft model. These results were obtained setting $m_q=3.63\times 10^{-4}\,\text{MeV}$ (near the chiral limit).}
\label{Fig:Msa0}
\end{figure}

\begin{figure}[ht!]
\centering
\includegraphics[width=7cm]{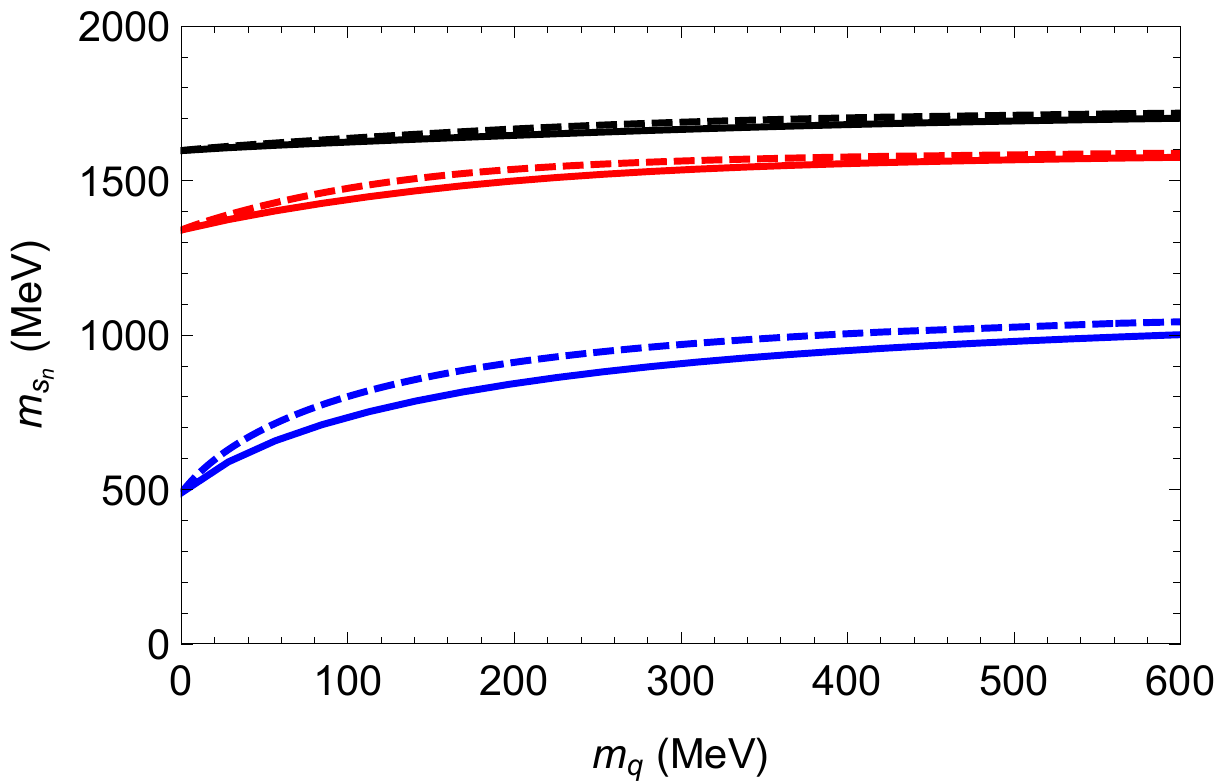}\hfill 
\includegraphics[width=7cm]{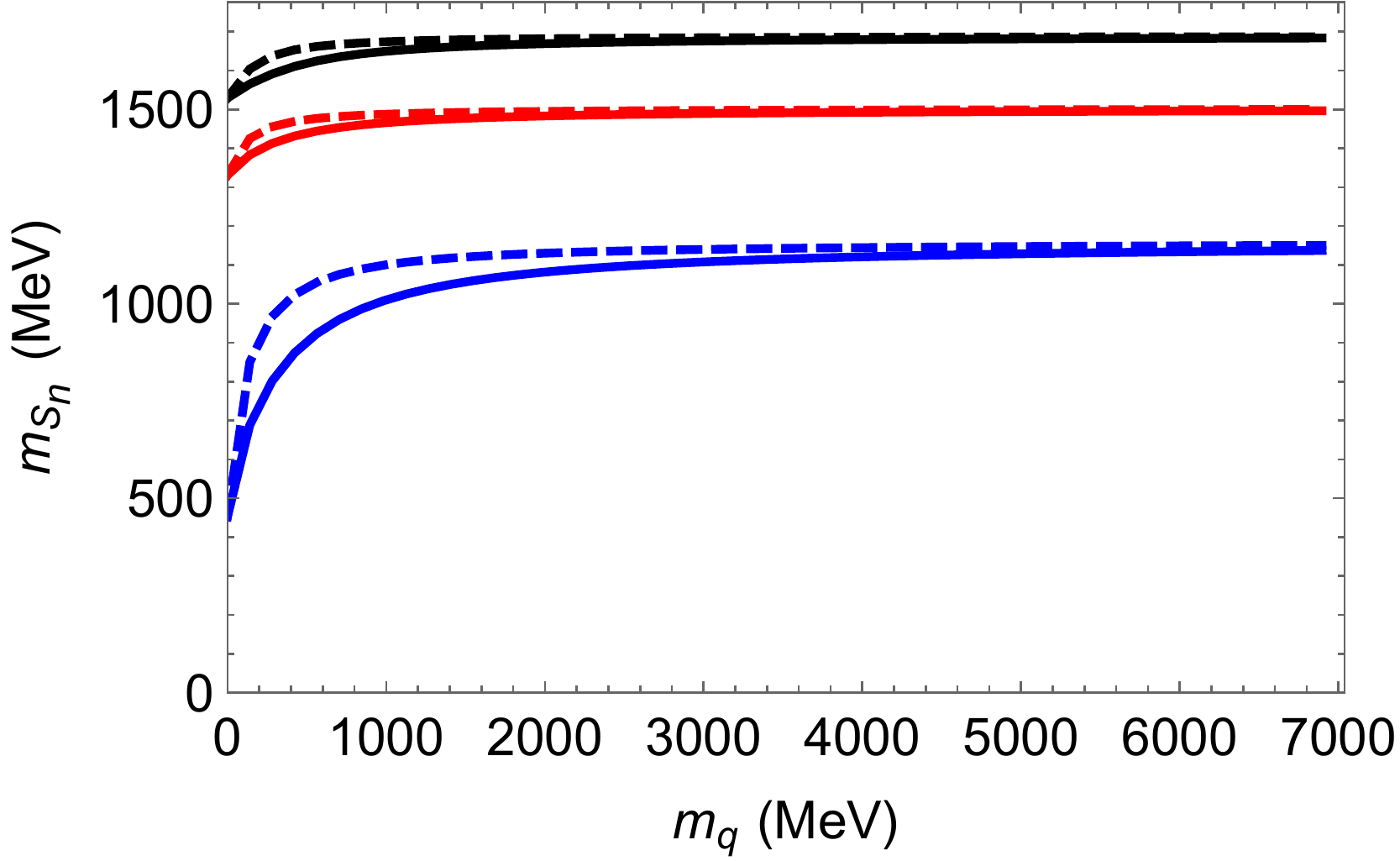}
\caption{
Left: Masses of scalar mesons as functions of the quark mass. Solid lines represent the results for Model IA ($a=b$) and  $\lambda=160$, while dashed lines for Model IB ($a\neq b$) and  $\lambda=380$, and we set $a_0=3.5$.
Right: Masses of scalar mesons as functions of the quark mass. Solid lines represent the results for Model IIA ($a=b$) and $\lambda=60$, while dashed lines for Model IIB ($a\neq b$) and  $\lambda=413$, and we set $a_0=6.5$.}
\label{Fig:MsDil2}
\end{figure}

A plot of the masses as functions of $a_0$ is displayed on the left panel of Fig.~\ref{Fig:Msa0} for $m_q=3.63\times 10^{-4}\,\text{MeV}$ (near the chiral limit), where solid lines are the results for Models IA and IB, while dashed lines represent the result of the linear soft wall model given by Eq.~\eqref{Eq:MassScalarMesonsLSW}. Note that the mass of the ground state decreases with the increasing of $a_0$ up to $a_{0_c}$, then, it increases with the increasing of $a_0$. Thus, to avoid any zero mode in the scalar sector of models of type I we set $a_0=3.5$, represented by the vertical dashed line. In turn, the numerical results of the masses as functions of the quark mass, for fixed $a_0$ and $\lambda$, are displayed on the left panel of Fig.~\ref{Fig:MsDil2}, where solid lines represent the results for Model IA , while dashed lines display the results for Model IB. As expected, for fixed $a_0$ and $\lambda$, the masses of scalar mesons increase with the increasing of the quark mass $m_q$. Note that the states obtained in Model IA are heavier than the states obtained in Model IB. 

The masses as functions of $a_0$ obtained in models of type II are displayed on the right panel of Fig.~\ref{Fig:Msa0}, for $m_q=3.63\times 10^{-4}\,\text{MeV}$ (near the chiral limit). The results for Model IIA are represented by solid lines whereas the results for Model IIB are represented by  dashed lines. Again, the mass of the ground state decreases with the increasing of $a_0$ up to $a_{0_c}$, then, it increases with $a_0$. This behavior motivated us to the choice $a_0=6.5$ to avoid the zero mode, as we did in models of type I. Finally, for fixed $a_0$ and $\lambda$, the masses as functions of the quark mass are displayed on the right panel of Fig.~\ref{Fig:MsDil2}, results for Model IIA are represented with solid line, while for Model IIB with dashed lines. As expected, the masses increase with the increasing of the quark mass. These results are in agreement with the results obtained in Ref.~\cite{Ballon-Bayona:2020qpq}, where the non-linear extension of the soft wall model was investigated.
We would like to remark we do not find instabilities in the physical region, i.e., $m_q>0$, as was previously reported in the spectrum of the scalar mesons in extensions of the linear soft wall model \cite{Li:2013oda} (see also \cite{Sui:2009xe, Ballon-Bayona:2020qpq}).

\begin{table}[tbp]
\centering
\begin{tabular}{l |c|c|c|c|c|l}
\hline 
\hline
 $n$ & Model IA & Model IB & Model IIA & Model IIB &
 BM \cite{Ballon-Bayona:2020qpq}&
$f_0$ experimental \cite{Zyla:2020zbs} \\
  & $(a=b)$ & $(a\neq b)$ & $(a=b)$ & $(a\neq b)$ &  & \\
\hline 
 $0$ & 526  & 519  & 539  &546  & 980   & $990\pm 20$  \\
 $1$ & 1351 & 1349 & 1348 &1350 & 1246  & $1350\pm 150$  \\
 $2$ & 1600 & 1599 & 1540 &1541 & 1466  & $1505\pm 6$ \\
 $3$ & 1755 & 1755 & 1718 &1719 & 1657  & $1724\pm 7$  \\
 $4$ & 1904 & 1904 & 1881 &1881 & 1829  & $1992\pm 16$  \\
 $5$ & 2048 & 2048 & 2032 &2032 & 1986  & $2103\pm 8$ \\
 $6$ & 2185 & 2185 & 2174 &2174 & 2132  & $2314\pm 25$ \\
 $7$ & 2315 & 2315 & 2313 &2313 & 2268  &  \\
\hline\hline
\end{tabular}
\caption{
Masses of scalar mesons (in MeV) obtained in the 
models of type I and models of type II, compared against the holographic model of Ref.~\cite{Ballon-Bayona:2020qpq}, and experimental data from PDG \cite{Zyla:2020zbs}. To get these results we have considered $\lambda=160$ for Model IA, and $\lambda=380$ for Model IB, while the quark mass is $m_q=9\,\text{MeV}$ and $m_q=4.7\,\text{MeV}$, respectively. For Model IIA we have considered $\lambda=60$, while the quark mass is $m_q=9.8\,\text{MeV}$, and for Model IIB $\lambda=413$ and $m_q=26.8\,\text{MeV}$.
}
\label{Tab:ScalarMesons}
\end{table}

 For the fixed set of parameters, described in Table \ref{tab:Parameters}, we may calculate the spectrum provided by the holographic models and compare the results against some results available in the literature. In Table \ref{Tab:ScalarMesons} we show our results for Model IA setting $m_q=9\,\text{MeV}$, Model IB setting $m_q=4.7\,\text{MeV}$, Model IIA setting $m_q=9.8\,\text{MeV}$, and Model IIB setting $m_q=26.8\,\text{MeV}$. Each particular choice of the quark masses will be justified below, and we will see that these values are in agreement with the pion mass in the pseudoscalar mesons.

\subsection{Axial-vector mesons}
\label{Sec:AV}

\begin{figure}[ht!]
\centering
\includegraphics[width=7cm]{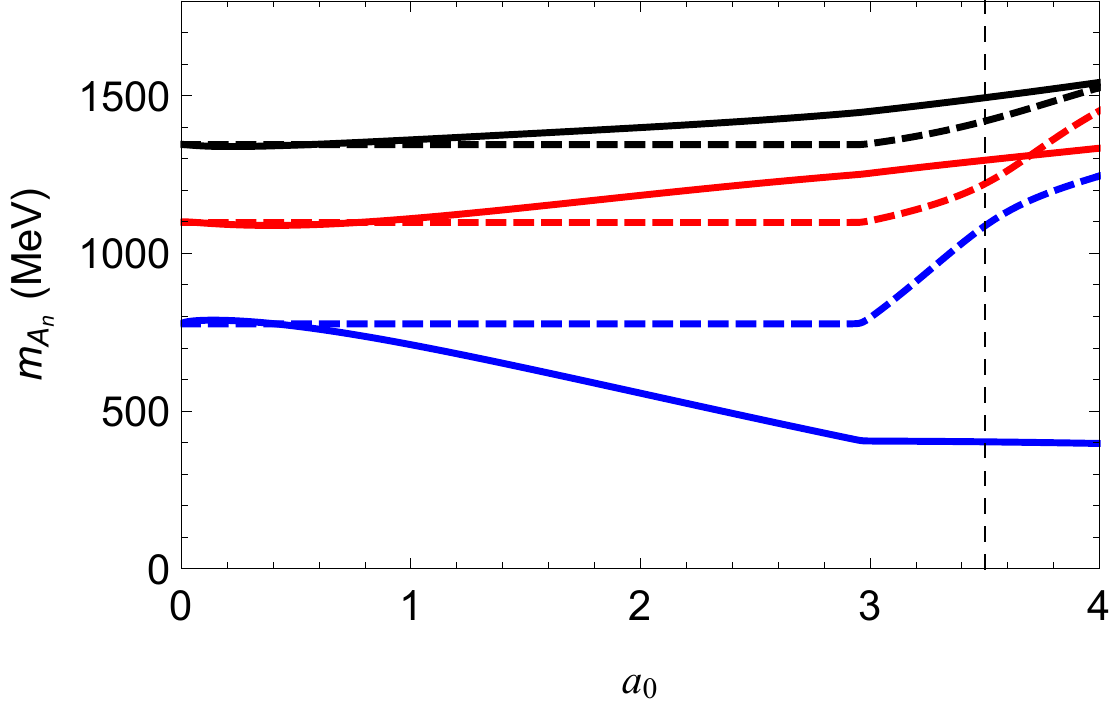}\hfill 
\includegraphics[width=7cm]{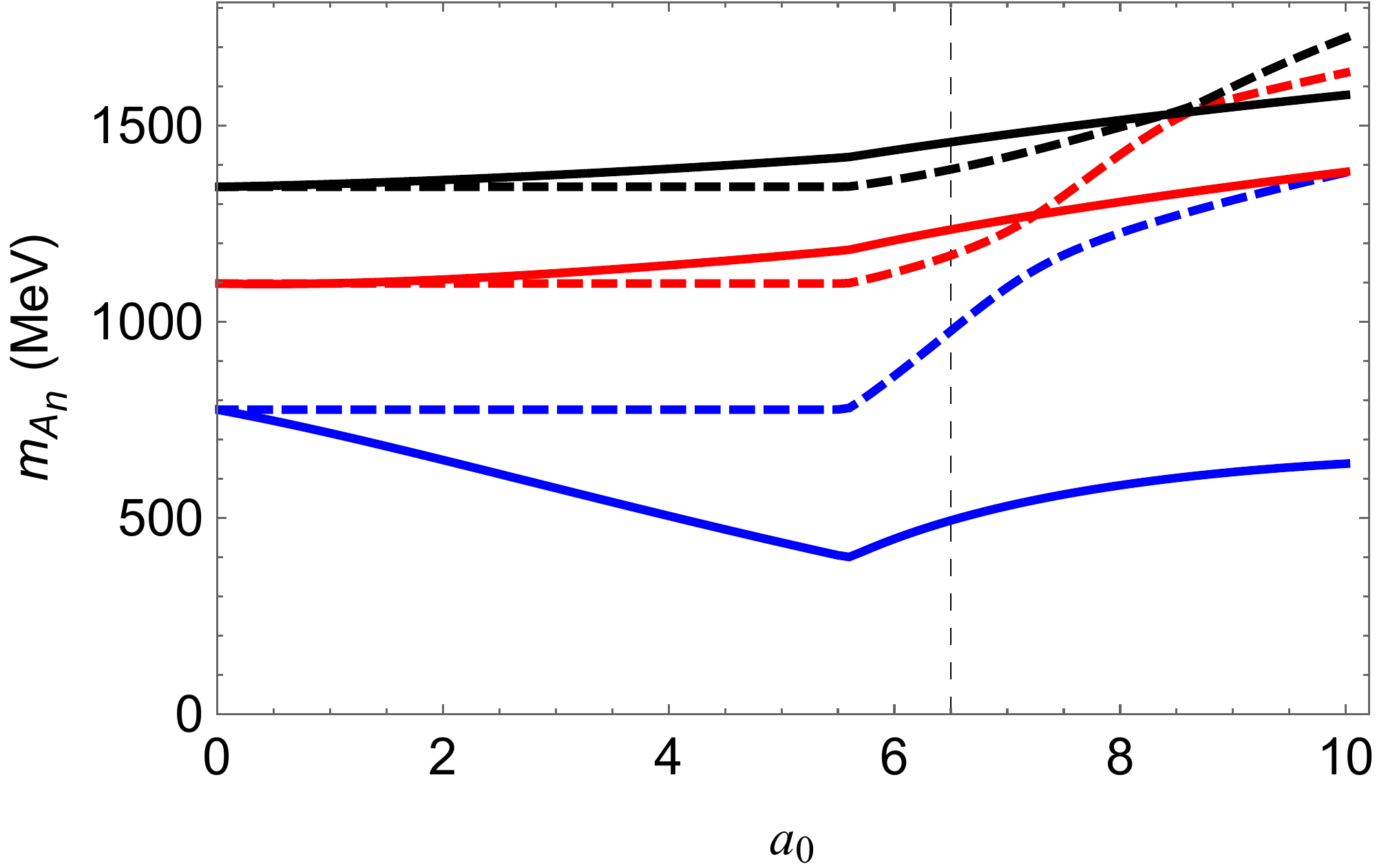}
\caption{
Left: Masses of axial-vector mesons as functions of $a_0$. Solid lines represent the results for Model IA ($a=b$) and  $\lambda=160$, while dashed lines for Model IB ($a\neq b$) and  $\lambda=380$. Right: Masses of axial-vector mesons as  functions of $a_0$. Solid lines represent the results for Model IIA ($a=b$) and  $\lambda=60$, while dashed lines for Model IIB ($a\neq b$) and  $\lambda=413$. These results were obtained setting $m_q=3.63\times 10^{-4}\,\text{MeV}$ (near the chiral limit).}
\label{Fig:MAVa0}
\end{figure}

The axial-vector mesons are described by Eq.~\eqref{Aeq}. Considering the Fourier transform on $A^{\mu}_{\perp}(x^{\mu},z)$,  $A^{\mu}_{\perp}(x^{\mu},z)\to A^{\mu}_{\perp}(k^{\mu},z)$, then, replacing  $\square\to m_{A_n}^2$, one may write the Schr\"odinger-like equation by redefining the axial-vector field as $A_{\mu}= \eta_{\mu} e^{-B_{A}} \psi_{a_n}$, where $2B_{A}=A_s-b$ and $\eta_{\mu}$ a polarisation vector, thus, we get
\noindent
\begin{equation}\label{Eq:AVSchroEq}
-\partial^2_z \psi_{a_n}+V_{A}\,\psi_{a_n}=m_{A_n}^2\,\psi_{a_n},
\end{equation}
\noindent
where the potential is given by
\begin{equation}\label{Eq:AVPotential}
V_{A}=\left(\partial_z B_{A}\right)^2+\partial^2_{z} B_{A}
+\beta,
\end{equation}
\noindent
and $\beta=g_5^2\, v^2\,e^{2A_s +b-a}$. Note that the potential depends explicitly on the functions $a$ and $b$ and depends on $\lambda$ and $m_q$ through the tachyonic field $v = \tilde v/\sqrt{\lambda}$. Analogous to what we have done before, we solve the eigenvalue problem numerically. Our numerical results for the masses as functions of $a_0$ at fixed  $m_q=3.63\times 10^{-4}\,\text{MeV}$ (near the chiral limit) are displayed on the left panel of Fig.~\ref{Fig:MAVa0} for Model IA ($a=b$) with solid lines and Model IB ($a\neq b$) with dashed lines . Similarly to the vectorial case, the mass of the ground state decreases with $a_0$. In the limit $(a_0,m_q) \to 0$ the function $\beta$  and the field $v$ go to zero. Hence, the Schr\"odinger potentials for the vectorial and axial sector coincide in that limit, i.e. $V_A=V_V$, meaning that the vector and axial-vector states are degenerate (chiral symmetry is restored). The masses as functions of $a_0$ for models of type II are displayed on the right panel of Fig.~\ref{Fig:MAVa0}, where solid lines are results for Model IIA, while dashed lines results for Model IIB. Those results were also obtained setting $m_q=3.63\times 10^{-4}\,\text{MeV}$ (near the chiral limit) and are similar to the ones found for models of type I.

\begin{figure}[ht!]
\centering
\includegraphics[width=7cm]{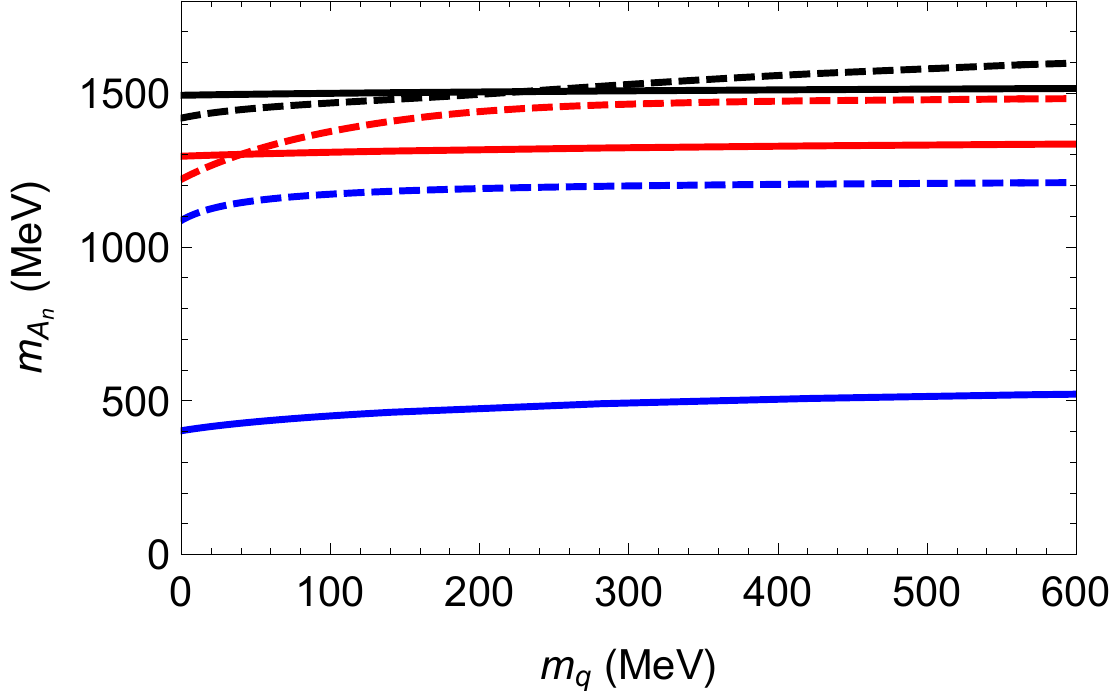}\hfill 
\includegraphics[width=7cm]{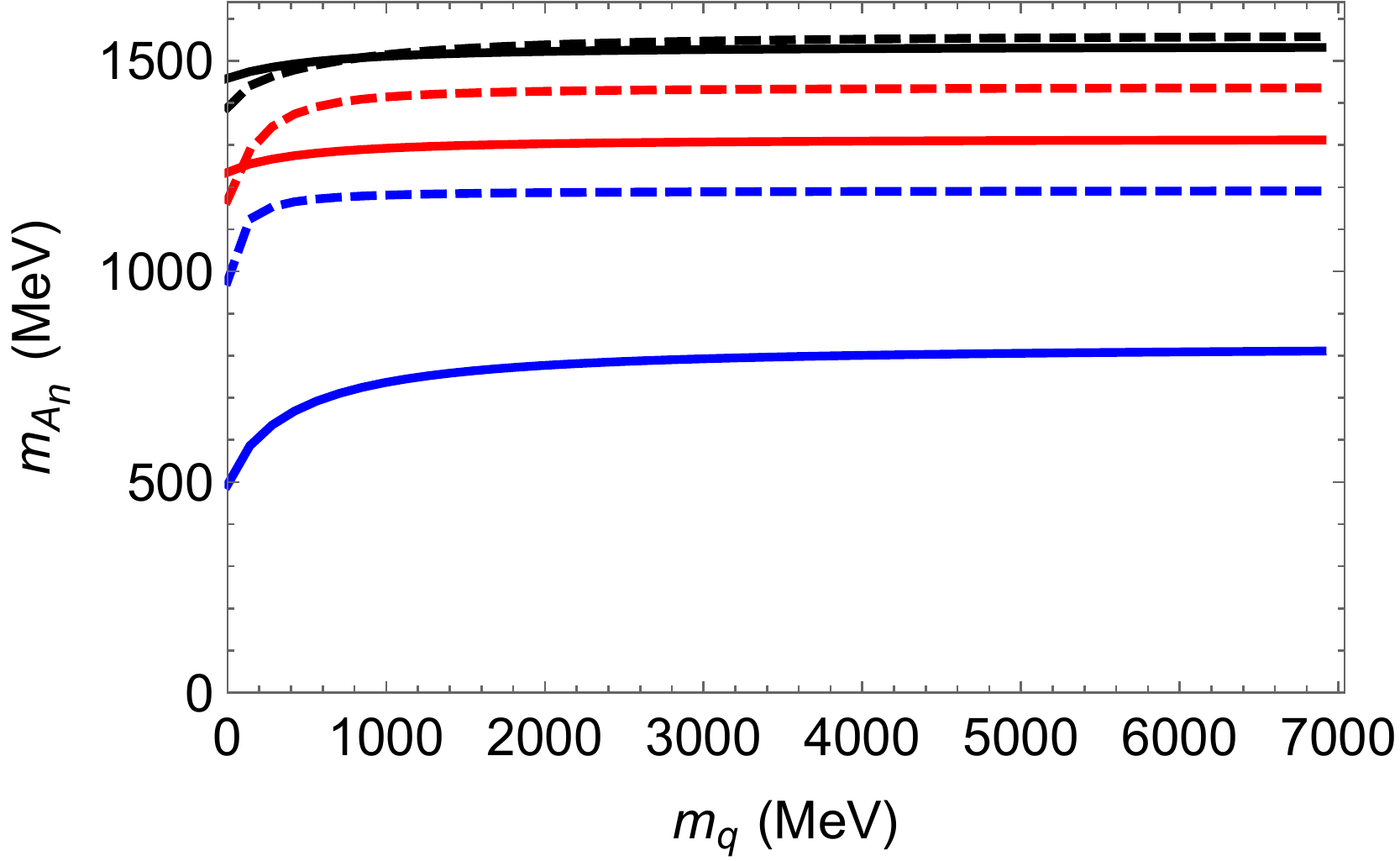}
\caption{
The mass of the axial-vector mesons as a function of the quark mass. Solid lines represent the results for Model IA ($a=b$) and  $\lambda=160$, while dashed lines for Model IB ($a\neq b$) and  $\lambda=380$, and we set $a_0=3.5$.
Right: The mass of the axial-vector mesons as a function of the quark mass. Solid lines represent the results for Model IIA ($a=b$) and $\lambda=60$, while dashed lines for Model IIA ($a\neq b$) and  $\lambda=413$, and we set $a_0=6.5$.}
\label{Fig:MAVDil2}
\end{figure}

Having fixed the parameter $a_0=3.5$ for models of type I, one may calculate the mass solving the eigenvalue problem numerically. The masses of axial-vector mesons as functions of the quark mass, for fixed $a_0$ and $\lambda$, are displayed on the left panel of Fig.~\ref{Fig:MAVDil2}, results for Model IA are represented with solid lines, while results for Model IB with dashed lines. As expected, the masses increases with the increasing of the quark mass. However, note that the resonances become less sensitive to the quark mass in the heavy quark regime. These conclusions are also true for the results obtained in models of type II, displayed on the right panel of Fig.~\ref{Fig:MAVDil2} for $a_0=6.5$ and selected values of $\lambda$. These results are in qualitative agreement with the results obtained in Ref.~\cite{Ballon-Bayona:2020qpq}.

Finally, for the set of parameters displayed in Table \ref{tab:Parameters}, we may calculate the spectrum provided by our models and compare them against the results available in the literature. In Table \ref{Taba:02} we write our results for Model IA setting $m_q=9\,\text{MeV}$, Model IB setting $m_q=4.7\,\text{MeV}$, Model IIA setting $m_q=9.8\,\text{MeV}$, and Model IIB setting $m_q=26.8\,\text{MeV}$. Each particlar choice of the quark masses will be justified bellow, and we will see that these values lead to a pion mass in agreement with experimental data. We observe that the results provided by models IB and IIB are closer to experimental data compared with the results provided by models IA and IIA.
\begin{table}[ht]
\centering
\begin{tabular}{l |c|c|c|c|c|l}
\hline 
\hline
 $n$ & Model IA & Model IB & Model IIA & Model IIB&
 GKK \cite{Gherghetta:2009ac}&
$a_1$ experimental \cite{Zyla:2020zbs} \\
  & $(a=b)$ & $(a\neq b)$ & $(a= b)$ & $(a\neq b)$ & \\
\hline 
 $0$ & 409  & 1098 & 525 &1105   & 1185   & $1230\pm 40$  \\
 $1$ & 1296 & 1231 &1242 &1261   & 1591  & $1647\pm 22$  \\
 $2$ & 1494 & 1423 &1463 &1431   & 1900  & $1930^{+30}_{-70}$ \\
 $3$ & 1669 & 1625 &1651 &1625   & 2101  & $2096\pm 122$  \\
 $4$ & 1828 & 1797 &1817 &1798  & 2279  & $2270^{+55}_{-40}$  \\
 $5$ & 1978 & 1950 &1970 &1954   &   &  \\
 $6$ & 2316 & 2092 &2114 &2100  &   &  \\
\hline\hline
\end{tabular}
\caption{
The masses of the axial-vector mesons (in MeV) obtained in the 
models of type I and models of type II, compared against the holographic model of \cite{Gherghetta:2009ac}  and experimental data \cite{Zyla:2020zbs}. We have considered $\lambda=160$ for Model IA, and $\lambda=380$ for Model IB, while the quark mass is $m_q=9\,\text{MeV}$ and $m_q=4.7\,\text{MeV}$, respectively. For Model IIA we have considered $\lambda=60$, while the quark mass is $m_q=9.8\,\text{MeV}$, and for Model IIB $\lambda=413$ and $m_q=26.8\,\text{MeV}$.
}
\label{Taba:02}
\end{table}

\subsection{Pseudoscalar mesons}

\begin{figure}[ht!]
\centering
\includegraphics[width=7cm]{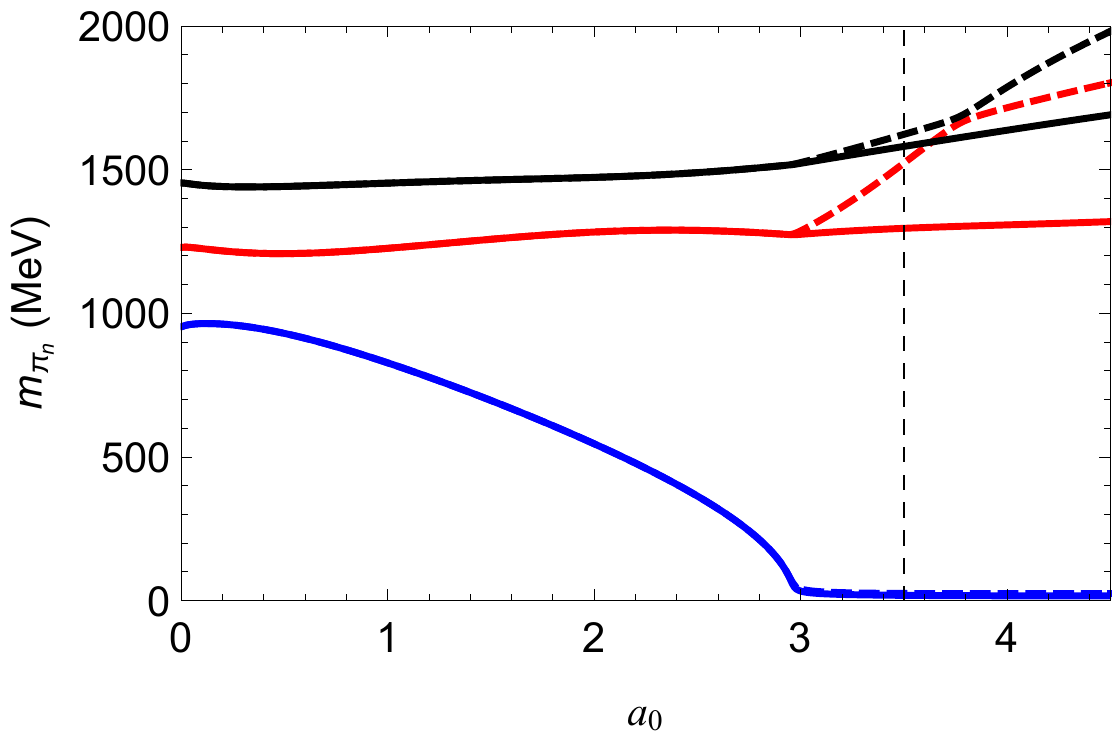}\hfill 
\includegraphics[width=7cm]{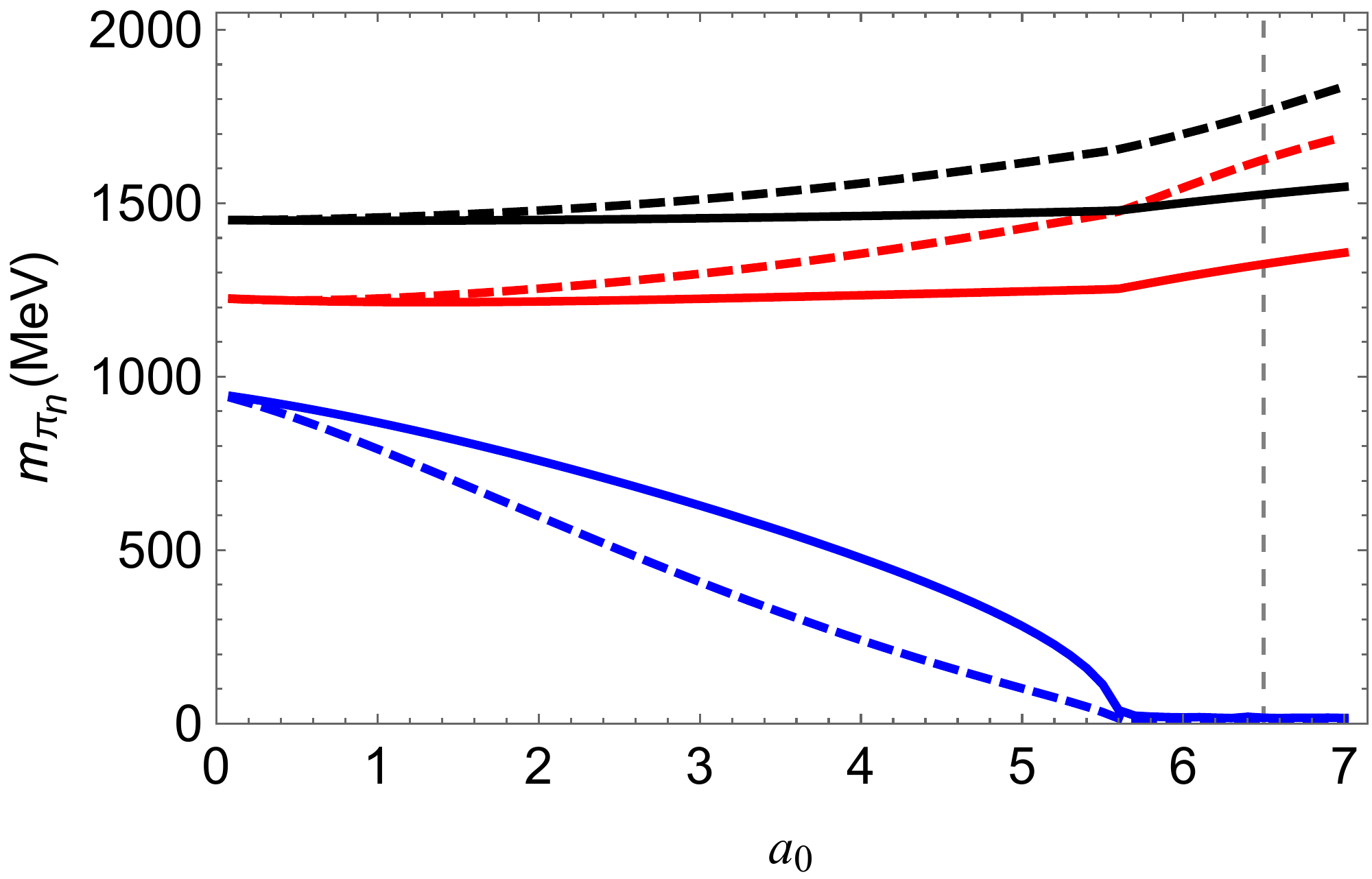}
\caption{
Left: The mass of the pseudoscalar mesons as a function of $a_0$. Solid lines represent the results for Model IA ($a=b$) and  $\lambda=160$, while dashed lines for Model IB ($a\neq b$) and  $\lambda=380$. Right: The mass of the pseudoscalar mesons as a function of $a_0$. Solid lines represent the results for Model IIA ($a=b$) and  $\lambda=60$, while dashed lines for Model IIB ($a\neq b$) and  $\lambda=413$. These results were obtained setting $m_q=3.63\times 10^{-4}\,\text{MeV}$.}
\label{Fig:Mpia0}
\end{figure}

We now focus in the pseudoscalar sector. As we did in previous subsections, we may rewrite the coupled differential equations \eqref{Pieq} in the Schr\"odinger-like form. However, we realized that the ground state is very sensitive numerically, for that reason we change our strategy and solve the coupled differential equations \eqref{Pieq} directly. Introducing the Fourier transform $\pi(x,z)\to \pi(k,z)$ and $\varphi(x,z)\to \varphi(k,z)$, then, replacing $\square\to m^2_{\pi_n}$ in Eq.~\eqref{Pieq} we get
\noindent
\begin{subequations}
\begin{align}
\Big [ \partial_z + A_s' - b' \Big ] \partial_z  \varphi +   \beta (  \pi  -  \varphi )  =&\, 0 \, , \nonumber \\
- m_{\pi_n}^2\partial_z \varphi + \beta  \, \partial_z \pi  = &\,0\,,
\end{align}
\end{subequations}
\noindent
where $\beta=g_5^2\,v^2\,e^{2A_s+b-a}$. The eigenvalue problem is solved numerically using the shooting method. Our results for the masses as functions of $a_0$ are displayed in the left panel of Fig.~\ref{Fig:Mpia0}, where results for Model IA are represented with solid lines, while results for Model IB with dashed lines, we set $m_q=3.63\times 10^{-4}\,\text{MeV}$ (near the chiral limit). Note that the mass of the ground state decreases with the increasing of $a_0$ up to $a_{0_c}$, then, it lies very close to zero, representing the Nambu-Goldstone boson arising for $a_0\geq a_{0}^c$ with $a_0^c \approx 2.97$ the same value found in subsection \ref{Subsec:SpontXSB} from the analysis of the chiral condensate.  In turn, the masses of pseudoscalar resonances display a different behavior with increasing of $a_0$. The same conclusions are true for models of type II, displayed on the right panel of Fig.~\ref{Fig:Mpia0}, where solid lines represent results for Model IIA, while dashed lines results for Model IIB. The Nambu-Goldstone state appears for $a_0\geq a_{0}^c$ with $a_0^c \approx 5.6$, the same value found in subsection \ref{Subsec:SpontXSB} from the analysis of the chiral condensate.

\begin{figure}[ht!]
\centering
\includegraphics[width=7cm]{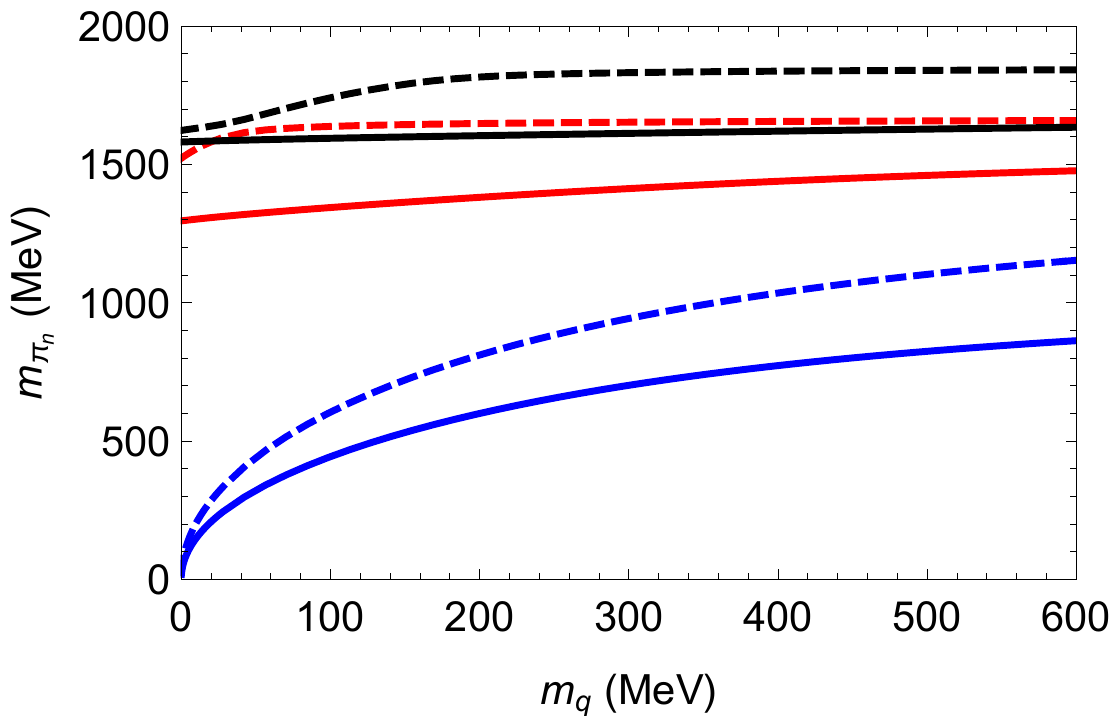}\hfill 
\includegraphics[width=7cm]{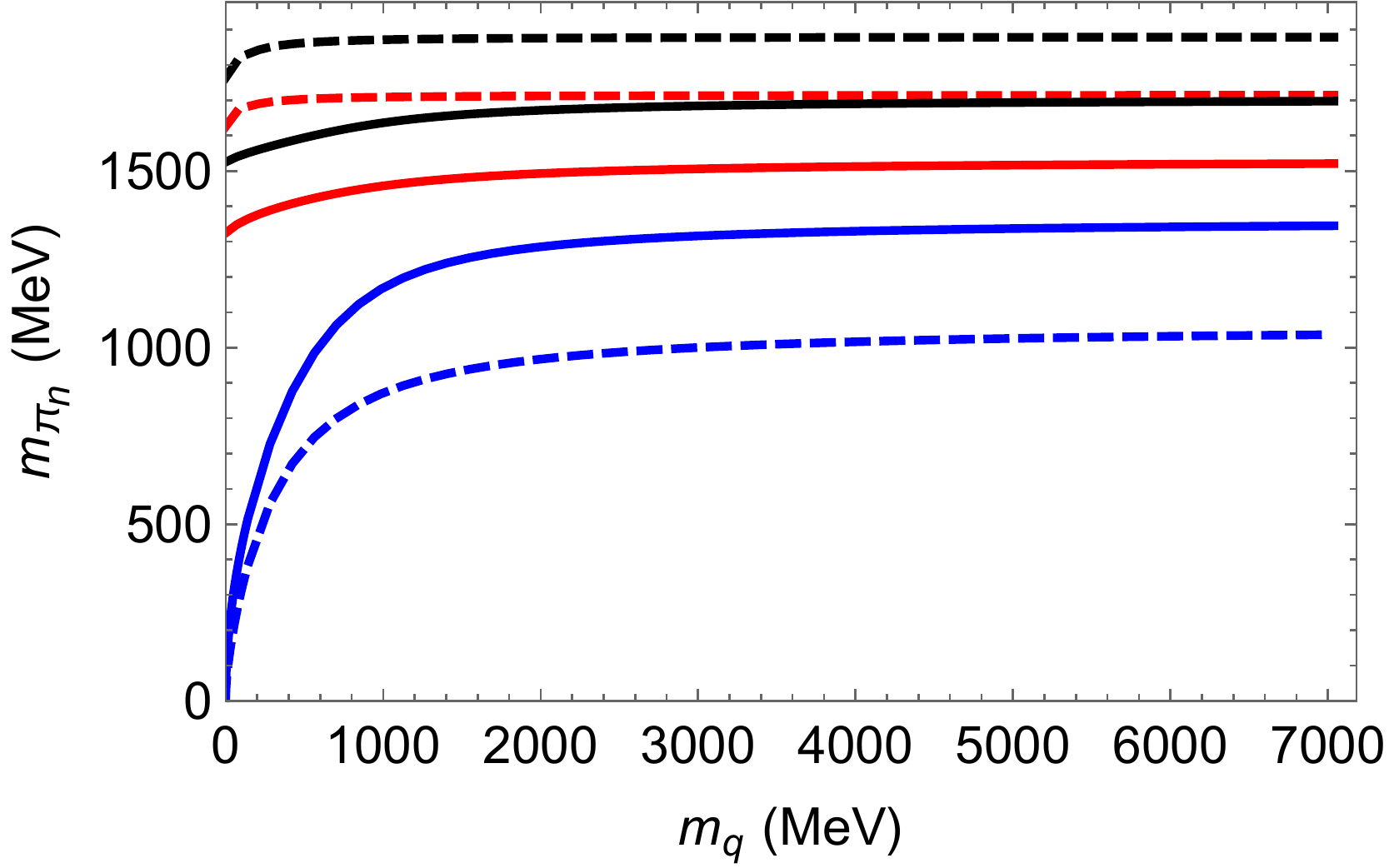}
\caption{
Masses of pseudoscalar mesons as functions of the quark mass. Solid lines represent the results for Model IA ($a=b$) and  $\lambda=160$, while dashed lines for Model IB ($a\neq b$) and  $\lambda=380$, and we set $a_0=3.5$.
Right: Masses of pseudoscalar mesons as functions of the quark mass. Solid lines represent the results for Model IIA ($a=b$) and $\lambda=60$, while dashed lines for Model IIA ($a\neq b$) and  $\lambda=413$, and we set $a_0=6.5$.}
\label{Fig:MpiDil2}
\end{figure}

Therefore, the choice of the parameter $a_0=3.5$ for models of type I and $a_0=6.5$ for models of type II is well justified by the scalar sector and pseusoscalar sectors. This choice allowed us to avoid a zero mode in the scalar sector and find a Nambu-Goldstone boson in the pseudoscalar sector in the chiral limit. Masses of pseudoscalar mesons as functions of the quark mass, for fixed $a_0$ and $\lambda$, are displayed on the left panel of Fig.~\ref{Fig:MpiDil2}, where results for Model IA are represented with solid lines, while results for Model IB are represented with dashed lines. As expected, the mass of the ground state approahes zero near the chiral limit, while the masses of the resonances approach finite values. All the masses increase with the increasing of the quark mass. However, note that the masses of pseudoscalar resonances are less sensitive to the quark mass in the heavy quark regime. These conclusions are also true for results obtained in models of type II, displayed on the right panel of Fig.~\ref{Fig:MpiDil2}.

Finally, for the set of parameters displayed in Table \ref{tab:Parameters}, we calculate the spectrum provided by our models and compare them against the results available in the literature. In Table \ref{Taba:Pseudoscalar} we show our results for Model IA setting $m_q=9\,\text{MeV}$, Model IB setting $m_q=4.7\,\text{MeV}$, Model IIA setting $m_q=9.8\,\text{MeV}$, and Model IIB setting $m_q=26.8\,\text{MeV}$. Each particular choice of the quark masses is justified by the fact that it allowed us to reproduce the experimental result for the pion mass. 
\begin{table}[ht]
\centering
\begin{tabular}{l |c|c|c|c|c|l}
\hline 
\hline
 $n$ & Model IA & Model IB & Model IIA & Model IIB &
 KBK \cite{Kelley:2010mu}&
$\pi$ experimental \cite{Zyla:2020zbs} \\
 & $(a=b)$ & $(a\neq b)$ & $(a= b)$ & $(a\neq b)$ & & \\
\hline 
 $0$ & 140  & 140  &140   &140 & 144    & $140$  \\
 $1$ & 1301 & 1539 &1338   &1675& 1557 & $1300\pm 100$  \\
 $2$ & 1582 & 1626 &1533   &1819& 1887 & $1816\pm 14$ \\
 $3$ & 1739 & 1794 &1713   &1945& 2090 & $2070$  \\
 $4$ & 1890 & 1945 &1877   &2183& 2270 & $2360$  \\
 $5$ & 2036 & 2083 & 2028  &2301& 2434 &  \\
 $6$ & 2175 & 2212 & 2170  &2422& 2586 &  \\
\hline\hline
\end{tabular}
\caption{
Masses of pseudoscalar mesons (in MeV) obtained in  
models of type I and models of type II, compared against the soft wall model \cite{Karch:2006pv}, the holographic model of \cite{Kelley:2010mu}  and experimental data \cite{Zyla:2020zbs}. We have considered $\lambda=160$ for models of type I, and $\lambda=380$ for models of type II, while the quark mass is $m_q=9\,\text{MeV}$ and $m_q=4.79\,\text{MeV}$, respectively. For Model IIA we have considered $\lambda=60$, while the quark mass is $m_q=9.8\,\text{MeV}$, and for Model IIB $\lambda=413$ and $m_q=26.8\,\text{MeV}$.
}
\label{Taba:Pseudoscalar}
\end{table}

So far, we have found a consistent description of spontaneous chiral symmetry breaking because the chiral condensate is nonzero in the chiral limit, and a Nambu-Goldstone state arises in the spectrum of  pseudoscalar mesons, which is consistent with the pion. There are other important tests of consistency that we want to show, the behaviour of meson decay constants and the Gell-Mann-Oakes-Renner (GOR) relation. This is done in the next section.

\section{Decay constants}
\label{Sec:DecayConstants}

To complement the analysis presented above, we now calculate the decay constants of vector, axial-vector, scalar and pseudo-scalar mesons. Details on the derivation of the holographic dictionary are presented in Appendix \ref{Sec:Decay} (see also \cite{Ballon-Bayona:2017bwk,Ballon-Bayona:2020qpq}). We will be particularly interested in the pseudoscalar sector where we expect to confirm the presence of Nambu-Goldstone bosons in the chiral limit. We will fulfill this expectation and will also reproduce the Gell-Mann-Oakes-Renner (GOR) relation near the chiral limit.

\subsection{Vector mesons}
\label{Subsec:Fvn}

\begin{figure}[ht!]
\centering
\includegraphics[width=7cm]{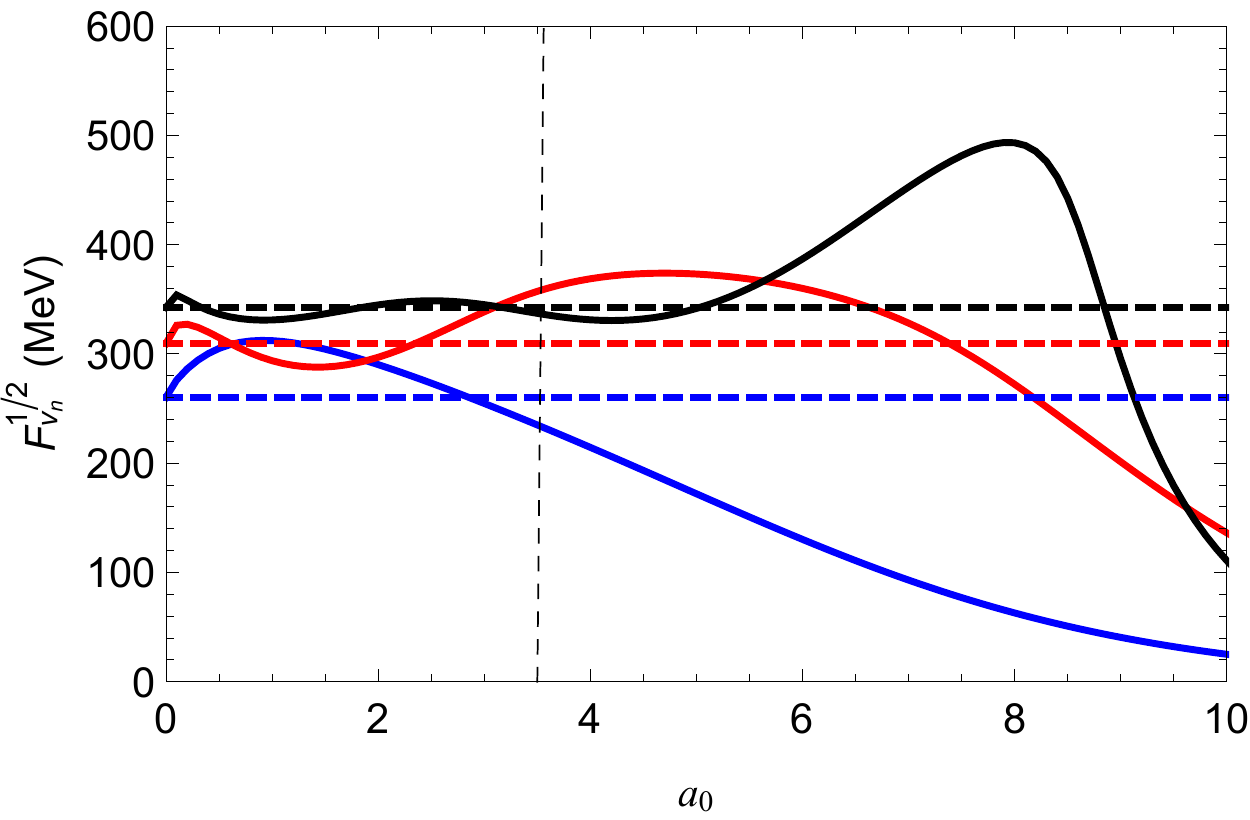}\hfill 
\includegraphics[width=7cm]{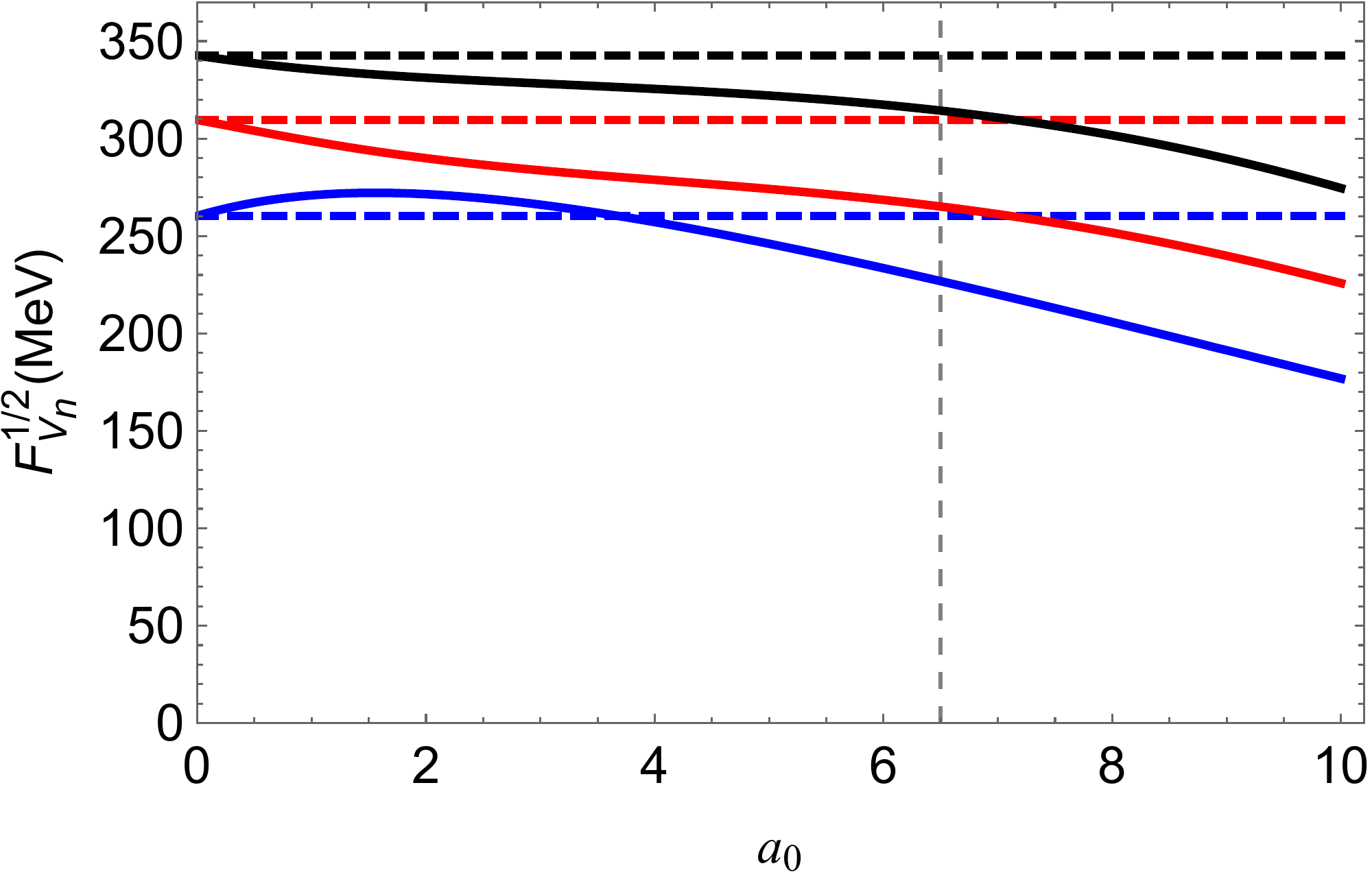}
\caption{
Left: Decay constants of the vector mesons as a function of $a_0$ obtained in models of type I, where solid lines represent results obtained in Model IA ($a=b$), while dashed lines represent results obtained Model IB ($a\neq b$). Right: The same as left panel for models of type II.
}
\label{Fig:DecayVDil2}
\end{figure}

In holographic models for QCD, the decay constants are related to the normalization constants arising from the normalization solutions of the eigenvalue problems. The normalization condition for the vector mesons is given by
\noindent
\begin{equation}
\int\,dz\, \psi_{v_m}(z)\psi_{v_n}(z)=\int dz\, e^{A_s-b}v_m(z)\,v_n(z)=\delta_{mn}, \label{Eq:Normvec}
\end{equation}
\noindent
where $A_s = - \ln z$ is the AdS warp factor, $b$ the dilaton coupling to the gauge fields, $\psi_{v_n}$ is the wave function of the Schrödinger equation \eqref{Eq:SchrodingerVector} and $v_n$ is the normalizable solution in the vectorial sector related to $\psi_{v_n}$ by $v_n(z)=\,e^{-B_{V}}\psi_{v_{n}}(z)$. As described in Appendix \ref{Sec:Decay}, the meson decay constants are given by Eq.~\eqref{Eq:DecayConstants}. For the vectorial sector, the decay constants reduce to the following formula:
\noindent
\begin{equation}
F_{v_n}=\lim_{\epsilon\to 0}\frac{e^{A_s-b}}{g_5}\,\partial_z v_n\bigg{|}_{z=\epsilon}=\frac{2}{g_5}N_{v_n},
\end{equation}
\noindent
where we have defined the normalisation constant $N_{v_n}$, as the coefficient which appears in the UV expansion of the vector mode, i.e.  $v_n(z)= N_{v_n} z^2 + \dots $. Hence, the decay constants are proportional to the normalisation constants. Therefore, solving numerically the Schr\"odinger-like equation and using the normalization condition we are able to calculate the decay constants of the vector mesons. We display our numerical results for the vector meson decay constants as functions of $a_0$ on the left panel of Fig.~\ref{Fig:DecayVDil2}, where solid lines represent results for Model IA, while dashed lines results for Model IB. As illustrated by the figure, the results for Model IA have a peculiar behavior changing the hierarchy with the increasing of $a_0$. In contrast, the results provided by Model IB do not change with $a_0$. In fact, the results for Model IB are the same as obtained in the linear soft wall model, because in that case the dilaton coupling $b$ is minimal. In that particular case, we find an analytic solution for the decay constants in the linear soft wall model, derived in Appendix \ref{MesonsSW}.

Our numerical results for models of type II are displayed on the right panel of Fig.~\ref{Fig:DecayVDil2}, where solid lines represent results for Model IIA, while dashed line results for Model IIB. From the figure we note that  the decay constants obtained in Model IIA are smoother than those obtained in Model IA, meaning that the decay constants are sensitive to the form of the interpolation functions \eqref{Eq:Dilaton1} and \eqref{Eq:Dilaton2}. Finally,  for the set of parameters displayed in Table \ref{tab:Parameters}, although in the vectorial case the results are independent of $\lambda$, we calculate the corresponding values of the vector meson decay constants. The results are displayed in Table \ref{Taba:DecayConstantVector}, compared with experimental results of Ref.~\cite{Donoghue:1992dd}. Note that there is a change in the hierarchy of vector meson decay constants $F_{V^n}$ in Model IA. The other models display the ordinary hierarchy $F_{V^0} < F_{V^1} < F_{V^2}$ found in previous works.
\begin{table}[ht]
\centering
\begin{tabular}{l |c|c|c|c|c|l}
\hline 
\hline
 & Model IA &Model IB & Model IIA & Model IIB & SW \cite{Karch:2006pv} &  Experimental \cite{Donoghue:1992dd} \\
  & $(a=b)$ & $(a\neq b)$ & $(a= b)$ & $(a\neq b)$ & & 
  ($F_{V^a}=g_{\rho})$\\
\hline 
 $F_{\scriptscriptstyle{V_0}}^{\scriptscriptstyle{1/2}}$ & 235 & 260 & 226 & 260 & 261 & $346.2\pm 1.4$  \\
 $F_{\scriptscriptstyle{V_1}}^{\scriptscriptstyle{1/2}}$ & 357 & 310 & 265 & 310 &  & $433\pm 13$ \\
 $F_{\scriptscriptstyle{V_2}}^{\scriptscriptstyle{1/2}}$ & 337 & 343 & 314 & 343 &  &  \\
\hline\hline
\end{tabular}
\caption{
The decay constants (in MeV) obtained in the models of type I and II, compared against the result obtained in the linear soft wall model ~\cite{Karch:2006pv} and experimental results of \cite{Donoghue:1992dd}. These results were found for the set of parameters displayed in Table \ref{tab:Parameters}.
}
\label{Taba:DecayConstantVector}
\end{table}

\subsection{Scalar mesons}
\label{Subsec:Fsn}

\begin{figure}[ht!]
\centering
\includegraphics[width=7cm]{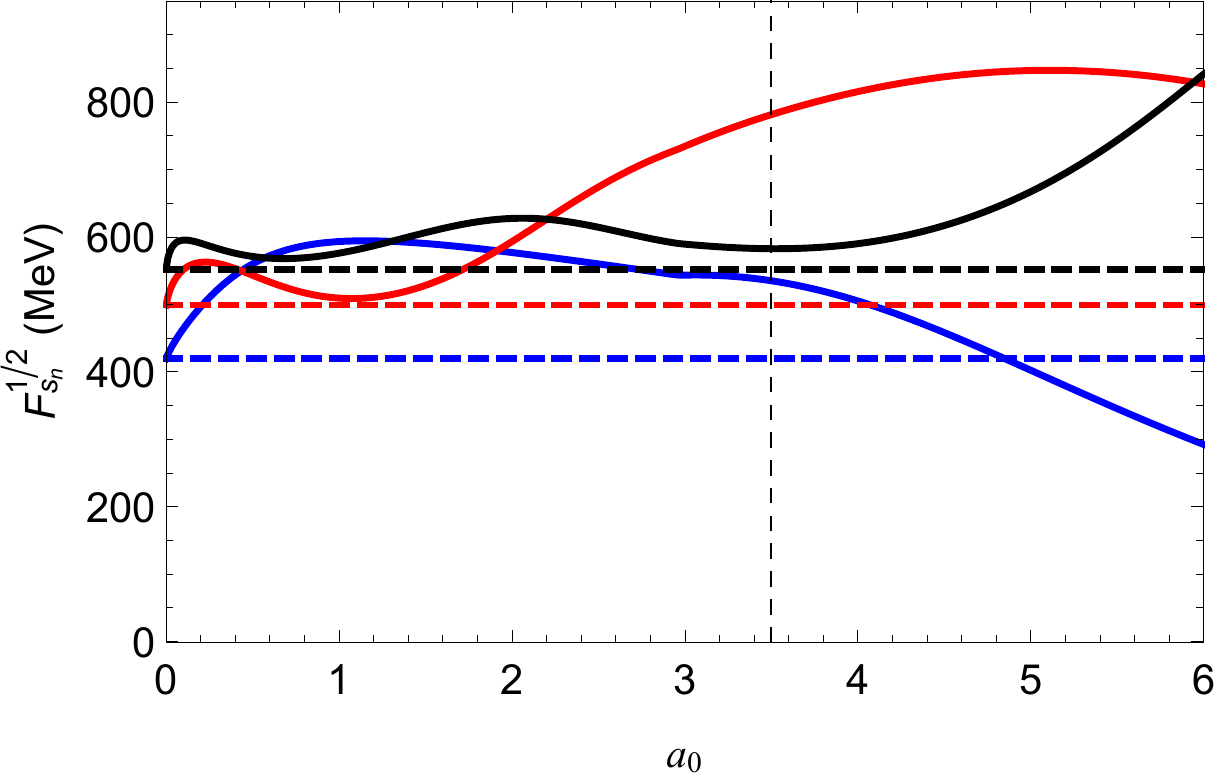}\hfill 
\includegraphics[width=7cm]{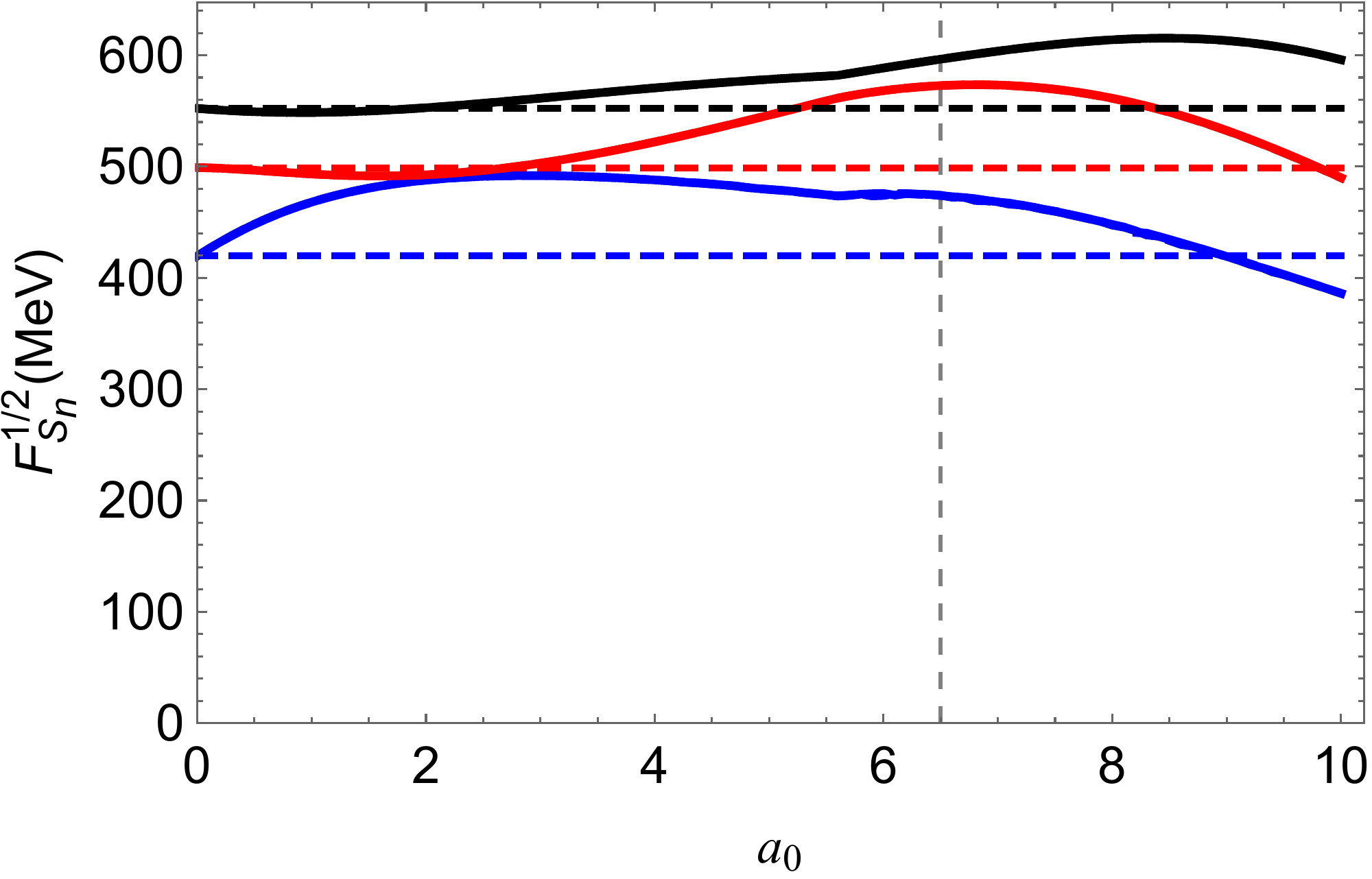}
\caption{
Left: Decay constants of scalar mesons as functions of $a_0$ for models of type I. Right: Decay constants of scalar mesons as  functions of $a_0$ for models of type II. We get these results for $m_q=3.63\times 10^{-4}\,\text{MeV}$ (near the chiral limit) in both type of models . Horizontal dashed lines represent the analytic results obtained in the linear soft wall model, see Appendix \ref{MesonsSW}. Vertical dashed lines represent the fixed values $a_0=3.5$ and $a_0=6.5$ for models of type I and II respectively.}
\label{Fig:DecayS}
\end{figure}

As we did in our previous subsection, we start with the normalization condition. For the scalar sector, it is given by the kinetic term related to scalar fluctuation of the Lagrangian \eqref{4dLag}, which is
\noindent
\begin{equation}\label{Eq:NormCondScalar}
\int dz\,\psi_{s_m}(z)\psi_{s_n}(z)=\int dz\, e^{3A_s-a}S_m(z)\,S_n(z)=\delta_{mn}.
\end{equation}
\noindent
Considering the UV expansion of the normalizable solution of $S_n(z)$, which is given by $S_n(z)=N_{s_n}z^3+\cdots$, here $N_{s_n}$ is the normalization constant calculated by plugging the solution in \eqref{Eq:NormCondScalar}. Thus, the decay constants are given by \eqref{Eq:DecayConstants}
\noindent
\begin{equation}
F_{s_n}=\zeta\,z\,e^{3A_s-a}\partial_zS_n\bigg{|}_{z=\epsilon}=3\zeta\,N_{s_n}.
\end{equation}
\noindent
As expected, the meson decay constants are proportional to the corresponding normalization constants. We display the  results for the scalar meson decay constants as functions of $a_0$ on the left panel of Fig.~\ref{Fig:DecayS}, for $m_q=3.63\times 10^{-4}\,\text{MeV}$ (near the chiral limit). In fact, in the scalar sector the results for Model IA and Model IB are the same, represented with solid lines in the figure. The figure also shows that the hierarchy in certain intervals of the parameter $a_0$ changes. We highlight the vertical dashed line at $a_0=3.5$, the fixed value used in this paper in models of type I. We also plot the analytic results obtained in the linear soft wall model (see Appendix \ref{MesonsSW}) with dashed lines. The corresponding results for models of type II are displayed on the right panel of Fig.~\ref{Fig:DecayS}. Interestingly, the hierarchy does not change with $a_0$, and the results are smoother than those obtained in models of type I; we also highlight the vertical dashed line for $a_0=6.5$, the fixed value used for models of type II.

\begin{figure}[ht!]
\centering
\includegraphics[width=7cm]{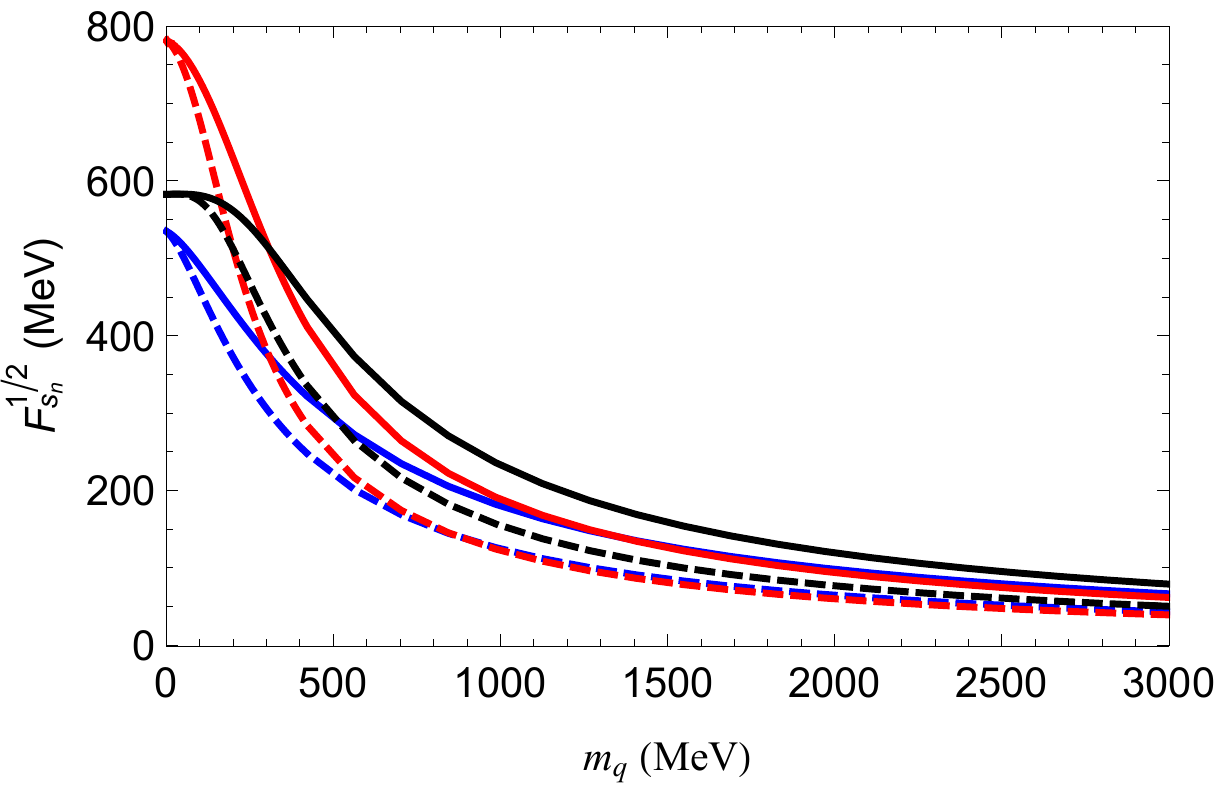}\hfill 
\includegraphics[width=7cm]{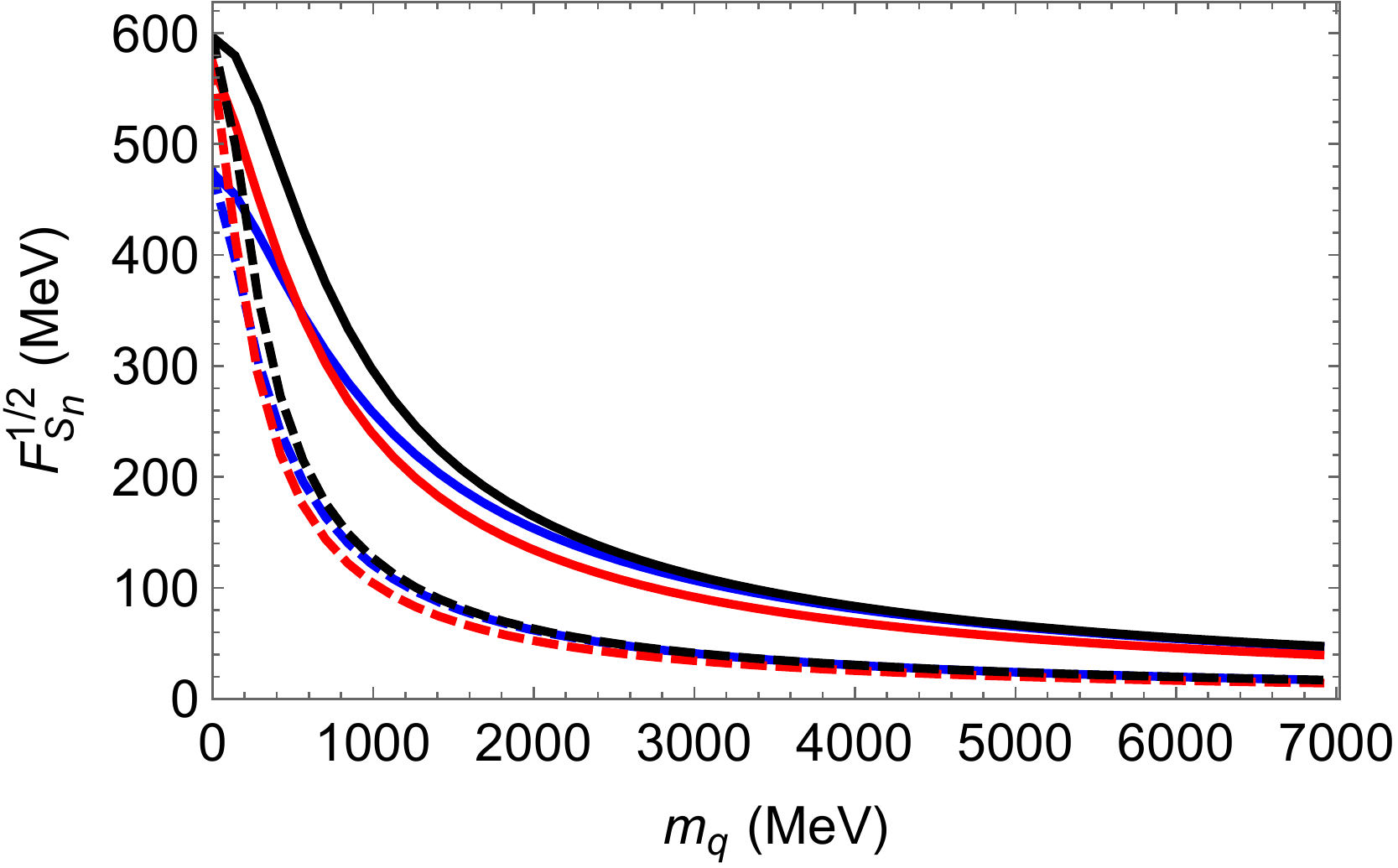}
\caption{
Left: Decay constants of scalar mesons as functions of $m_q$ in models of type I with $a_0=3.5$; solid lines represent the results for Model IA ($\lambda=160$), while dashed dashed lines for Model IB ($\lambda=380$).
Right: Decay constants of scalar mesons as functions of $m_q$ for models of type II with $a_0=6.5$; solid lines represent the results for Model IIA ($\lambda=60$), while dashed dashed lines for Model IIB ($\lambda=413$).}
\label{Fig:DecayS2}
\end{figure}

The scalar meson decay constants as functions of the quark mass are displayed on the left panel of Fig.~\ref{Fig:DecayS2}, where solid lines represent results for Model IA, while dashed lines represent results for Model IB. Note the change in the hierarchy of decay constants near the chiral limit, this is consistent with the left panel of Fig.~\ref{Fig:DecayS} (see vertical dashed line). In both models of type I, the scalar meson decay constants decrease with the increasing of the quark mass, and the hierarchy is restored in the regime of heavy quarks. The corresponding results for models of type II are displayed on the right panel of Fig.~\ref{Fig:DecayS2}, where solid lines represent results for Model IIA, while dashed line results for Model IIB. As the figures illustrates,  the scalar meson decay constants for models of type IIdecreases with the increasing of the quark mass and there is a change in the hierarchy in the regime of heavy quarks. 

Finally, we show our results for the decay constants at the specific values of the parameters given in Table \ref{tab:Parameters}. The results are displayed in Table \ref{Taba:DecayConstantS}. The quark masses were fixed as  $m_q=9\,\text{MeV}$, $m_q=4.7\,\text{MeV}$,  $m_q=9.8\,\text{MeV}$ and $m_q=26.8\,\text{MeV}$ for models IA, IB, IIA and IIB respectively.

\begin{table}[ht]
\centering
\begin{tabular}{l |c|c|c|c|c|c}
\hline 
\hline
 & Model IA &Model IB & Model IIA & Model IIB & SW &  QCD \\
 & $(a=b)$ & $(a\neq b)$ & $(a= b)$ & $(a\neq b)$ & \cite{Karch:2006pv} & Results \cite{Gokalp:2001} \\
\hline 
 $F_{\scriptscriptstyle{s_0}}^{\scriptscriptstyle{1/2}}$  & 532.7 & 533.2 & 470.4 &426.6 & 420 & 425.3  \\
 $F_{\scriptscriptstyle{s_1}}^{\scriptscriptstyle{1/2}}$  & 779.7 & 780.1 & 561.1 & 465.7& 499 &  \\
 $F_{\scriptscriptstyle{s_2}}^{\scriptscriptstyle{1/2}}$  & 582.5 & 582.5 & 595.8 &544.5 & 552 &  \\
\hline\hline
\end{tabular}
\caption{
Decay constants of scalar mesons (in MeV) obtained in models of type I and II for the set of parameters given in Table \ref{tab:Parameters}, compared against the result obtained in the linear soft wall model ~\cite{Karch:2006pv} and the result obtained in QCD \cite{Gokalp:2001}. The quark masses were fixed as  $m_q=9\,\text{MeV}$, $m_q=4.7\,\text{MeV}$,  $m_q=9.8\,\text{MeV}$ and $m_q=26.8\,\text{MeV}$ for models IA, IB, IIA and IIB respectively.}
\label{Taba:DecayConstantS}
\end{table}

\subsection{Axial-vector mesons}
\label{Subsec:Fan}

\begin{figure}[ht!]
\centering
\includegraphics[width=7cm]{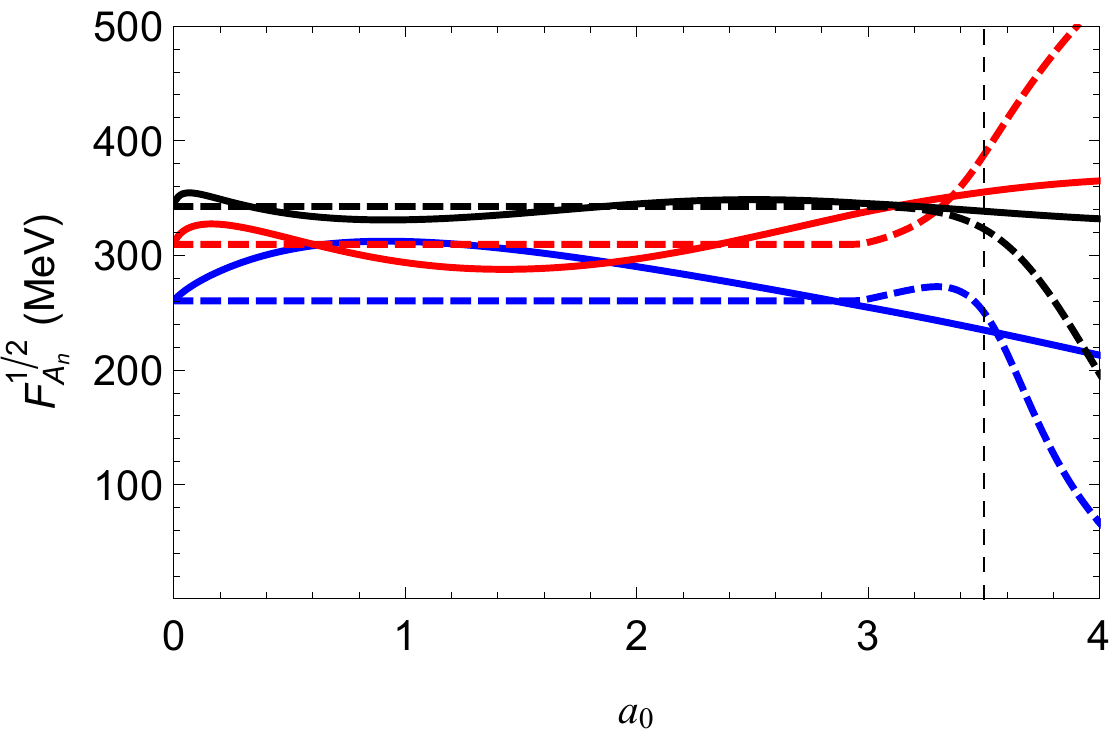}\hfill 
\includegraphics[width=7cm]{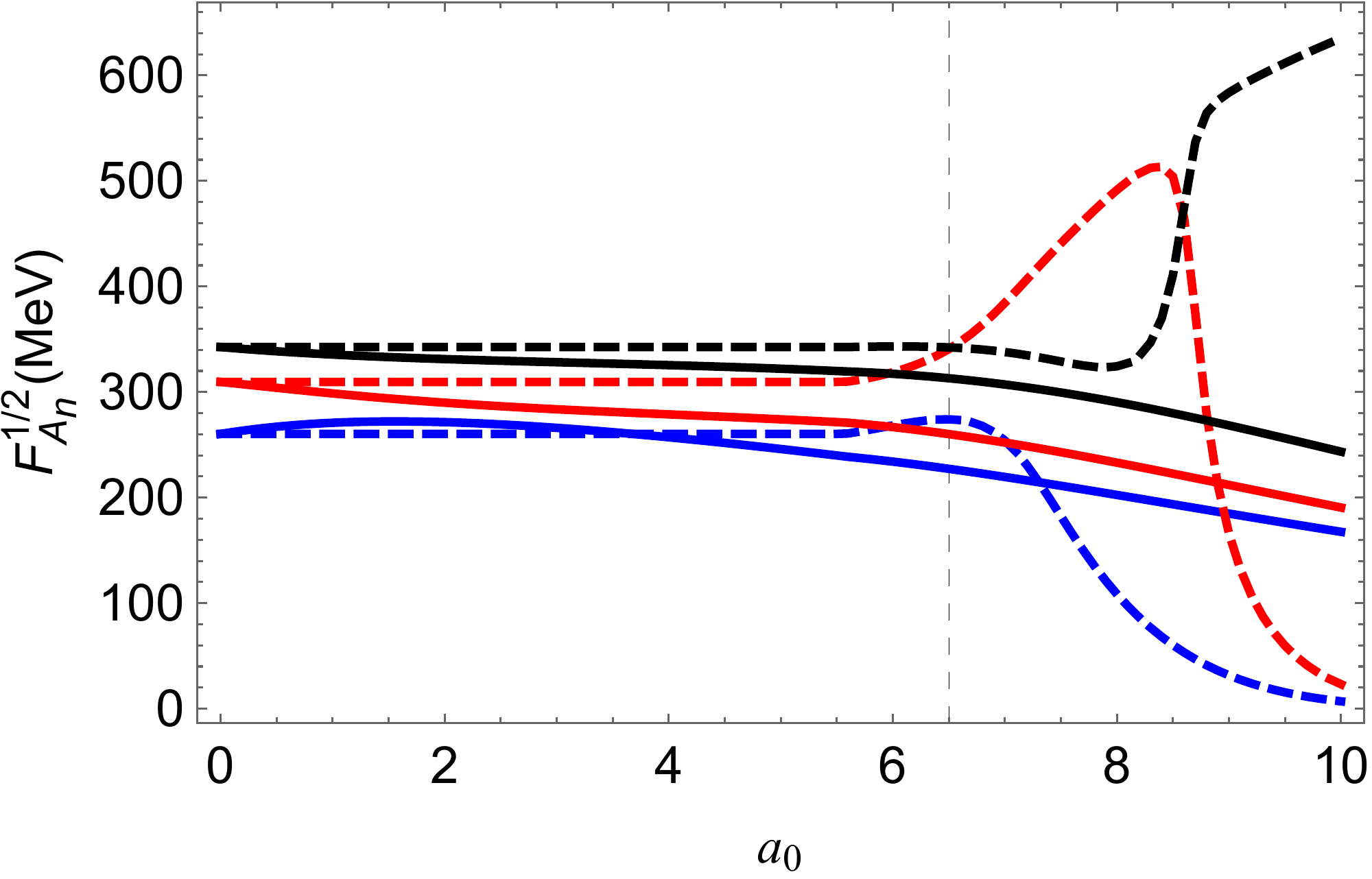}
\caption{
Left: Decay constants of axial-vector mesons as functions of $a_0$ for models of type I, where solid lines represent Model IA ($\lambda=160$), while dashed lines Model IB ($\lambda=380$). Right: Decay constants of axial-vector mesons as functions of $a_0$ for models of type II, where solid lines represent Model IIA ($\lambda=60$), while dashed lines Model IIB ($\lambda=413$). We obtained these results for $m_q=3.63\times 10^{-4}\,\text{MeV}$ in all the models (near the chiral limit).
}
\label{Fig:DecayAV}
\end{figure}

We follow the same procedure described above this time for the axial-vector mesons, where the normalization condition is given by
\noindent
\begin{equation}
\int dz\,\psi_{a_m}(z)\psi_{a_n}(z)=\int dz\, e^{A_s-b}a_m(z)\,a_n(z)=\delta_{mn}, \label{Eq:NormAxial}
\end{equation}
\noindent
where $a_n(z)=\,e^{-B_{A}}\psi_{a_{n}}(z)$ is the axial-vector normalizable solution and $\psi_{a_{n}}(z)$ is the corresponding wave function of the Schrödinger-like equation \eqref{Eq:AVSchroEq}. Plugging the UV expansion of $a_n(z)$, which is given by $a_n(z)=N_{a_n} z^2 + \dots $, where $N_{a_n}$ is the normalization constant, in  \eqref{Eq:DecayConstants} we obtain the axial-vector meson decay constants
\noindent
\begin{equation}
F_{a_n}=\lim_{\epsilon\to 0}\frac{e^{A_s-b}}{g_5}\,\partial_z a_n\bigg{|}_{z=\epsilon}=\frac{2}{g_5}N_{a_n}.
\end{equation}
\noindent
Thus, the problem of finding decay constants has been reduced to the calculation of normalization constants. Our numerical results for the axial-vector meson decay constants for $m_q=3.63\times 10^{-4}\,\text{MeV}$ (near the chiral limit) as functions of $a_0$ are displayed on the left panel of Fig.~\ref{Fig:DecayAV}, where solid lines represent results for Model IA, while dashed lines results for Model IB. As illustrated by the figure, the decay constants in Model IA have a peculiar behavior changing the hierarchy when $a_0$ increases. The decay constants in Model IB do not change with $a_0$ up to $a_0=a_{0_c}$, then, they vary with $a_0$ faster than the decay constants in Model IA. The results provided by models of type II are displayed on the right panel of Fig.~\ref{Fig:DecayAV}, where results for Model IIA (IIB) are represented by solid lines (dashed lines). Note that the variation of decay constants with $a_0$ in Model IIA is smoother than the variation in Model IA. On the other hand, the decay constants in Models IB and IIB change rapidly for $a_0\geq a_{0_c}$.

\begin{figure}[ht!]
\centering
\includegraphics[width=7cm]{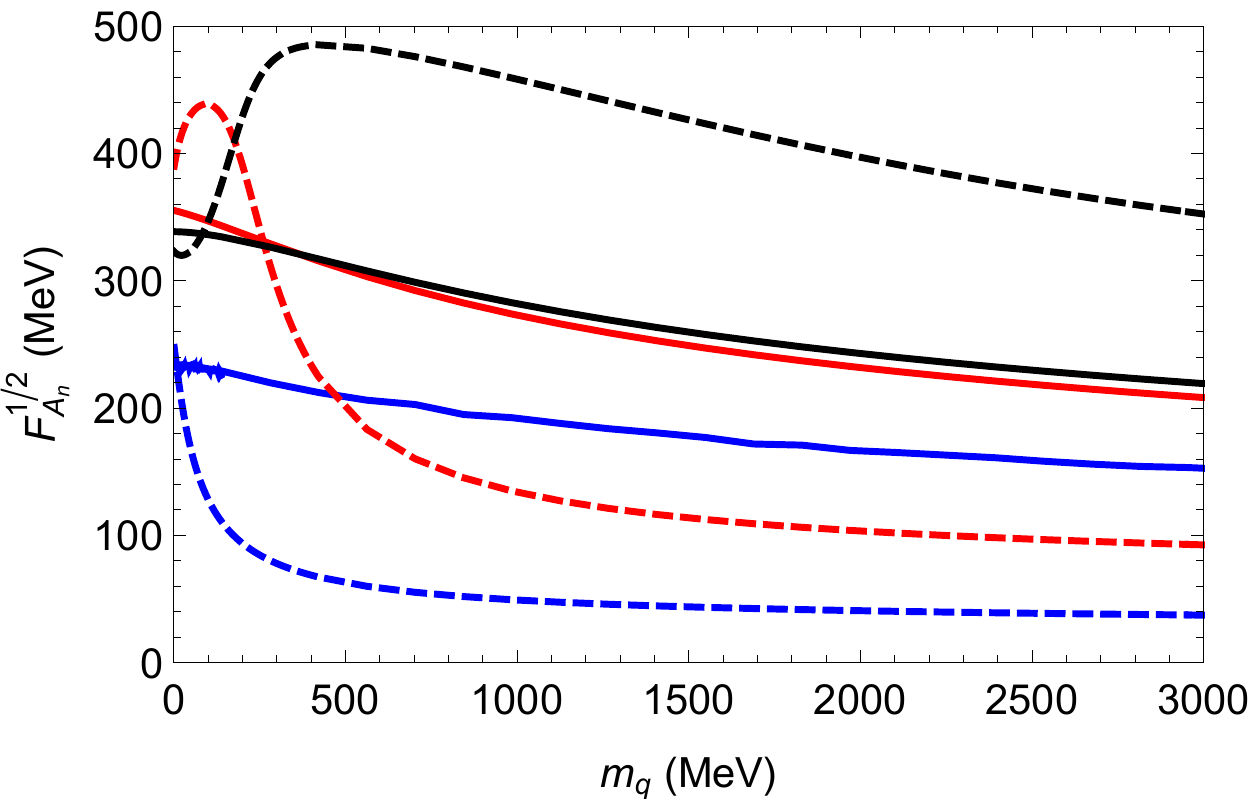}\hfill 
\includegraphics[width=7cm]{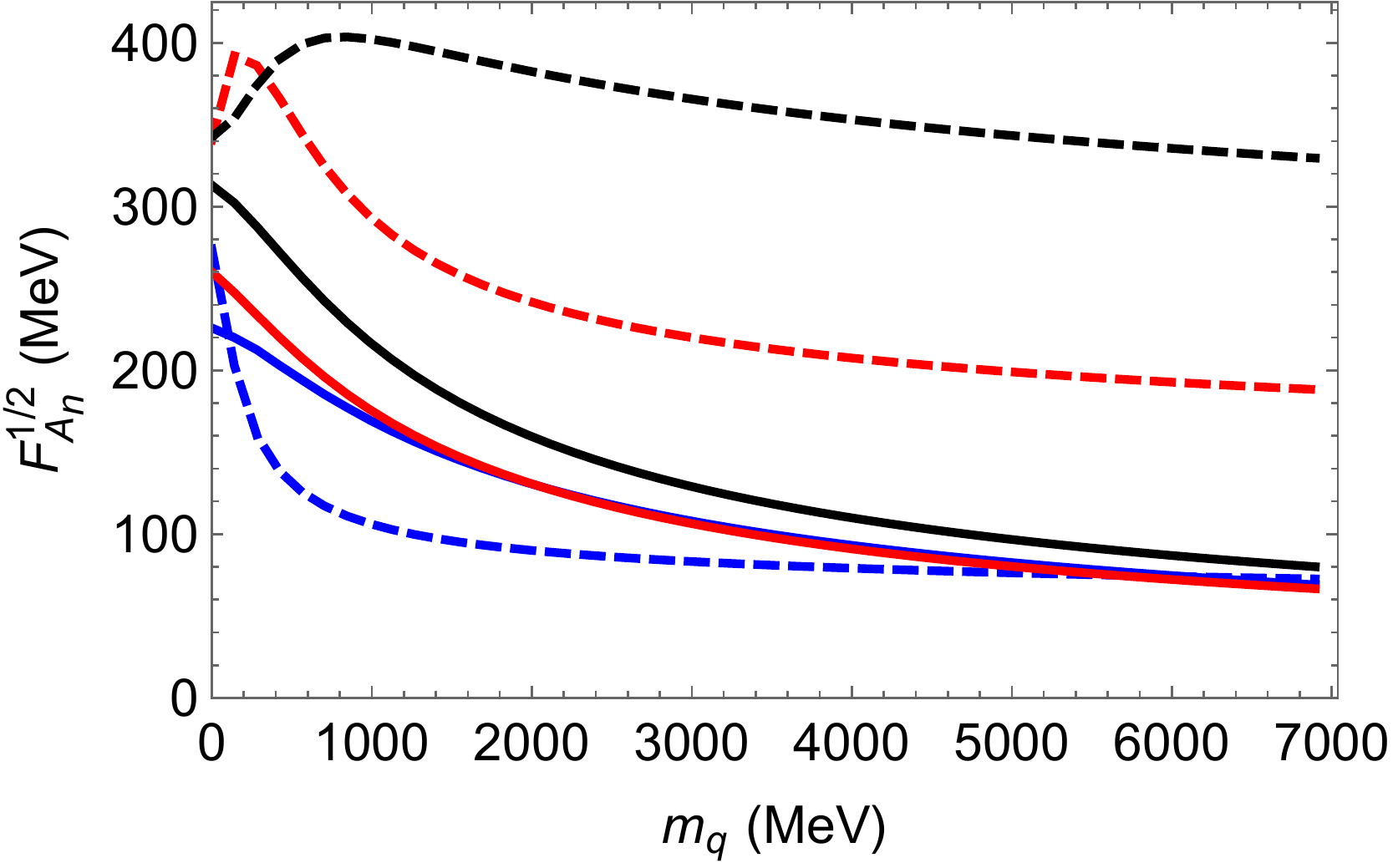}
\caption{
Left: Decay constants of axial-vector mesons as functions of $m_q$ in models of type I with $a_0=3.5$, where solid lines represent the results for Model IA ($\lambda=160$), while dashed dashed lines for Model IB ($\lambda=380$).
Right: Decay constants of axial-vector mesons as functions of $m_q$ in models of type II with $a_0=6.5$, where solid lines represent the results for Model IIA ($\lambda=60$), while dashed dashed lines for Model IIB ($\lambda=413$).
}
\label{Fig:DecayAVDil2}
\end{figure}

Having fixed the parameter $a_0=3.5$ for models of type I, we calculated the decay constants of axial-vector mesons as for different values of the quark mass. Our numerical results are displayed in the left panel of Fig.~\ref{Fig:DecayAVDil2}, where solid lines (dashed lines) represent results for Model IA (IB). As shown in the figure, the behavior of the decay constants is different for each state in the region of small quark mass. There is a change in hierarchy between the second and third states near the chiral limit, which is consistent with the results displayed on the left panel of Fig.~\ref{Fig:DecayAV} (see the vertical dashed line). The hierarchy is restored in the regime of heavy quarks, where we have $F_{A_0}^{1/2}<F_{A_1}^{1/2}<F_{A_2}^{1/2}$; this result is in qualitative agreement with the results reported in Ref.~\cite{Ballon-Bayona:2020qpq}. The results obtained in models of type II are displayed on the right panel of Fig.~\ref{Fig:DecayAVDil2}, where solid lines (dashed lines) represent results for Model IIA (IIB). In model IIA the hierarchy between decay constants is preserved near the chiral limit but slightly changes in the heavy quark regime. In Model IIB the hierarchy is not preserved near the chiral limit but it is restored in the regime of heavy quarks, where we have $F_{A_0}^{1/2}<F_{A_1}^{1/2}<F_{A_2}^{1/2}$.

For the fixed set of parameters displayed in Table \ref{tab:Parameters}, we calculate the corresponding values of axial-vector meson decay constants and display the results in Table \ref{Taba:DecayConstantAV}. The quark masses where fixed as  $m_q=9\,\text{MeV}$,  $m_q=4.7\,\text{MeV}$,  $m_q=9.8\,\text{MeV}$, and $m_q=26.8\,\text{MeV}$ for models of type IA, IB, IIA and IIB respectively.

\begin{table}[ht]
\centering
\begin{tabular}{l |c|c|c|c|c|l}
\hline 
\hline
 & Model IA &Model IB & Model IIA & Model IIB & SW \cite{Karch:2006pv} &  Experimental  \\
 & $(a=b)$ & $(a\neq b)$ & $(a= b)$ & $(a\neq b)$ & &  ($F_{A^c}=f_{a_1}/\sqrt{2}$) \cite{Isgur:1988vm} \\
\hline 
 $F_{\scriptscriptstyle{A_0}}^{\scriptscriptstyle{1/2}}$  & 141.82 & 241 &226.07 &224.24 & 261 & $433\pm 13$ \\
 $F_{\scriptscriptstyle{A_1}}^{\scriptscriptstyle{1/2}}$  & 324.10 & 395 &257.12 &386.50 &  &  \\
 $F_{\scriptscriptstyle{A_2}}^{\scriptscriptstyle{1/2}}$  & 295.17 & 322 &310.91 &348.54 &  &  \\
\hline\hline
\end{tabular}
\caption{
Decay constants of axial-vector mesons (in MeV) obtained in models of type I and II, compared against the result obtained in the linear soft wall model ~\cite{Karch:2006pv} and experimental results of \cite{Isgur:1988vm}. These results were obtained for the set of parameters displayed in table \ref{tab:Parameters} and the quark masses were fixed as $m_q=9\,\text{MeV}$,  $m_q=4.7\,\text{MeV}$,  $m_q=9.8\,\text{MeV}$, and $m_q=26.8\,\text{MeV}$ for models of type IA, IB, IIA and IIB respectively.
}
\label{Taba:DecayConstantAV}
\end{table}

\subsection{Pseudo-scalar mesons}
\label{Subsec:Fpin}

For the pseudoscalar mesons the normalization condition in terms of the normalizable solution $\varphi_n(z)$ is given by
\noindent
\begin{equation}\label{Eq:NormPi}
\int_{\epsilon}^{\infty} dz\,\frac{e^{A_s-b}}{\beta(z)}\left(\partial_z\varphi_m\right)\left(\partial_z\varphi_n\right)=\frac{\delta_{mn}}{m_{\pi_n}^2}.
\end{equation}
\noindent
We follow in this work the prescription for calculating decay constants for the pion and their resonances  developed in \cite{Ballon-Bayona:2014oma,Ballon-Bayona:2020qpq}). Details on the derivation are given in Appendix \ref{Sec:Decay}. The decay constants of pseudoscalar mesons are calculated using the following holographic dictionary, cf. Eq.~\eqref{Eq:DecayConstants}, 
\noindent
\begin{equation}\label{Eq:DecayPi1}
f_{\pi_n}=-\lim_{\epsilon\to 0}\frac{e^{A_s-b}}{g_5}\partial_z\varphi_n(z)\bigg{|}_{z=\epsilon}.
\end{equation}
\noindent
Considering the UV expansion of the field $\varphi_n(z)$, which is given by $\varphi_n=-N_{\pi_n}(z^2+\cdots)$, where $N_{\pi_n}$ is the normalization constant obtained from \eqref{Eq:NormPi}, and plugging this expression into \eqref{Eq:DecayPi1} we get the result
\noindent
\begin{equation}\label{Eq:DecayPi2}
f_{\pi_n}=\frac{2}{g_5}\,N_{\pi_n}.
\end{equation}
\noindent

\begin{figure}[ht!]
\centering
\includegraphics[width=7cm]{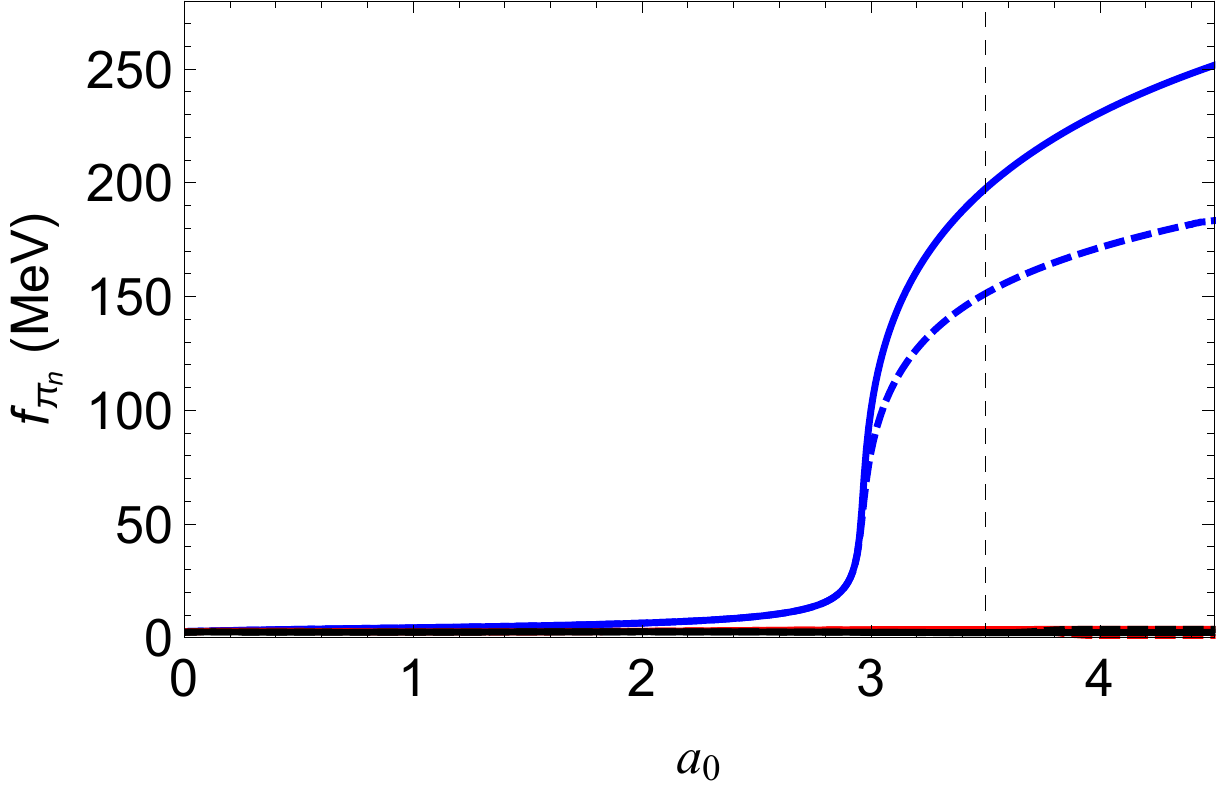}\hfill 
\includegraphics[width=7cm]{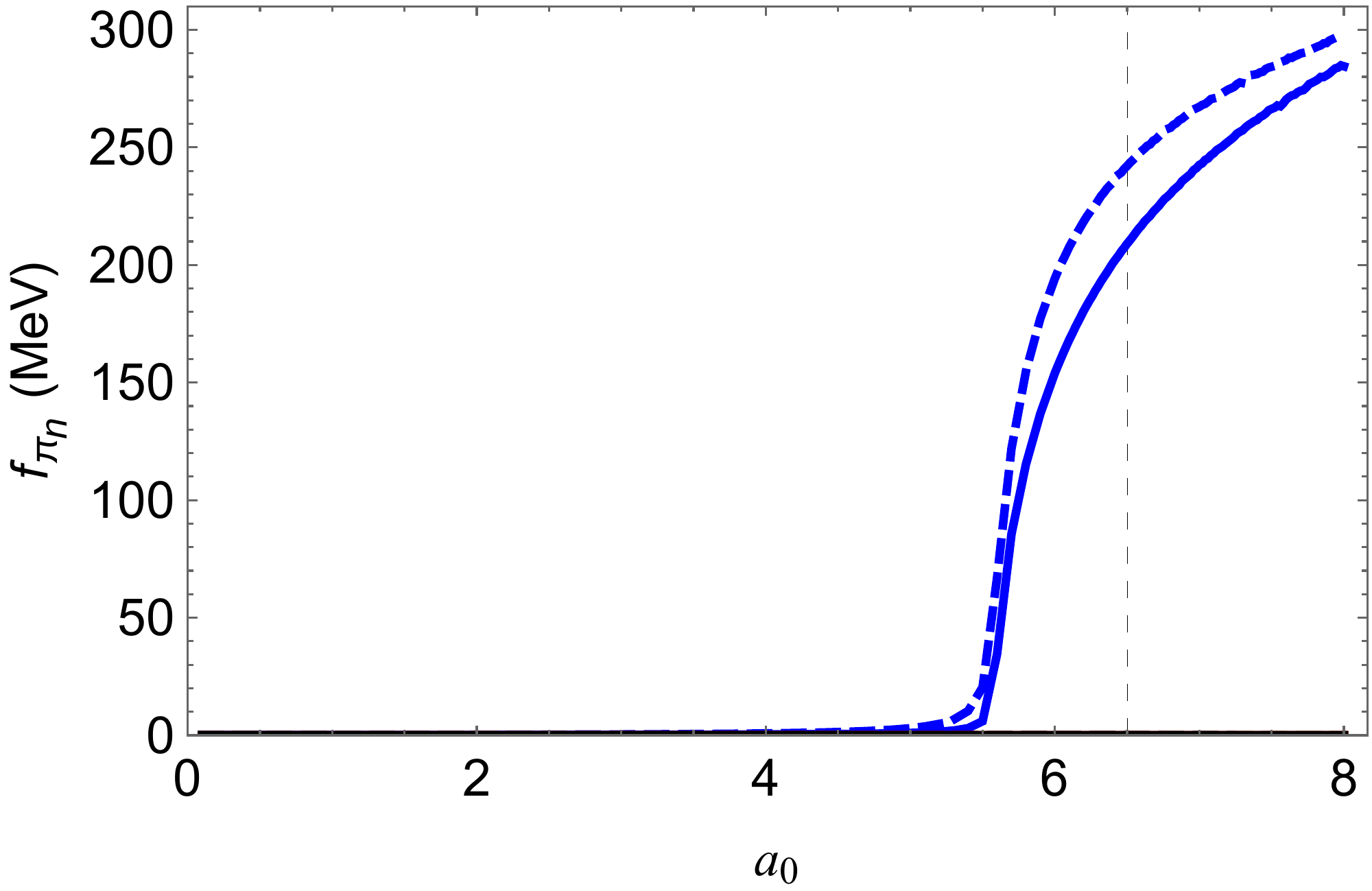}
\caption{
Left: The decay constants of the pseudoscalar mesons as a function of $a_0$ for models of type I, where solid lines represent Model IA ($\lambda=160$), while dashed lines Model IB ($\lambda=380$). Right: The decay constants of the pseudoscalar mesons as a function of $a_0$ for models of type II, where Solid lines represent Model IIA ($\lambda=60$), while dashed lines Model IIB ($\lambda=413$). We get these results for $m_q=3.63\times 10^{-4}\,\text{MeV}$ in both Models.
}
\label{Fig:DecayPi}
\end{figure}

Once again, the problem of finding decay constants has been reduced to the calculation of normalization constants. Our numerical results for the decay constants in models of type I as functions of $a_0$ are displayed on the left panel of Fig.~\ref{Fig:DecayPi}, where solid lines (dashed lines) represent results for Model IA (IB). Similarly, the results in models of type II are displayed on the right panel of Fig.~\ref{Fig:DecayPi}, where solid lines (dashed lines) represent results for Model IIA (IIB). All these results were obtained fixing the quark mass as $m_q=3.63\times 10^{-4}\,\text{MeV}$ (near the chiral limit)

Note that the decay constants of the excited states in all models are close to zero for $a_0< a_{0_c}$. This is consistent with the interpretation of these excited states in terms of pion resonances and is in qualitative agreement with the results obtained in Ref.~\cite{Ballon-Bayona:2020qpq}, where no Nambu-Goldstone boson was found.  Note, however, that the decay constant of the ground state starts to grow up for values close to the critical value $a_0\sim a_{0_c}$, meaning that it becomes finite for $a_0\geq a_{0_c}$. This is true for Models I and II. These results support our choice of $a_0=3.5$ for models of type I and $a_0=6.5$ for models of type II.

\begin{figure}[ht]
\centering
\includegraphics[width=7cm]{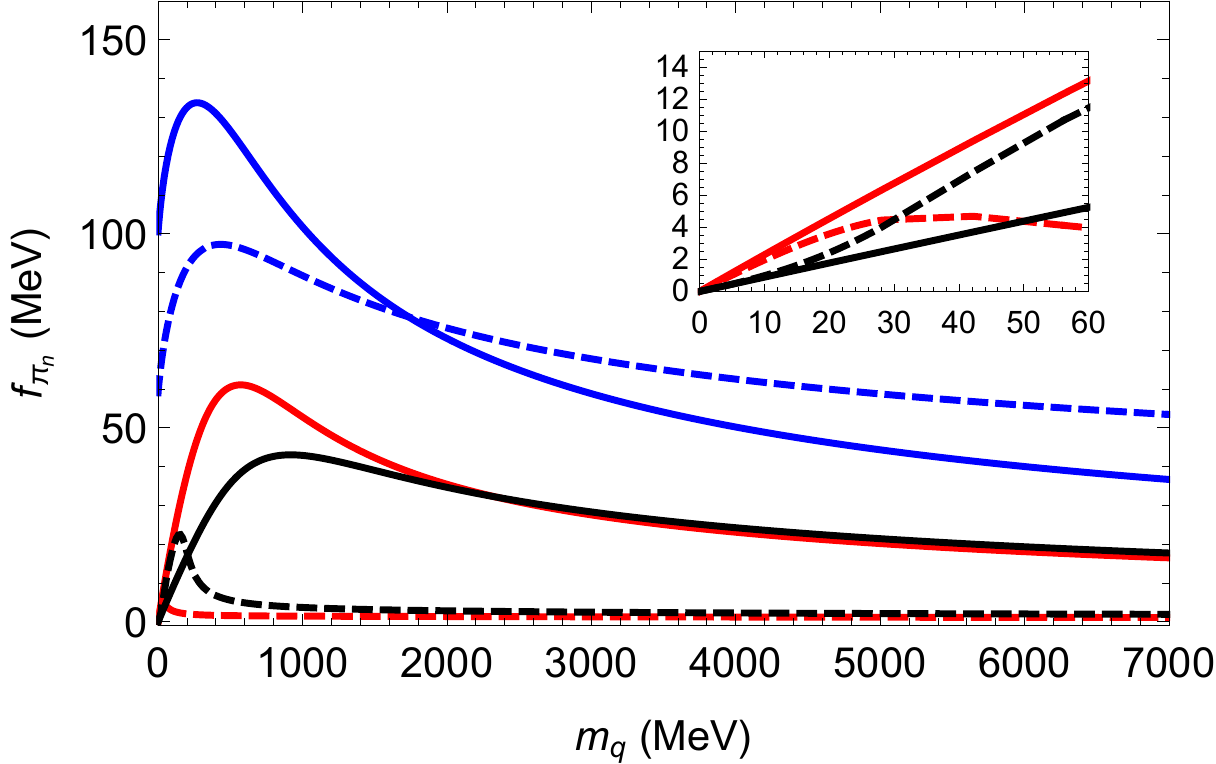}\hfill 
\includegraphics[width=7cm]{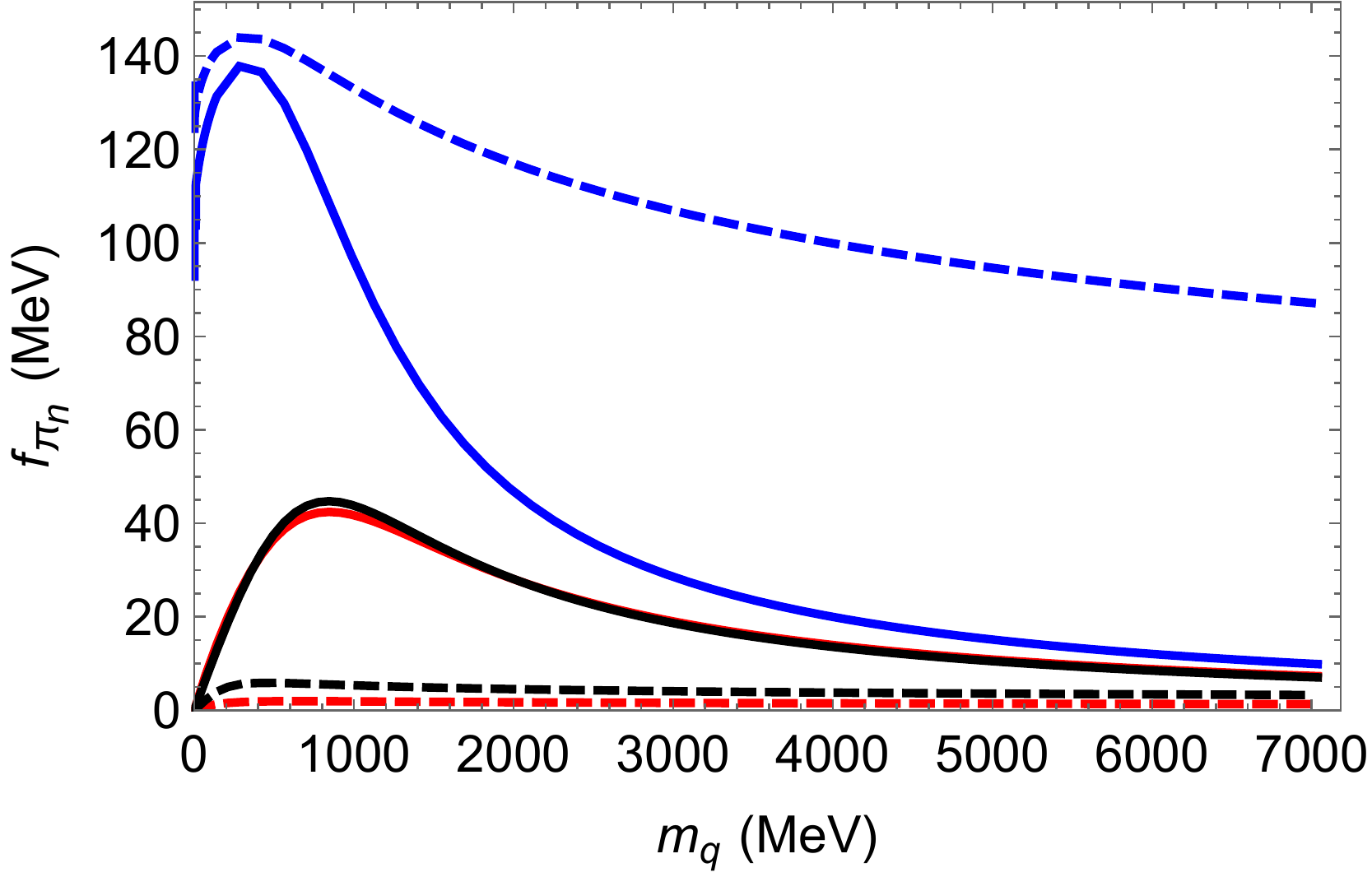}
\caption{
Left: Decay constants of  pseudoscalar mesons as functions of $m_q$, where solid lines represent the results for Model IA ($\lambda=160$), while dashed dashed lines for Model IB ($\lambda=380$).
Right: Decay constants of pseudoscalar mesons as functions of $m_q$, where solid lines represent the results for Model IIA ($\lambda=60$), while dashed dashed lines for Model IIB ($\lambda=413$).
}
\label{Fig:DecayPiDil2}
\end{figure}

The decay constants of pseudoscalar mesons in models of type I, with $a_0=3.5$, as functions of the quark mass are displayed on the left panel of Fig.~\ref{Fig:DecayPiDil2}, where results for Model IA (IB) are represented by solid (dashed) lines. As expected, the decay constant of the ground state is finite in the chiral limit, which supports our conclusion that the ground state is the pion. Note also that the decay constants of the resonances are zero in this limit. This result was previously observed in the holographic hard wall model \cite{Ballon-Bayona:2014oma}. We observe from the figure that the decay constants are non-monotonic functions of the quark mass up, growing in the regime of light quarks and then decreasing in the regime of heavy quarks. We also observe that the hierarchy between decay constants changes to  $f_{\pi_0}>f_{\pi_1}>f_{\pi_2}$ in the regime of light quarks, and becomes $f_{\pi_0}>f_{\pi_2}>f_{\pi_1}$ in the regime of heavy quarks. The results obtained for the resonances are in qualitative agreement with the results obtained in Ref.~\cite{Ballon-Bayona:2020qpq}. The results obtained for models of type II, displayed on the right panel of Fig.~\ref{Fig:DecayPiDil2}, are qualitatively similar to the results obtained in models of type I.

\noindent
\noindent

So far we have observed that the decay constant of the ground state can be related to the pion. To support this statement, we need the pion arising in the spectrum, as in fact we observed in Table \ref{Taba:Pseudoscalar}. Having fixed the model parameters in Table \ref{tab:Parameters}, 
we calculate the corresponding values of the pseudoscalar meson decay constants and display the results in Table \ref{Tab:DCPion}. The quark masses were fixed as $m_q=9\,\text{MeV}$,  $m_q=4.7\,\text{MeV}$, $m_q=9.8\,\text{MeV}$, and $m_q=26.8\,\text{MeV}$ for models of type IA, IB, IIA and IIB respectively.

\begin{table}[ht]
\centering
\begin{tabular}{l |c|c|c|c|l}
\hline 
\hline
 & Model IA &Model IB & Model IIA & Model IIB &  Experimental \\
  & $(a=b)$ & $(a\neq b)$ & $(a=b)$ & $(a\neq b)$ & ($f_{\pi^{+}}/\sqrt{2})$ \cite{Zyla:2020zbs}  \\
\hline 
 $f_{\pi_0}$  & 104.3 & 60.9 &118.3 &138.68& $92.1 \pm 0.8$  \\
 $f_{\pi_1}$  & 2.05  & 0.95 &3.94 &1.04&  \\
 $f_{\pi_2}$  & 0.79  & 0.42 &3.37 &2.97&  \\
\hline\hline
\end{tabular}
\caption{
Decay constants (in MeV) obtained in models of type I and models of type II for the set of parameters given in Table \ref{tab:Parameters}, compared against the experimental results of PDG \cite{Zyla:2020zbs}. The quark masses were fixed as $m_q=9\,\text{MeV}$,  $m_q=4.7\,\text{MeV}$, $m_q=9.8\,\text{MeV}$, and $m_q=26.8\,\text{MeV}$ for models of type IA, IB, IIA and IIB respectively.
}
\label{Tab:DCPion}
\end{table}

\subsection{The GOR relation}
\label{Subsec:GOR}

The final step to check if the lightest state arising in the pseudoscalar sector is the pion, it must satisfy the Gell-Mann-Oakes-Renner (GOR) relation. In the following analysis we follow the procedure implemented in Ref.~\cite{Abidin:2009aj} adapted to the holographic model we are working with. The idea is to use the normalization condition \eqref{Eq:NormPi} together with the definition of the decay constant \eqref{Eq:DecayPi1}. It is worth mentioning that the functions $a(\Phi)$ and $b(\Phi)$ are subleading close to the boundary, where the leading term is the warp factor $A_s=-\ln{z}$. Thus, we may rewrite \eqref{Eq:NormPi} in the form (setting $m=n$)
\noindent
\begin{equation}\label{eq:NormalCondGOR}
g_5^2\,m_{\pi_n}^2\int_{\epsilon}^{\infty} dz\,\frac{e^{-A_s-b}}{\beta(z)}\left(\frac{e^{A_s}}{g_5}\partial_z\varphi_n\right)\left(\frac{e^{A_s}}{g_5}\partial_z\varphi_n\right)=1 \, ,
\end{equation}
\noindent
or 
\begin{equation}\label{eq:NormalCondGORv2}
\,m_{\pi_n}^2\int_{\epsilon}^{\infty} dz\,\frac{e^{-3A_s-b}}{v^2(z) } (e^{a-b}) \left(\frac{e^{A_s}}{g_5}\partial_z\varphi_n\right)\left(\frac{e^{A_s}}{g_5}\partial_z\varphi_n\right)=1 \, ,
\end{equation}
where $\beta$ was defined in \eqref{Eq:Beta}. Near the chiral limit, i.e $m_q \to 0$, it can be shown that the term $e^{-3A_s-b}/v^2(z)$ in the integral becomes highly peaked near the boundary $z=\epsilon$. If the terms in parenthesis are nonzero in that limit they can be moved outside of the integral. This is true only for the ground state because it is the only state that has a nonzero decay constant in the chiral limit, see Eq.~\eqref{Eq:DecayPi1} and Fig.~\ref{Fig:DecayPiDil2}. 

Then for the ground state we find the following result near the chiral limit:
\noindent
\begin{equation}\label{Eq:NorCondition2}
f_{\pi_0}^2m_{\pi_0}^2\int_{\epsilon}^{\infty} \frac{e^{-3A_s-b}}{v^2(z)}dz=1.
\end{equation}
\noindent
Near the chiral limit the integrand $e^{-3A_s-b}/v^2(z)$ has its relevant contribution close to the boundary, i.e., $e^{-3A_s-b}/v^2(z)\sim z/(c_1+c_3z^2)^2$. 
Hence, the integral provides the result
\noindent
\begin{equation}\label{Eq:NorCondition3}
\begin{split}
\int_{\epsilon}^{\infty} \frac{e^{-3A_s-b}}{v^2(z)}dz\approx\,&\frac{1}{2 c_1c_3}.
\end{split}
\end{equation}
\noindent
This result is valid only for small $c_1$. Then, plugging \eqref{Eq:NorCondition3} in \eqref{Eq:NorCondition2}, and considering that $c_1=m_q\,\zeta$ and $c_3={\Sigma}/(2\zeta)$, we get the Gell-Mann-Oakes-Renner (GOR) relation
\noindent
\begin{equation}
f_{\pi_0}^2m_{\pi_0}^2=m_q\Sigma.
\end{equation}
\noindent
One may rewrite the last equation using the relation $\Sigma=2\sigma$, where $\sigma$ is the up and down quark condensates $\langle \bar{u}u\rangle=\langle \bar{d}d\rangle=\sigma$,
\noindent
\begin{equation}
f_{\pi_0}^2m_{\pi_0}^2=2m_q\sigma.
\end{equation}
\noindent

\begin{figure}[ht]
\centering
\includegraphics[width=7cm]{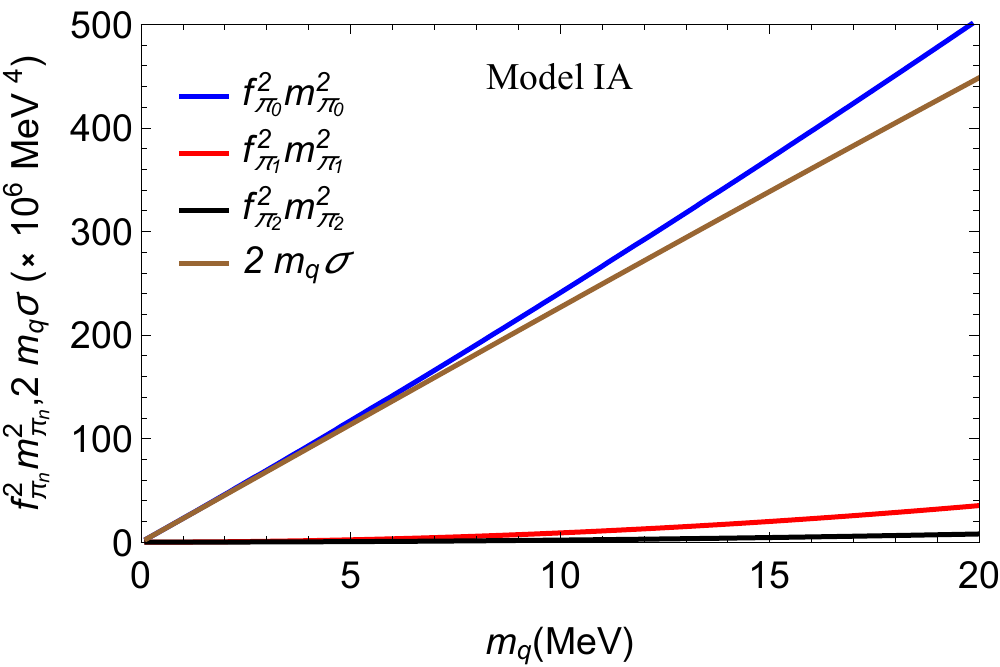}\hfill 
\includegraphics[width=7cm]{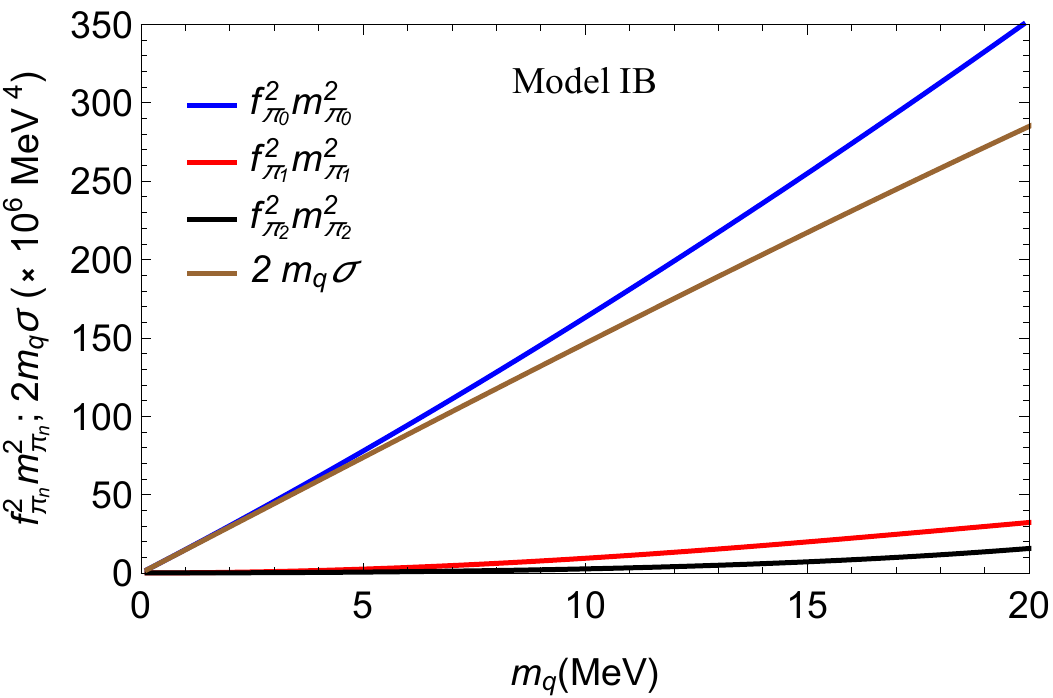}\\
\includegraphics[width=7cm]{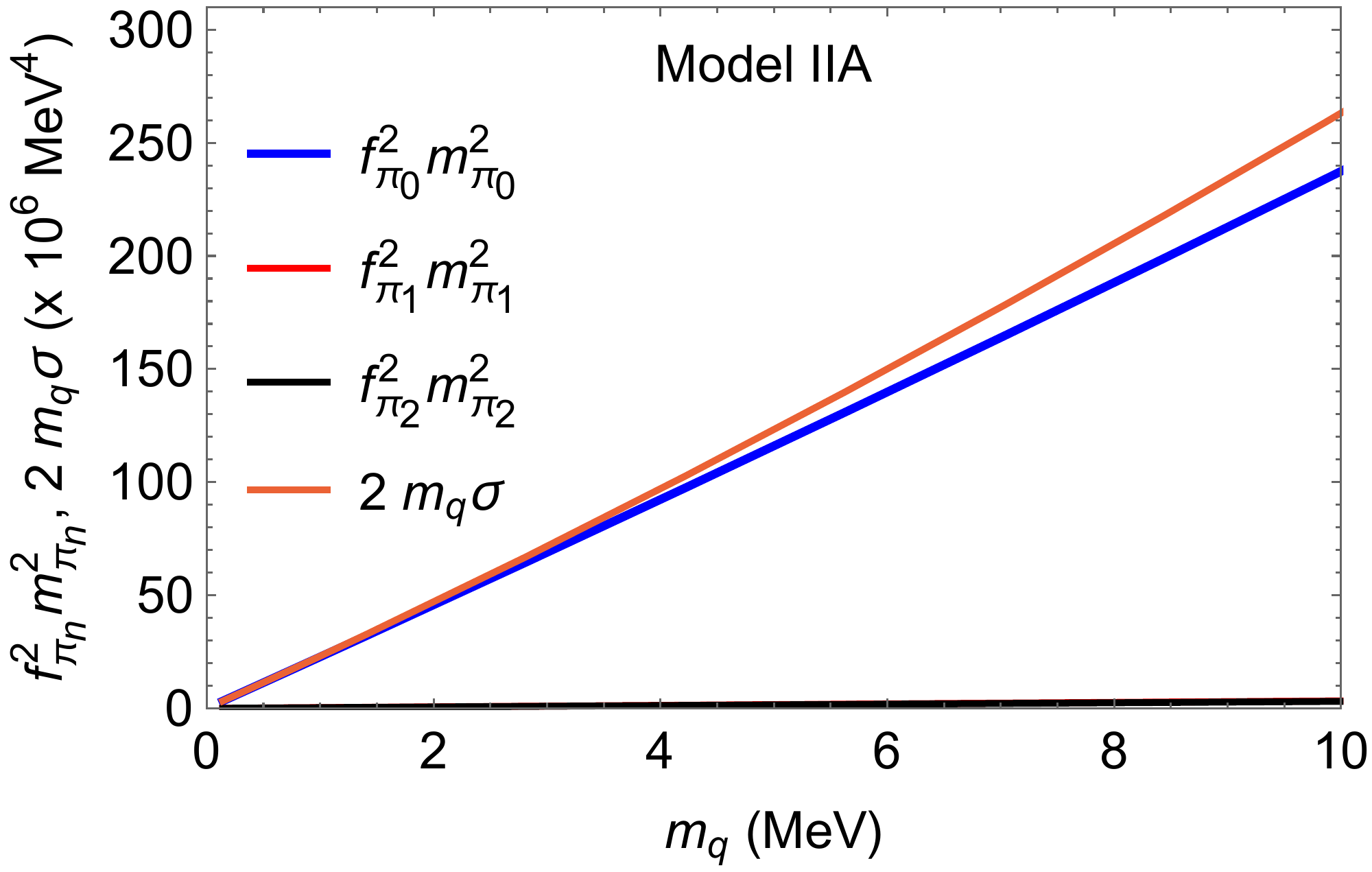}\hfill 
\includegraphics[width=7cm]{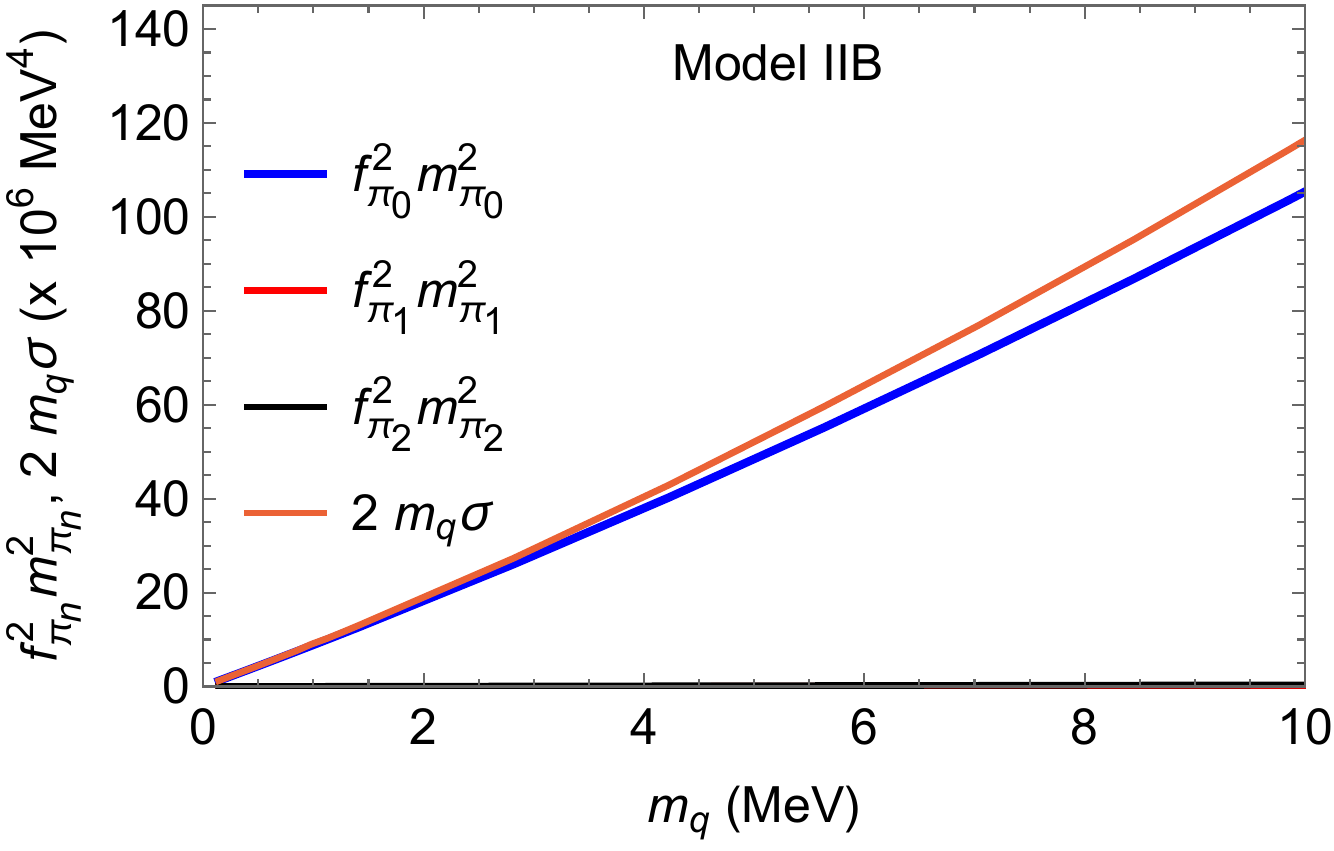}
\caption{
The product $f_{\pi_{n}}^2m_{\pi_n}^2$ and $2\,m_q\sigma$ as a function of the quark mass for models of type I (top panel) and models of type II (bottom panel). As can be seen, the GOR relation arises for the ground state in the regime of light quarks.
}
\label{Fig:GOR}
\end{figure}

To support the last procedure we calculated the GOR relation numerically. Our results are displayed in Fig.~\ref{Fig:GOR}. As can be seen, the ground state satisfies the GOR relation in the region of light quarks.

To understand better the approximation we have done in the integral \eqref{Eq:NorCondition3}, we refer the reader to Appendix \ref{Sec:GORToyModel}, where we discuss the derivation of the GOR relation using a toy model.

\noindent
\noindent

\section{Discussion and conclusions}
\label{Sec:Conclusions}

In this paper, we have extended the original soft wall model  considering a Higgs potential for the tachyonic field and non-minimal dilaton couplings $\exp(a(\Phi))$ and $\exp(b(\Phi))$ for the tachyonic and gauge field respectively. As in the original soft wall model, the 5d metric was fixed as the AdS spacetime in Poincar\'e coordinates and the dilaton $\Phi$ was fixed to be quadratic in the radial coordinate $z$. In order to guarantee good properties for the meson spectrum such as approximate linear behaviour with the excitation number and absence of spurious massless modes we imposed that at large $\Phi$ (far from the boundary) the non-minimal couplings reduce to the minimal case, i.e. $a(\Phi)$ and $b(\Phi)$ reduce to $\Phi$. 

Our work was inspired by previous works on soft wall models such as \cite{deTeramond:2009xk,Zuo:2009dz,Gherghetta:2009ac,Chelabi:2015gpc} that realised that considering an effective dilaton field, which in our notation corresponds to a particular choice for the non-minimal couplings $a(\Phi)=b(\Phi)=\tilde \Phi$, that is negative in some region it is possible to incorporate spontaneous chiral symmetry. 
In order to avoid the massless mode found in Ref.~\cite{Karch:2010eg} and instabilities in the background we considered dilaton couplings $a(\Phi)$ and $b(\Phi)$ that satisfy the IR constraint $a(\Phi)=b(\Phi)=\Phi$.  To simplify the analysis we considered two possibilities for the dilaton coupling $b(\Phi)$, namely $b(\Phi)=a(\Phi)$ (models of type A) and $b(\Phi)=\Phi$ (models of type B). For the dilaton coupling $a(\Phi)$ we considered two interpolations,  namely $a_I(\Phi)$ (models of type I) and $a_{II}(\Phi)$ (models of type II). The two interpolations $a_I(\Phi)$ and $a_{II}(\Phi)$, displayed in Fig. \ref{Fig:Interpolations}, share the feature that are non-monotonic and become negative in some region. The main difference between them is that $a_I(\Phi)$ reduces to $-a_0 \Phi$ at small $\Phi$ (near the boundary) whereas $a_{II}(\Phi)$ reduces to $\Phi$.  The four different type of models IA, IB, IIA and IIB were described in table \ref{tab:Models}.

Is it worth pointing out that soft wall models introduce an ``artificial'' confinement because the 5d metric is fixed to AdS spacetime and therefore the warp factor in the string frame does not have a minimum, required by confinement \cite{Kinar:1998vq}. It turns, however, that taking into account the backreaction of the dilaton field into the 5d metric within the framework of Einstein-dilaton theories the dilaton quadratic behaviour predicted by the soft wall model at large $z$ remains intact, see e.g. \cite{Gursoy:2007er,Ballon-Bayona:2017sxa}.

The non-minimal couplings introduced in this work brought a new parameter $a_0$ which controls the emergence of spontaneous chiral symmetry breaking, as shown in Fig.~\ref{Fig:PhiC3}. We realized that the addition of a Higgs potential for the tachyonic field was not enough to induce spontaneous symmetry breaking, i.e., the emergence of the pion as a Nambu-Goldstone boson, as was previously shown in Ref.~\cite{Ballon-Bayona:2020qpq}.  It seems to us that the reason behind the requirement of non-minimal couplings is due to the fact that the Higgs potential, usually introduced in 4d Minkowski spacetime to describe spontaneous symmetry breaking, has been extended to 5d AdS spacetime. The non-minimal dilaton couplings somehow compensate the deviations from the Mexican Hat potential due to AdS spacetime. Moreover, we have verified that a scalar perturbation around the trivial solution $v(z)=0$ can be written in an AdS form where the effective 5d mass $m_X^2$ evolves with the radial direction $z$ violating the BF bound for some values of $z$. This BF violation was caused by the presence of the non-minimal dilaton couplings and signifies  the instability of the trivial solution. This, in turn, indicates the presence of a non-trivial solution for $v(z)$ that is stable. This is the solution that we find in this work and it is the gravity dual of a non-perturbative vacuum that breaks chiral symmetry spontaneously. We conclude that the addition of non-minimal couplings with an additional parameter is well justified in order to induce spontaneous chiral symmetry breaking in the chiral limit and the emergence of pions as Nambu-Goldstone bosons. 

Once the non-minimal couplings were fixed, the tachyon profile was obtained numerically. We noticed a scaling symmetry for the tachyon solution that allowed us to find the solution $v(z)$  for any value of the Higgs coupling $\lambda$ in terms of the rescaled field $\tilde v(z) = \sqrt{\lambda} v(z)$.
By solving the differential equations of the field perturbations we were able to find the spectrum of vector, scalar, axial-vector and pseudo-scalar mesons. It turns out that the differential equation for the scalar sector depends only on $\tilde v(z)$ and therefore in the chiral limit the scalar meson masses depend only on the parameter $a_0$ associated with the non-minimal coupling. Interestingly, the lightest scalar meson becomes massless exactly at the same critical value $a_0^c$ where the transition between explicit and spontaneous chiral symmetry breaking takes place. This strongly suggests that conformal symmetry was also spontaneously broken for that critical value.  For $a_0>a_0^c$ we are in the regime of spontaneous chiral symmetry breaking and the lightest scalar meson grows monotonically, as shown in Fig. \ref{Fig:Msa0}. Considering these results we fixed the parameter $a_0$ at some value above $a_0^c$ in order to guarantee the absence of a massless scalar meson and the presence of spontaneous chiral symmetry breaking. We fixed $a_0=3.5$ in models of type I and $a_0=6.5$ in models of type II. In turn, the parameter $\lambda$ of the Higgs potential was fixed separately for each model in order to guarantee axial-vector meson masses $m_{A_n}$ and pseudo-scalar meson masses $m_{\pi^n}$ that do not present any crossing between the fundamental and excited states. Finally, the parameter $\phi_{\infty}$  of the dilaton field was fixed to the usual value used in soft wall models. The choice of parameters in the four different type of models considered in this work was summarized in table \ref{tab:Parameters}. For those particular parameter choices we analyzed the evolution of meson masses with the quark mass in the scalar, axial and pseudo-scalar sector (the vectorial sector is independent of the quark mass in this framework).  We found that all the meson masses grow monotonically with the quark mass and that as we approach the chiral limit massless modes emerge in the pseudo-scalar sector, identified as the Nambu-Goldstone bosons of spontaneous chiral symmetry breaking. Finally, fixing the quark mass we were able to compare our results for the meson masses against experimental data and previous results in soft wall models.

We also calculated the decay constants for all the mesons. 
We found a peculiar behavior for the decay constants of vector, scalar and axial-vector mesons, changing the hierarchy in certain intervals of $a_0$. This peculiar behavior is almost absent in models of type II, see figures \ref{Fig:DecayVDil2}, \ref{Fig:DecayS} and \ref{Fig:DecayAV}.  The decay constants of the pseudoscalar sector as a function of $a_0$ shows the emergence of the pion for $a_0\geq a_{0_c}$, see Fig.~\ref{Fig:DecayPi}.
For the fixed values of $a_0$, $\lambda$ and $\phi_{\infty}$ given in table \ref{tab:Parameters} we investigated the evolution of the decay constants of axial-vector, scalar and pseudo-scalar mesons with the quark mass. We found that the hierarchy between fundamental and excited states change as the quark mass evolves from the near chiral limit to the heavy quark limit. In any case, all the decay constants decrease with the quark mass as we approach the heavy quark limit, which is in qualitative agreement with expectations from perturbative QCD. We showed that the emergence of the pions as Nambu-Goldstone bosons associated with spontaneous chiral symmetry breaking is supported by the decay constants in the pseudo-scalar sector, obtained in Fig.~\ref{Fig:DecayPiDil2}. The results near the chiral limit are qualitatively similar to the results obtained in the hard-wall model \cite{Ballon-Bayona:2014oma}. As a check of consistency we showed that our models satisfy the Gell-Mann-Oakes-Renner (GOR) relation, see Fig.~\ref{Fig:GOR}. Finally, fixing the quark masses we were able to compare our results for the decay constants against experimental data and previous results in the literature.

In conclusion, we have shown in this work that spontaneous chiral symmetry breaking can be described using a holographic set up that extends the original soft wall model by including a Higgs potential for the tachyon and non-minimal dilaton couplings for the tachyon and gauge fields. Our numerical results show no instabilities in the spectrum. We realized that an additional parameter was needed in order to control the transition between spontaneous and explicit chiral symmetry breaking. The non-minimal dilaton couplings  can be justified by the fact that we are describing the dynamics of a Higgs potential for a scalar field (the tachyon) in a curved space (AdS). These non-minimal couplings, combined with the Higgs potential, allowed us to find solutions for the tachyonic field with a VEV coefficient that remains finite in the limit where the source coefficient goes to zero (chiral limit). It is possible that the requirement of non-minimal couplings for describing spontaneous chiral symmetry breaking is an artifact of soft wall models in the sense that the background is not obtained by solving the Einstein's equations. We will test this hypothesis  investigating chiral symmetry breaking in backgrounds  obtained by solving the Einstein-dilaton equations, see for instance \cite{Ballon-Bayona:2017sxa, Ballon-Bayona:2018ddm, Gursoy:2007er, Li:2013oda}. Since those backgrounds describe color confinement in a consistent way it would be interesting to investigate the connection between confinement and spontaneous chiral symmetry breaking. 

Further extensions of this project include finite temperature effects of mesons and quarks in a non-conformal plasma \cite{Mamani:2013ssa, Miranda:2009uw,Bayona:2009qe,Mamani:2018uxf, Bartz:2016ufc} as well as its relation to the spectrum of quasinormal modes (QNMs). Backreaction effects of the tachyonic and gauge fields would also be an interesting problem in order to find the quark contribution to the thermodynamics  and transport coefficients of a non-conformal plasma (for related work in the top-down approach see \cite{Bigazzi:2009bk,Bigazzi:2009tc}) . We will address some of these problems in the future.

\section*{Acknowledgments}
The work of A.B-B is partially funded by Conselho Nacional de Desenvolvimento Cient\'\i fico e Tecnol\'ogico (CNPq, Brazil), grants No. 306528/2018-5 and No. 434523/2018-6 and Coordena\c{c}\~ao de Aperfei\c{c}oamento do Pessoal de N\'ivel Superior (CAPES, Brazil), Finance Code 001. L.~A.~H.~M. has financial support from Coordena\c{c}\~ao de Aperfei\c{c}oamento do Pessoal de N\'ivel Superior - Programa Nacional de P\'os-Doutorado (PNPD/CAPES, Brazil). D. M. R. is supported by grant \raisebox{\depth}{\#}2021/01565-8, São Paulo Research Foundation (FAPESP).

\appendix

\section{Numerical analysis}
\label{Sec:NumericalAnalysis}

\begin{figure}[ht!]
\centering
\includegraphics[width=7cm]{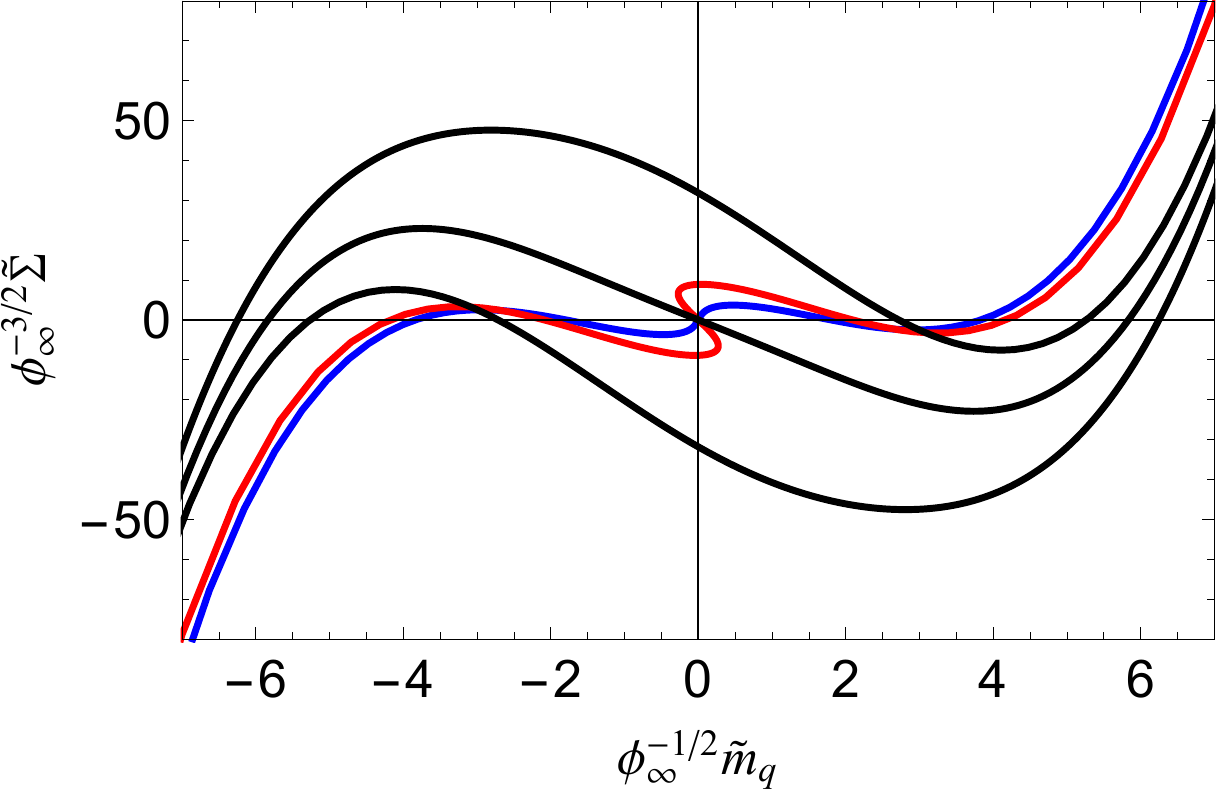}
\hfill
\includegraphics[width=7cm]{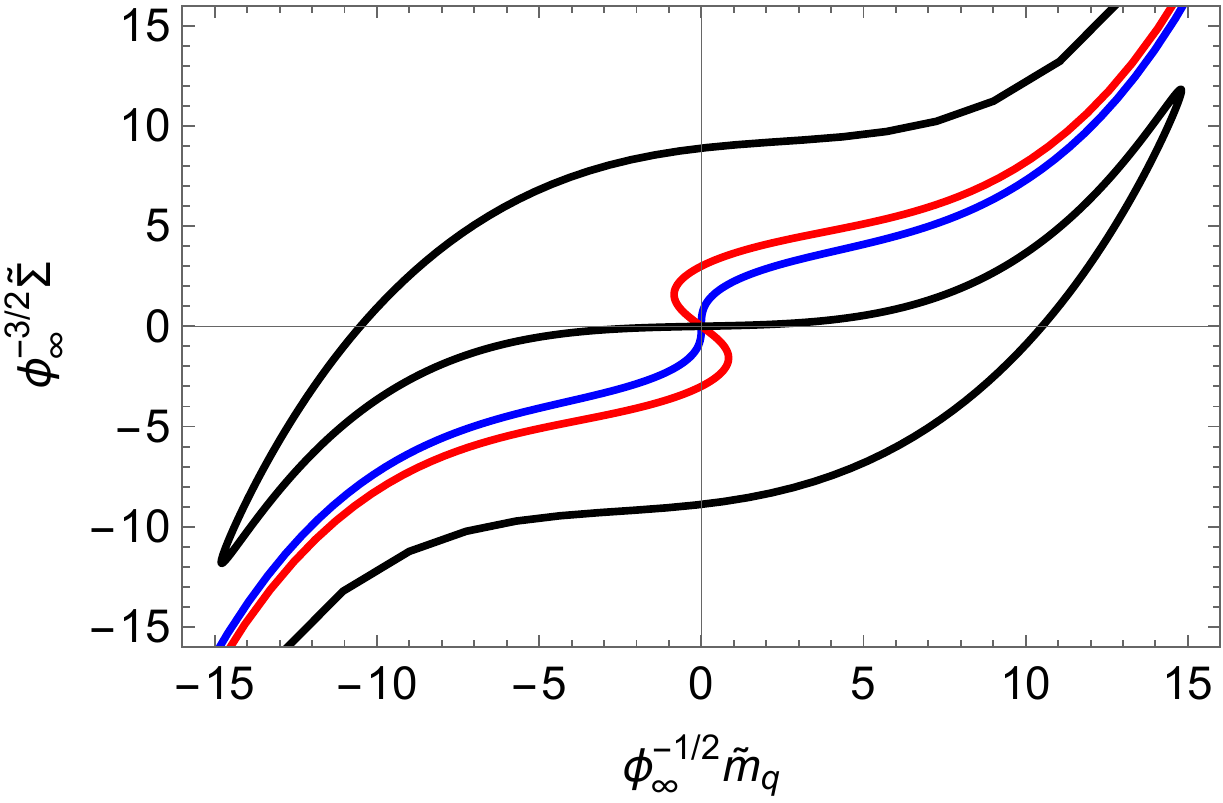}
\caption{
Left: Dimensionless condensate $\phi_{\infty}^{-3/2}\widetilde{\Sigma}$ as a function of the dimensionless quark mass $\phi_{\infty}^{-1/2}\widetilde{m}_q$ for models of type I with different values of $a_0$: $a_0=2.8$ (blue line), $a_0=3.5$ (red line), and $a_0=6$ (black lines).
Right: Dimensionless condensate $\phi_{\infty}^{-3/2}\widetilde{\Sigma}$ as a function of the dimensionless quark mass $\phi_{\infty}^{-1/2}\widetilde{m}_q$ for models of type II with different values of $a_0$: $a_0=5.6$ (blue line), $a_0=6.5$ (red line), and $a_0=10$ (black line).
}
\label{Fig:C1C3InterDil}
\end{figure}

Here, we write details on the numerical procedure. We realized that the numerical results obtained solving the tachyon differential equation \eqref{Eq:TachyonDE} can be split up in three regions depending on the value for the parameter $a_0$. Firstly, for $0\leq a_0< a_{0_c}$, one may vary $\widetilde{C}_0$ continuously in the interval $-\sqrt{6}\leq \widetilde{C}_0 \leq\sqrt{6}$. In this region, there is explicit chiral symmetry breaking because $\widetilde{c}_3$ is zero in the chiral limit, i.e., $\widetilde{m}_q$=0, note that $\widetilde{m}_q\propto\widetilde{C}_0$ in this region. Moreover, $a_{0_c}$ depends on the form for the interpolation function. Here we work with Models I and II. The numerical results for $\phi_{\infty}^{-3/2}\widetilde{\Sigma}$ as a function of $\phi_{\infty}^{-1/2}\widetilde{m}_q$ are displayed in Fig.~\ref{Fig:C1C3InterDil} with blue lines for both Models. As can be seen, the parameter $\phi_{\infty}^{-3/2}\widetilde{\Sigma}$ goes to zero in the chiral limit for $a_0<a_{0_c}$.

Secondly, for $a_{0_c}\leq a_0\leq a_{0_{\text{max}}}$, $\widetilde{C}_0$ still vary in the interval $-\sqrt{6}\leq \widetilde{C}_0 \leq\sqrt{6}$. Nevertheless, there is spontaneous chiral symmetry breaking because $\phi_{\infty}^{-3/2}\widetilde{\Sigma}$ is nonzero in the chiral limit. $a_{0_{\text{max}}}\sim 5.954$ for models of type I. In this region, the parameter $a_0$ controls the value of the condensate. Our numerical results for Models I and II are displayed in Fig.~\ref{Fig:C1C3InterDil} with red lines, left panel shows $\phi_{\infty}^{-3/2}\widetilde{\Sigma}$ as a function of $\phi_{\infty}^{-1/2}\widetilde{m}_q$ for light quarks. However, as can be seen in the right panel of this figure, $\phi_{\infty}^{-3/2}\widetilde{\Sigma}$ becomes negative in the intermediate region. May be this behavior is suggesting that the model with negative dilaton in the UV is not enough to describe a monotonic increasing function for $\phi_{\infty}^{-3/2}\widetilde{\Sigma}$ expected in real world QCD. In turn, models of type II provides improved results compared against the results obtained in models of type I, as can be seen in the right panel of Fig.~\ref{Fig:C1C3InterDil}. This means that the negative dilaton in the intermediate region is enough to guarantee a monotonic increasing function for $\phi_{\infty}^{-3/2}\widetilde{\Sigma}$ as a function of $\phi_{\infty}^{-1/2}\widetilde{m}_q$.

Thirdly, for $a_0>a_{0_{\text{max}}}$, $\widetilde{C}_0$ is constrained to take values only close to the minimum of the Higgs potential $\widetilde{C}_0<\sqrt{6}-\delta_1$, $\widetilde{C}_0>-\sqrt{6}+\delta_1$, and close to the trivial solution $-\delta_2 \leq \widetilde{C}_0=0 \leq \delta_2$, where $\delta_1$ and $\delta_2$ are small numbers. This means that the solution splits up into three branches. The numerical results for models of type I, where $a_0\sim 5.954$, are displayed in the left panel of Fig.~\ref{Fig:C1C3InterDil} with black lines. As can be seen, each branch corresponds to the solutions mentioned above. Note again that the results for models of type I shows a negative condensate in the intermediate region. In turn, numerical results for models of type II are displayed in the right panel of this figure. Note also that the condensate $\phi_{\infty}^{-3/2}\widetilde{\Sigma}$ is a monotonically increasing function of $\phi_{\infty}^{-1/2}\widetilde{m}_q$, see the right panel.

\begin{figure}[ht!]
\centering
\includegraphics[width=7cm]{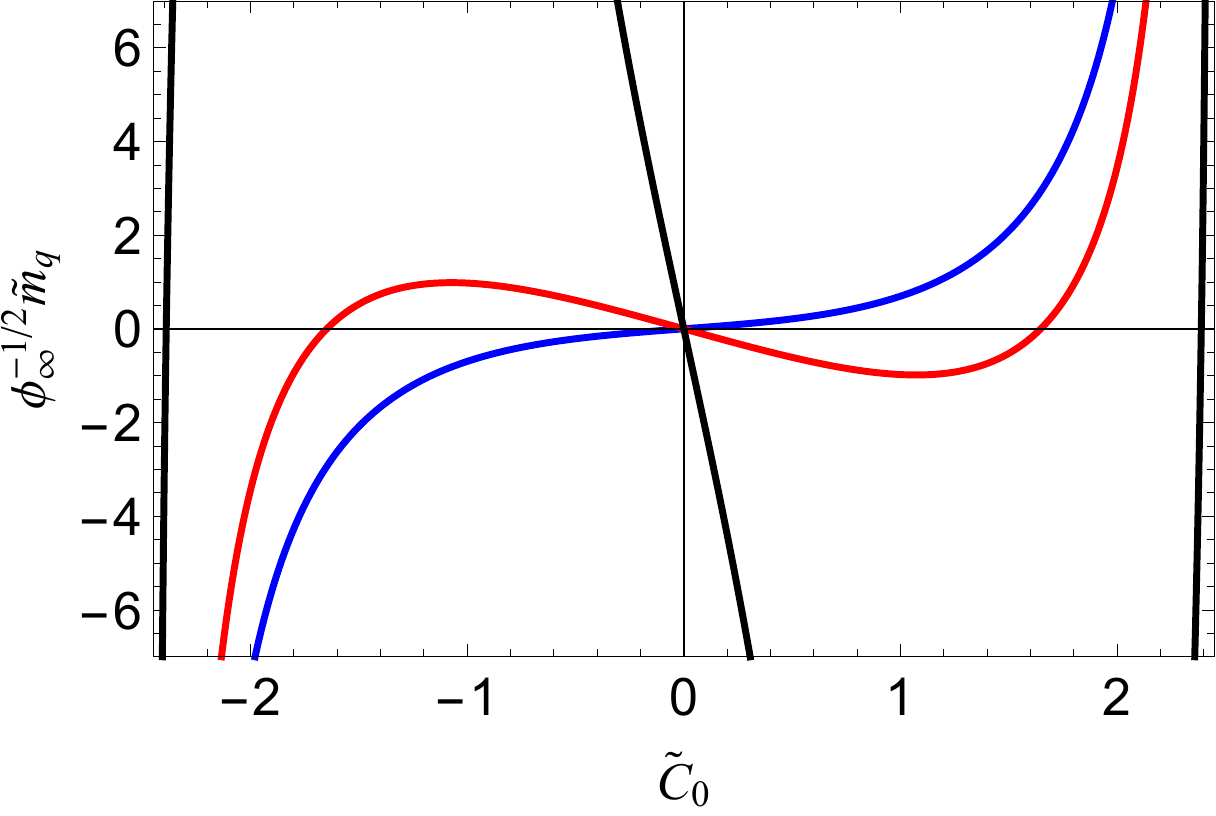}
\hfill
\includegraphics[width=7cm]{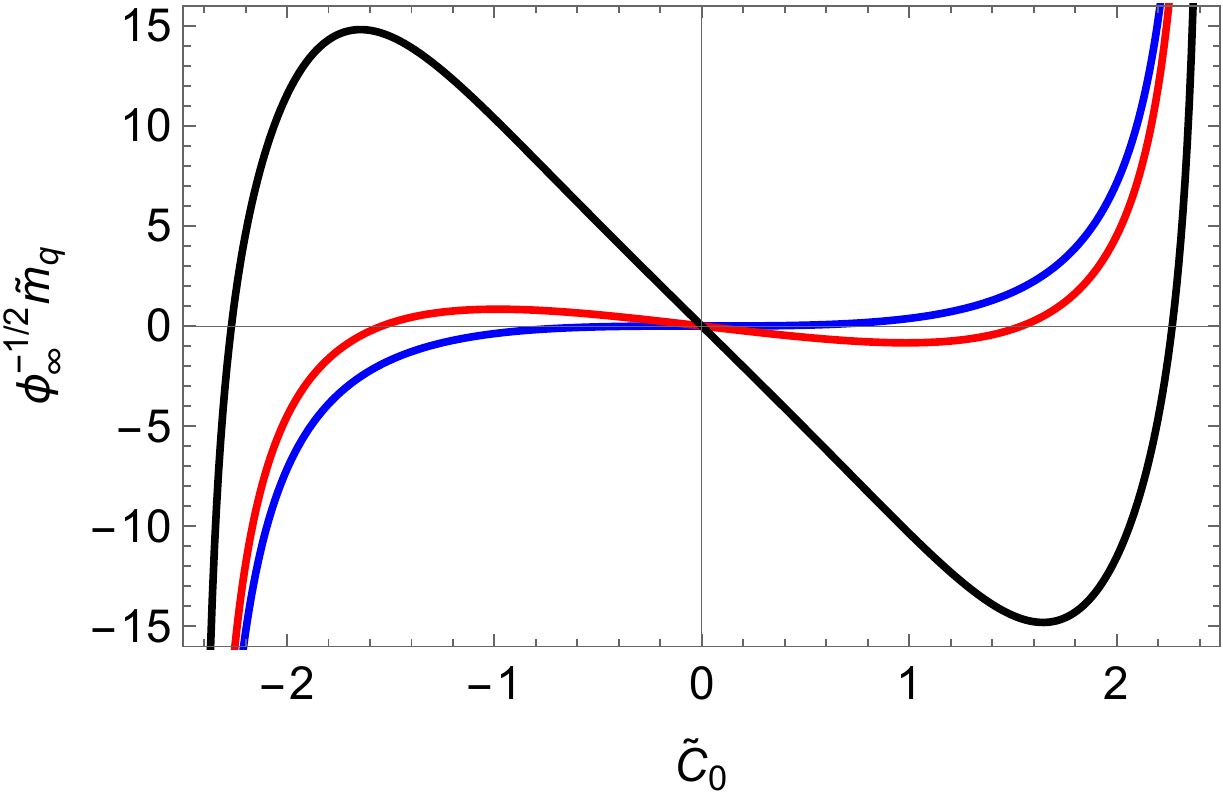}
\caption{
Left: Dimensionless quark mass $\phi_{\infty}^{-1/2}\widetilde{m}_q$ as a function of $\widetilde{C}_0$ for models of type I with different values of $a_0$: $a_0=2.8$ (blue line), $a_0=3.5$ (red line), and $a_0=6$ (black line).
Right: Dimensionless quark mass $\phi_{\infty}^{-1/2}\widetilde{m}_q$ as a function of $\widetilde{C}_0$ for models of type II with different values of $a_0$: $a_0=5.6$ (blue line), $a_0=6.5$ (red line), and $a_0=10$ (black line).
}
\label{Fig:C0C1InterDil}
\end{figure}

On the other hand, a complementary analysis for models of type I shows the evolution of the dimensionless quark mass $\phi_{\infty}^{-1/2}\widetilde{m}_q$ as a function of $\widetilde{C}_0$ in the left panel of Fig.~\ref{Fig:C0C1InterDil}. As can be seen, $\widetilde{C}_0$ is constrained to take values close to the minimum of the Higgs potential, as well as close to the trivial solution for $a_0>a_{0_{\text{max}}}$. In turn, the right panel of this figure shows the results for models of type II.
\noindent

Finally, we also get the dimensionless condensate $\phi_{\infty}^{-3/2}\widetilde{\Sigma}$ as a function of $\widetilde{C}_0$. Our numerical results for models of type I are displayed in the left panel of Fig.~\ref{Fig:C0C1InterDil}. As can be seen, $\widetilde{C}_0$ is constrained to take values close to the minimum of the Higgs potential, as well as close to the trivial solution for $a_0>a_{0_{\text{max}}}$. In turn, the right panel of this figure shows the corresponding results for models of type II.

\begin{figure}[ht!]
\centering
\includegraphics[width=7cm]{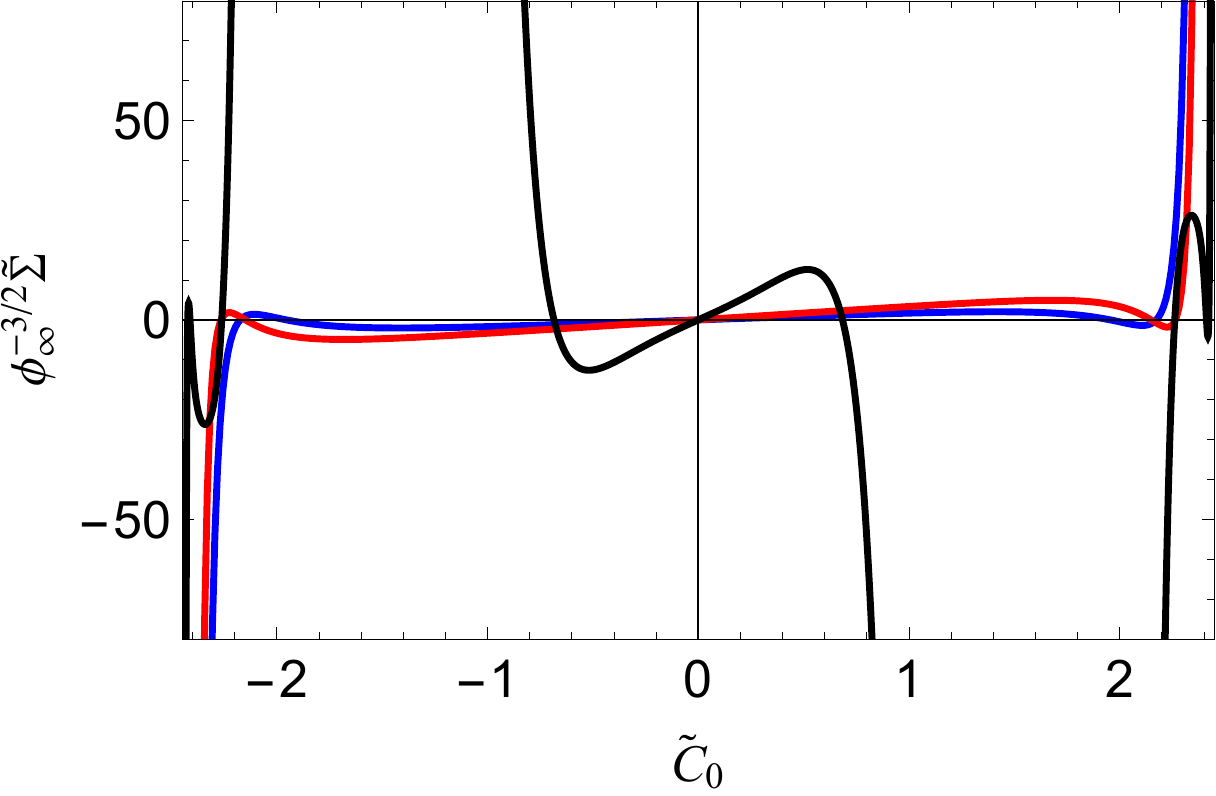}
\hfill
\includegraphics[width=7cm]{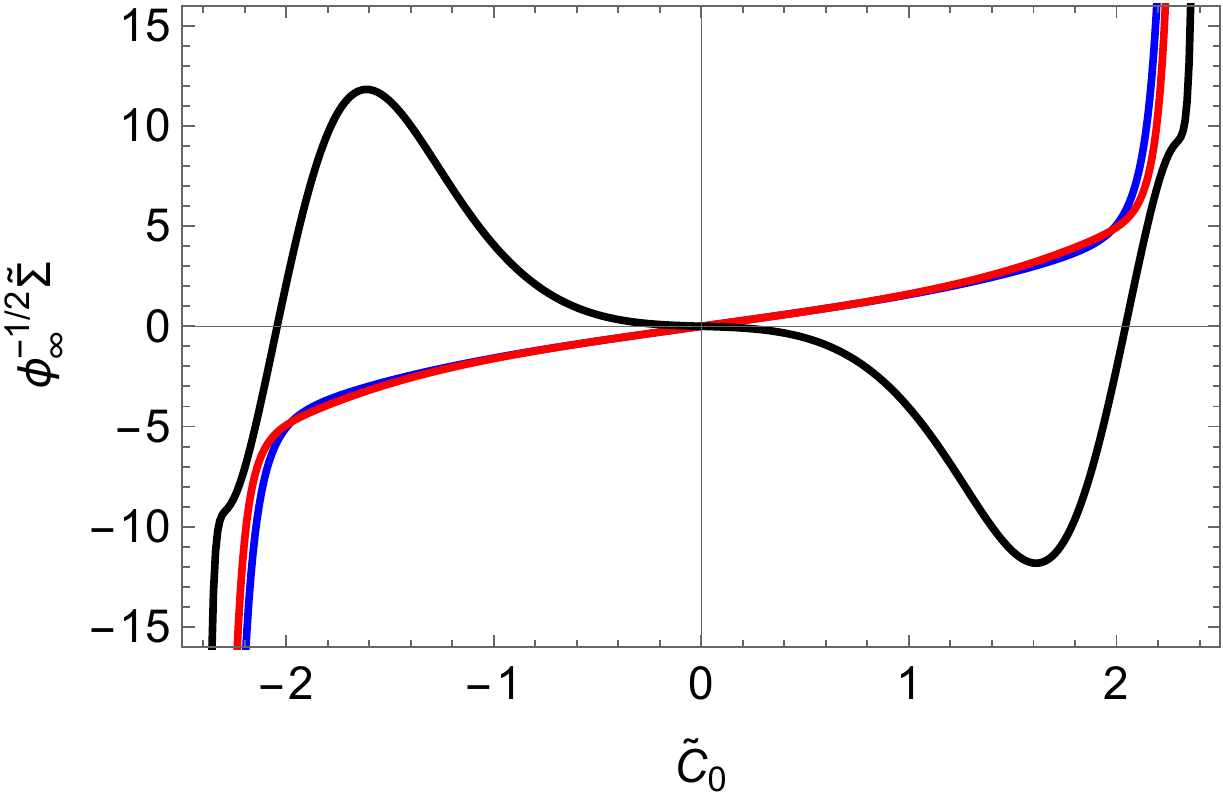}
\caption{
Left: Dimensionless condensate $\phi_{\infty}^{-3/2}\widetilde{\Sigma}$ as a function of $\widetilde{C}_0$ for models of type I with different values of $a_0$: $a_0=2.8$ (blue line), $a_0=3.5$ (red line), and $a_0=6$ (black lines).
Right: Dimensionless condensate $\phi_{\infty}^{-3/2}\widetilde{\Sigma}$ as a function of $\widetilde{C}_0$ for models of type II with different values of $a_0$: $a_0=5.6$ (blue line), $a_0=6.5$ (red line), and $a_0=10$ (black lines).
}
\label{Fig:C0C3InterDil}
\end{figure}

\section{The Kaluza-Klein expansion}
\label{app:KKexp}

The starting point is the 5d action
\begin{align}
S_2 &= - V_4 \int dz \, \Big \{   e^{3 A_s -a}  \Big [ \frac12 (\partial_{\hat m} S )^2   + \frac12 e^{2 A_s} \frac{d^2 U}{dv^2}S^2 
 \Big ] \nonumber \\
&+ \frac{1}{g_5^2} e^{A_s - b} \,
\Big [ \frac14 {v_{\hat m \hat n}^c}^2 + \frac14 {a_{\hat m \hat n}^c}^2 + \frac{\beta}{2} (\partial_{\hat m} \pi^c - A_{\hat m}^c)^2 \Big ] \Big \} \, ,\label{S2}
\end{align}
that dictates the dynamics of the field perturbations. 
We decompose the 5d gauge fields   $V_{\hat m} = (V_z , V_{\hat \mu})$ , $A_{\hat m} = (A_z , A_{\hat \mu})$ and also the derivatives $\partial_{\hat m} = (\partial_z , \partial_{\hat \mu})$. The vectorial gauge symmetry and the residual axial gauge symmetry allows us to set $V_z$ and $A_z$ to zero respectively. Then the action \eqref{S2} takes the form
\begin{align}
S_2 &= - V_4 \int dz \, \Big \{   e^{3 A_s -a}  \Big [ \frac12 (\partial_{\hat \mu} S )^2 + \frac12 (\partial_z S )^2   + \frac12 e^{2 A_s} \frac{d^2 U}{dv^2}S^2 
 \Big ] \nonumber \\
&+ \frac{1}{g_5^2} e^{A_s - b} \,
\Big [ \frac14 {v_{\hat \mu  \hat \nu}^c}^2  + \frac12 (\partial_z V_{\hat \mu}^c)^2 + \frac14 {a_{\hat \mu \hat \nu}^c}^2 
+ \frac12 (\partial_z A_{\hat \mu}^c)^2 \nonumber \\
&+ \frac{\beta}{2} (\partial_{\hat \mu} \pi^c - A_{\hat \mu}^c)^2
+ \frac{\beta}{2} (\partial_{z} \pi^c)^2  \Big ] \Big \} \, .\label{S2v2}
\end{align}
Implementing the Lorentz decomposition
\begin{align}
V_{\hat \mu,a} &= V_{\hat \mu,a}^{\perp} + \partial_{\hat \mu} \xi^a \quad , \quad A_{\hat \mu,a} = A_{\hat \mu,a}^{\perp} + \partial_{\hat \mu} \varphi^a \,, 
\end{align}
\noindent
the action \eqref{S2v2} becomes
\begin{align}
S_2 &= - V_4 \int dz \, \Big \{   e^{3 A_s -a}  \Big [ \frac12 (\partial_{\hat \mu} S )^2 + \frac12 (\partial_z S )^2   + \frac12 e^{2 A_s} \frac{d^2 U}{dv^2}S^2 
 \Big ] \nonumber \\
&+ \frac{1}{g_5^2} e^{A_s - b} \,
\Big [ \frac14 {v_{\hat \mu  \hat \nu}^{\perp, c}}^2  + \frac12 \Big (\partial_z V_{\hat \mu}^{\perp,c} + \partial_{\hat \mu} \partial_z \xi^c \Big )^2 + \frac14 {a_{\hat \mu \hat \nu}^{\perp, c}}^2 +
\frac12 \Big (\partial_z A_{\hat \mu}^{\perp,c} + \partial_{\hat \mu} \partial_z \varphi^c \Big )^2 \nonumber \\
&+ \frac{\beta}{2} (\partial_{\hat \mu} \pi^c - \partial_{\hat \mu} \varphi^c  - A_{\hat \mu}^{\perp,c})^2
+ \frac{\beta}{2} (\partial_{z} \pi^c)^2  \Big ] \Big \} \, .\label{S2v3}
\end{align}
The fields $\xi^c$ can be set consistently to zero and we do that in the following \footnote{Non-zero $\xi^c$ would correspond to scalar mesons in the ${\bf 3}$ representation  with zero mass, not expected in QCD.}.

Now we introduce the Kaluza-Klein expansions
\begin{align}
V_{\hat \mu}^{\perp,c}(x,z) &= g_5 \, v_n^c(z) \hat V_{\hat \mu}^{c,n} (x) \, ,\nonumber \\
A_{\hat \mu}^{\perp,c}(x,z) &= g_5 \, a_n^c(z) \hat A_{\hat \mu}^{c,n} (x) \, , \nonumber \\
S(x,z) &= s_n (z) \hat S_n(x)  \, , \nonumber \\
\varphi^c(x,z) &= g_5 \, \varphi_n^c(z) \hat \Pi_n^c(x) \, , \nonumber \\
\pi^c(x,z) &= g_5 \, \pi_n^c(z) \hat \Pi_n^c(x) \, , \label{KKexp}
\end{align}
where the sum $\sum_{n=0}^{\infty}$ is implicit. Plugging \eqref{KKexp} into \eqref{S2v3}, with $\xi^c=0$, we arrive at the 4d action $S_2 = - \int d^4 x {\cal L}_2$  with 
\begin{align}
{\cal L}_2 &=   \frac12 \Delta_{mn}^{(S)} (\partial_{\hat \mu} \hat S_m) (\partial^{\hat \mu} \hat S_n) 
+ \frac12 M_{mn}^{(S)} \hat S_m \hat S_n  \nonumber \\
&+ \frac14 \Delta_{mn,cd}^{(V)} \hat v_{\hat \mu \nu}^{c,m} \hat v^{\hat \mu \hat \nu}_{d,n}  
+ \frac12 M_{mn,cd}^{(V)} \hat V_{\hat \mu}^{c,m} \hat V^{\hat \mu}_{d,n} 
  \nonumber \\ 
 &+ \frac14 \Delta_{mn,cd}^{(A)} \hat a_{\hat \mu \nu}^{c,m} \hat a^{\hat \mu \hat \nu}_{d,n}  
+ \frac12 M_{mn,cd}^{(A)} \hat A_{\hat \mu}^{c,m} \hat A^{\hat \mu}_{d,n} \nonumber \\
&+ \frac12 \Delta_{mn,cd}^{(\Pi)} (\partial_{\hat \mu} \hat \Pi_m^c) (\partial^{\hat \mu} \hat \Pi_n^d) 
 + \frac12 M_{mn,cd}^{(\Pi)} \hat \Pi_m^c \hat \Pi_n^d 
 + G_{mn,cd}^{(A \Pi)} \hat A_{\hat \mu}^{c,m} \partial^{\hat \mu} \hat \Pi^{d,n}   \, , \label{4dLag}
\end{align}
and we have defined the 4d coefficients
\begin{align}
\Delta_{mn}^{(S)} &= \int dz \,  e^{3 A_s - a} s_m s_n \quad , \quad
M_{mn}^{(S)} =  \int dz \, e^{3 A_s - a} \Big [ (\partial_z s_m) (\partial_z s_n) + e^{2A_s} \frac{d^2 U}{d v^2} s_m s_n \Big ] \, , \nonumber \\ 
\Delta_{mn,cd}^{(V)} &= \delta^{cd} \int dz \, e^{A_s - b} v_m^c v_n^d \quad , \quad 
M_{mn,cd}^{(V)} =  \delta^{cd} \int dz \, e^{A_s -b} (\partial_z v_m^c) (\partial_z v_n^d)  \, , \nonumber \\
\Delta_{mn,cd}^{(A)} &= \delta^{cd} \int dz \, e^{A_s - b} a_m^c a_n^d \quad , \quad 
M_{mn,cd}^{(A)} =  \delta^{cd} \int dz \, e^{A_s -b} \Big [ (\partial_z a_m^c) (\partial_z a_n^d) + \beta \, a_m^c a_n^d  \Big ]\, ,
\nonumber \\
\Delta_{mn,cd}^{(\Pi)} &= \delta^{cd} \int dz \, e^{A_s - b} \Big [ 
(\partial_z \varphi_m^c) (\partial_z \varphi_n^d) + \beta
(\pi_m^c-\varphi_m^c) (\pi_n^d - \varphi_n^d) \Big ] \, , \nonumber \\
M_{mn,cd}^{(\Pi)} &= \delta^{cd} \int dz \, e^{A_s - b} \beta (\partial_z \pi_m^c) (\partial_z \pi_n^d) \, , \nonumber \\
G_{mn,cd}^{(A \Pi)} &= \delta^{cd} \int dz \, e^{A_s - b} \Big [ 
(\partial_z a_m^c) (\partial_z \varphi_n^d) - \beta a_m^c (\pi_n^d - \varphi_n^d) \Big ] \,. 
\end{align}
\noindent
The Lagrangian \eqref{4dLag} becomes the standard Lagrangian for mesons if we impose the following conditions
\begin{align}
&\Delta_{mn}^{(S)} = \delta_{mn}  \qquad , \quad  \Delta_{mn,cd}^{(V)} = \Delta_{mn,cd}^{(A)} = \Delta_{mn,cd}^{(\Pi)} = \delta^{cd} \delta_{mn} \, , \label{KKnormalization}
\end{align}
 for the normalization coefficients and the following conditions
\begin{align}
& M_{mn}^{(S)} = m_{s_n}^2 \delta_{mn} \quad , \quad 
M_{mn,cd}^{(V)} = m_{v_n}^2 \delta^{cd} \delta_{mn} \nonumber \\
M_{mn,cd}^{(A)} &= m_{a_n}^2 \delta^{cd} \delta_{mn} \quad , \quad 
M_{mn,cd}^{(\Pi)} = m_{\pi_n}^2 \delta^{cd} \delta_{mn}
 \nonumber \\
& G_{mn,cd}^{(A \Pi)} = 0 \,,  \label{Couplings}
\end{align}
for the mass coefficients and couplings. In fact, the conditions \eqref{Couplings} are automatically satisfied if the Kaluza-Klein modes satisfy the following differential equations 
\begin{align}
&\Big \{ \Big [ \partial_z + 3 A_s' - a' \Big ] \partial_z 
-e^{2A_s} \frac{d^2 U}{dv^2} \Big \} s_n = - m_{s_n}^2 s_n 
\quad  \text{(scalar sector)} \, , \nonumber \\
& \Big [ \partial_z + A_s' - b' \Big ] \partial_z v_n^c = - m_{v_n}^2 v_n^c \quad  \text{(vectorial sector)} \, , \nonumber \\
& \Big \{ \Big [ \partial_z + A_s' - b' \Big ] \partial_z - \beta \Big \} a_n^c  = - m_{a_n}^2 a_n^c 
\quad \text{(axial sector)}\, , \nonumber \\
& \Big [ \partial_z + A_s' - b' \Big ] \partial_z \varphi_n^c = - \beta (\pi_n^c - \varphi_n^c) \, , \nonumber \\
& \beta \, \partial_z \pi_n^c = m_{\pi_n}^2 \partial_z \varphi_n^c  \quad  \text{(pseudo-scalar sector)} \, ,
\label{KKmodeseqs}
\end{align}
and are normalized according to \eqref{KKnormalization}. The differential equations \eqref{KKmodeseqs} satisfied by the Kaluza-Klein modes are compatible with the field equations \eqref{Seq}-\eqref{pieq} for the 5d fields. The main difference  between those equations is that \eqref{KKmodeseqs} are on-shell conditions involving only the $z$ dependence of the 5d fields whereas \eqref{Seq}-\eqref{pieq} are on-shell conditions for the $z$ and $x$ dependence.  We see that the Kaluza-Klein expansions allows us to consider the 4d fields in \eqref{4dLag} off-shell.

Note that the normalization conditions for the scalar, vectorial and axial sector can be written as
\begin{align}
&\Delta_{mn}^{(S)} = \int dz \, \psi_{s_m} \psi_{s_n} = \delta_{mn} \, ,  \nonumber \\
&\Delta_{mn}^{(V)} = \int dz \, \psi_{v_m} \psi_{v_n} = \delta_{mn} \quad , \quad
\Delta_{mn}^{(A)} = \int dz \, \psi_{a_m} \psi_{a_n} = \delta_{mn} \, ,
\end{align}
where $\psi_{s_n}$, $\psi_{v_n}$ and $\psi_{a_n}$ are the wave functions associated with the Schr\"odinger problem in the scalar, vectorial and axial sector defined previously. For simplicity, we have omitted the flavour index in the vectorial and axial sector. We conclude that the Schr\"odinger problem in those sectors is well posed in the sense that the wave functions form an orthonormal basis. 

\section{Meson operators and decay constants}
\label{Sec:Decay}

From \eqref{deltaS2bdy} we find that the on-shell variation of $S_2$ can be written as
\begin{align}
\delta S_2^{o-s} &= \int d^4 x \Big [ e^{3 A_s -a} \partial_z S \delta S 
+ \frac{1}{g_5^2} e^{A_s - b} v^{\hat z \hat \mu}_c  \delta V_{\hat \mu}^c \nonumber \\
&+ \frac{1}{g_5^2} e^{A_s - b} a^{\hat z \hat \mu}_c \delta A_{\hat \mu}^c 
+ \frac{1}{g_5^2} \beta e^{A_s -b} (\partial^{\hat z} \pi^c - A^{\hat z,c}) \delta \pi^c \Big ]_{z= \epsilon} \, ,
\end{align}
where all the 5d fields satisfy the equations of motion. 
The 5d scalar field $S(x,z)$ behaves near the boundary as 
\begin{equation}
S(x,z) =  S_1 (x) z + T_3(x) z^3 \ln z + S_3 (x) z^3 + \dots \, ,
\end{equation}
where $S_1(x) = \zeta^{-1} m_q(x)$ is the source coefficient. The VEV of the scalar operator responsible for creation of scalar mesons can be obtained from the holographic dictionary:
\begin{align}
\langle \bar q (x) q(x) \rangle &= \frac{ \delta S_2^{o-s}}{\delta m_q(x)} = \zeta \frac{ \delta S_2^{o-s}}{\delta S_1(x)} 
 \nonumber \\
&= \zeta \Big [ z \, e^{3 A_s -a} \partial_z S \Big ]_{z= \epsilon}  
\label{ScalarOp}
\end{align}
The vectorial and axial gauge fields behave near the boundary as
\begin{align}
V_{\hat \mu,c}(x,z) &= V_{\hat \mu,c}^{(0)} (x) + W_{\hat \mu,c}^{(2)}(x) z^2 \ln z + V_{\hat \mu,c}^{(2)}(x) z^2 + \dots \, , \nonumber \\
A_{\hat \mu,c}(x,z) &= A_{\hat \mu,c}^{(0)} (x) + B_{\hat \mu,c}^{(2)}(x) z^2 \ln z + A_{\hat \mu,c}^{(2)}(x) z^2 + \dots \, ,
\end{align}
with $V_{\hat \mu,c}^{(0)} (x)$ and $A_{\hat \mu,c}^{(0)} (x)$ 4d external sources. The corresponding vectorial and axial currents are then given by 
\begin{align}
\langle \bar q(x) \gamma^{\mu} T^c q(x) \rangle 
&= \frac{ \delta S_2^{o-s}}{\delta V_{\hat \mu,c}^{(0)}(x)}
=   \frac{1}{g_5^2} \Big [ e^{A_s - b} v^{\hat z \hat \mu}_c  \Big ]_{z=\epsilon} \nonumber \\
\langle \bar q(x) \gamma^{\mu} \gamma^5 T^c q(x) \rangle 
&= \frac{ \delta S_2^{o-s}}{\delta A_{\hat \mu,c}^{(0)}(x)}
=   \frac{1}{g_5^2} \Big [ e^{A_s - b} a^{\hat z \hat \mu}_c  \Big ]_{z=\epsilon} \, , \label{Currents}
\end{align}
Plugging the Kaluza-Klein expansions \eqref{KKexp} into \eqref{ScalarOp} and \eqref{Currents} we obtain a mode expansion for the scalar operator and currents:
\begin{align}
 \langle \bar q (x) q(x) \rangle &= F_{s^n} \hat S_n (x) \, , \nonumber \\
 \langle \bar q(x) \gamma^{\mu} T^c q(x) \rangle 
&= F_{v^{c,n}} \hat V_{\hat \mu}^{c,n} (x) \, , \nonumber \\
\langle \bar q(x) \gamma^{\mu} \gamma^5 T^c q(x) \rangle 
&= F_{a^{c,n}} \hat A_{\hat \mu}^{c,n} (x) 
- f_{\pi^{c,n}} \partial^{\hat \mu} \hat \pi^{c,n}(x)  \, ,
\end{align}
where the sum $\sum_{n=0}^{\infty}$ is implicit and the Kaluza-Klein coefficients are given 
\begin{align}\label{Eq:DecayConstants}
F_{s^n} &= \zeta \Big [ z \, e^{3 A_s -a} \partial_z s_n \Big ]_{z=\epsilon} \, , \nonumber \\
F_{v^{c,n}} &=  \frac{1}{g_5} \Big [ e^{A_s -b} \partial_z v_{c,n} \Big ]_{z=\epsilon} \, , \nonumber \\
F_{a^{c,n}} &=  \frac{1}{g_5} \Big [ e^{A_s -b} \partial_z a_{c,n} \Big ]_{z=\epsilon} \, , \nonumber \\
f_{\pi^{c,n}} &= - \frac{1}{g_5} \Big [ e^{A_s -b} \partial_z \varphi_{c,n} \Big ]_{z=\epsilon} \, .
\end{align}
The coefficients $F_{s^n}$, $F_{v^{c,n}}$, $F_{a^{c,n}}$ and $f_{\pi^{c,n}}$ are identified with the decay constants of the scalar, vectorial, axial and pseudo-scalar mesons. 

\section{Vector and scalar mesons in the linear soft wall model}
\label{MesonsSW}

In the limit $a_0 \to 0$ (minimal coupling) the differential equation for the vectorial mode $v^{n,c}$ in all models (IA, IB, IIA and IIB) reduce to 
\begin{equation}
\Big [ \partial_z + A_s' - \Phi' \Big ] \partial_z v_n^c = - m_{v_n}^2 v_n^c  \, .
\end{equation}
Since $A_s = - \ln z$ and $\Phi(z) = \phi_{\infty} z^2$, this is the equation found in the original linear soft wall model \cite{Karch:2006pv}. This equation can be solved in terms of the associated Laguerre polynomials. The normalized solutions can be written as \cite{Karch:2006pv}
\begin{equation}
v_n(z) = \sqrt{\frac{2}{n+1}} \phi_{\infty} z^2 L_n^1 (\phi_{\infty} z^2) \qquad n=0,1,\dots \,,
\end{equation}
and the corresponding masses take the form
\begin{equation}
m_{v_n}^2 = 4 \phi_{\infty}  (n +1 ) \qquad n=0,1,\dots \, ,
\end{equation}
which organize into a linear Regge trajectory. 
The vector meson decay constants in the linear soft wall model take the form
\begin{equation}
F_{v^n} =  \frac{1}{g_5} \Big [ e^{A_s -\Phi} \partial_z v_{c,n} \Big ]_{z \to 0} = \frac{\phi_{\infty}}{g_5} \sqrt{8 (n+1)} \,.
\end{equation}
Note that $F_{v^n}^2$ also organize into a linear Regge trajectory. 

For the differential equation of the scalar mode $s_n(z)$, all the models reduce to the linear soft wall model if we take the limits $a_0 \to 0$ (minimal coupling) and $m_q \to 0$ (chiral limit). 
The resulting differential equation can be written as
\begin{equation}
\Big \{ \Big [ \partial_z + 3 A_s' - \Phi' \Big ] \partial_z 
+ 3 e^{2A_s} \Big \} s_n = - m_{s_n}^2 s_n \, ,
\end{equation}
with normalized solutions \cite{Colangelo:2008us}
\begin{equation}
s_n(z) = \sqrt{\frac{2}{n+1}} \phi_{\infty} z^3 L_n^1(\phi_{\infty} z^2) \qquad n=0,1,\dots \, ,
\end{equation}
and masses
\begin{equation}\label{Eq:MassScalarMesonsLSW}
m_{s_n}^2 =  \phi_{\infty} (4 n + 6 ) \qquad n=0,1,\dots \, ,
\end{equation}
also organized into a linear Regge trajectory. 
The scalar meson decay constant in the linear soft wall model takes the form
\begin{equation}
F_{s^n} = \zeta \Big [ z \, e^{3 A_s - \Phi} \partial_z s_n \Big ]_{z \to 0} = \zeta \phi_{\infty}  \sqrt{18 (n+1)} \, . 
\end{equation}
Note that $F_{s^n}^2$ also organize into a linear Regge trajectory. 

\section{Linear soft wall model with negative dilaton}
\label{Sec:LinearDilNeg}

The negative dilaton, proposed in Refs.~\cite{deTeramond:2009xk, Zuo:2009dz}, was introduced as an alternative to describe hadronic spectrum and spontaneous chiral symmetry breaking in the holographic soft wall model for QCD. However, there is a debate about the consequences arising with the negative dilaton profile. It was shown by the authors of Ref.~\cite{Karch:2010eg}, that the negative dilaton will drive to the emergence of one massless scalar (unphysical) state in the spectrum of the vector mesons, besides of that, the negative dilaton does not allow to get the generalized relation valid for higher spin fields
\noindent
\begin{equation}
m^2_{n,S}=4\phi_{\infty}(n+S),
\end{equation}
\noindent
where $n$ is the radial excitation number and $S$ the spin. In the negative dilaton scenario, $\Phi=\Theta\,\phi_{\infty}z^2$, where $\Theta$ is a negative number. For higher spin fields the corresponding results with negative dilaton become
\noindent
\begin{equation}
m^2_{n,S}=4|\Theta|\phi_{\infty}(n+1),
\end{equation}
\noindent
\noindent
representing unphysical states. Despite of these caveats, in the following we present the derivation of the solution of the tachyon field considering the negative dilaton, then, we will show that the description of spontaneous chiral symmetry breaking is possible within this scenario. The negative dilaton is given by
\noindent
\begin{equation}
\Phi=-\phi_{\infty}\,z^2,
\end{equation}
\noindent
where we have set $\Theta=-1$. To calculate the solution of the tachyon differential equation \eqref{Eq:TachyonDE} we write it in the form (setting $a=\Phi$ and $\lambda=0$)
\noindent
\begin{equation}\label{Eq:Tachyon5D}
\left(z^2\partial_z^2-\left(3-2\phi_{\infty}z^2\right)z\partial_z-m_{X}^2\right)v(z)=0.
\end{equation}
\noindent
Introducing the new variable $x=\phi_{\infty}z^2$, the last equation becomes
\noindent
\begin{equation}
\left(x^2\partial_x^2-\left(1-x\right)x\partial_x-\frac{m_{X}^2}{4} \right)v(x)=0.
\end{equation}
\noindent
This equation can be rewritten in a convenient form using the transformation $v(x)=x^{\beta}\,\mathcal{V}(x)$
\noindent
\begin{equation}\label{Eq:KummerLike}
\left(x\partial_x^2+\left(2\beta-1+x\right)\partial_x+\beta+\frac{\beta^2-2\beta-\frac{m_{X}^2}{4}}{x}\right)\mathcal{V}=0.
\end{equation}
The last differential equation becomes a Kummer-like differential equation  under the condition
\noindent
\begin{equation}\label{Eq:beta}
\beta^2-2\beta-\frac{m_{X}^2}{4}=0,\quad \rightarrow \quad \beta=1\pm\sqrt{1+\frac{m_X^2}{4}}=\frac{\Delta_{\pm}}{2}.
\end{equation}
\noindent
To get the standard form of Kummer's differential equation we must introduce an additional transformation in \eqref{Eq:KummerLike}, $\hat{v}(x)=e^{-\alpha x}\,\mathcal{V}(x)$, such that the differential equation becomes
\noindent
\begin{equation}
\left(x\partial_x^2+\left(2\beta-1+(1-2\alpha)x\right)\partial_x+\beta+\alpha(1-2\beta)+\alpha(\alpha-1)x\right)\hat{v}(x)=0.
\end{equation}
\noindent
Kummer's differential equation is obtained by setting $\alpha=1$,
\noindent
\begin{equation}\label{Eq:Kummer}
\left(x\partial_x^2+\left(2\beta-1-x\right)\partial_x+1-\beta\right)\hat{v}(x)=0.
\end{equation}
\noindent
Then, the general solution is a linear combination of $U(a,b,;x)$, and $x^{1-b}M(1+a-b,2-b;x)$, the Tricomi and Kummer functions, respectively. From Eq.~\eqref{Eq:Kummer} we identify $a=\beta-1$ and $b=2\beta-1$,\footnote{Do not confuse the constants $a$ and $b$ with the non-minimal couplings defined in previous sections.} after restoring the original function and variable, $v(z)$, we get the complete solution
\noindent
\begin{equation}\label{Eq:TachyonSolPhiNeg}
v=e^{-\phi_{\infty} z^2}\left(\mathcal{C}_1 z^{2\beta} U\left(\beta-1,2\beta-1;\phi_{\infty} z^2\right)+\mathcal{C}_3 z^{2(2-\beta)}M\left(1-\beta,3-2\beta;\phi_{\infty} z^2\right)\right),
\end{equation}
\noindent
where $\mathcal{C}_1$ and $\mathcal{C}_3$ are constants. Plugging $m_X^2=-3$ in Eq.~\eqref{Eq:beta} we get $\Delta_-=1$, $\Delta_+=3$, and $\beta=1/2$. Thus, we may expand the solution close to the boundary and read off the leading coefficients, i.e., the source and vacuum expectation value,
\noindent
\begin{equation}\label{Eq:LinearTachyonUV}
v=c_1 z+\left(\sqrt{\pi}\,C_0\,\phi_{\infty}^{3/2}-\frac{c_1\phi_{\infty}}{2}\left(1+\gamma\right)\right)z^3-\frac{c_1\,\phi_{\infty}}{2}z^3\log{\left(\frac{\phi_{\infty}z^2}{4}\right)}+\cdots,
\end{equation}
\noindent
where we have defined $\mathcal{C}_1=\sqrt{\pi}\,c_1$, $\mathcal{C}_3=\phi_{\infty}^{3/2}\,C_0$, and $\gamma$ is Euler's constant. In turn, considering the asymptotic form of the Tricomi and Kummer functions in the IR, $U\left(a,b;x\right)\sim x^{-a}$ and $M\left(a,b;x\right)\sim e^{x}\,x^{a-b}$, the leading terms are $U\sim z$ and $M\sim e^{\phi_{\infty}z^2}z^{-3}$, then, expanding $v(z)$ in the IR
\noindent
\begin{equation}\label{Eq:LinearTachyonIR}
\begin{split}
v=&C_0\left(1+\frac{3}{4\phi_{\infty}z^2}+\frac{45}{32\phi_{\infty}^2z^4}+\cdots\right)\\&+C_2\,e^{-\phi_{\infty}z^2}\left(\phi_{\infty}z^2+\frac{1}{4}-\frac{3}{32\phi_{\infty}z^2}+\cdots\right),
\end{split}
\end{equation}
\noindent
where we have introduced the parameter $C_2=c_1\,\sqrt{\pi}/\sqrt{\phi_{\infty}}$. In AdS/CFT, the asymptotic expansion of the field close to the boundary gives us the source and vacuum expectation value of the corresponding dual operator. Thus, the general form of the asymptotic expansion of the tachyon is given by $v=c_1z+c_3z^3$, where $c_1\propto m_q$ (quark mass) and $c_3\propto \langle \bar{q} q\rangle$ (chiral condensate).
From Eq.~\eqref{Eq:LinearTachyonUV} we may extract the constant $c_3$
\noindent
\begin{equation}
c_3=C_0\,\sqrt{\pi}\,\phi_{\infty}^{3/2}-\frac{c_1\phi_{\infty}}{2}\left(1+\gamma\right).
\end{equation}
\noindent
As can be seen, the parameter $c_3$ depends on two independent parameters $C_0$ and $c_1$. In the chiral limit $c_1=0$, and $c_3\neq 0$ because $C_0\neq 0$, hence, $\langle\bar{q}q\rangle\neq 0$. Therefore, the scenario with negative dilaton allows a description of spontaneous chiral symmetry breaking. However, there is, at least, one issue that called our attention. The Hamiltonian depends on the derivative of the tachyon field in the IR (see for instance Ref.~\cite{Ballon-Bayona:2020qpq}) and it will diverge because of the presence of $C_0$ in Eq.~\eqref{Eq:LinearTachyonIR}. This suggest that $C_0$ must be zero in order to get an finite Hamiltonian in the IR. Finally, it is worth mentioning that an equivalent solution to  Eq.~\eqref{Eq:TachyonSolPhiNeg} was obtained in Ref.~\cite{Zuo:2009dz}.

\section{Non-linear soft wall with a negative dilaton}
\label{Se:non-linearNegativeDil}

Let us solve the non-linear differential equation of the tachyon considering a negative dilaton
\noindent
\begin{equation}\label{Eq:NonLinTachyon5DNegDil}
\left(z^2\partial_z^2-\left(3-2\phi_{\infty}z^2\right)z\partial_z-m_{X}^2-\frac{\lambda}{2}v^2(z)\right)v(z)=0.
\end{equation}
\noindent
It is expected that the presence of the non-linear term will change the solutions obtained in Eq.~\eqref{Eq:TachyonSolPhiNeg}. However, we may solve the non-linear differential equation asymptotically. Thus, the asymptotic solution close to the boundary may be determined through the ansatz
\noindent
\begin{equation}\label{Eq:TachyonNegDilUV}
v=c_1 z+d_3z^3\ln{\left(\frac{\phi_{\infty}z^2}{4}\right)}+c_3 z^3+d_5z^5\ln{\left(\frac{\phi_{\infty}z^2}{4}\right)}+c_5z^5+\cdots
\end{equation}
\noindent
where 
\noindent
\begin{equation}\label{Eq:UVSolnon-linearNegDil}
\begin{split}
d_3=&\frac{c_1}{8}\left(c_1^2\lambda-4\phi_{\infty}\right),\qquad d_5=\frac{3 c_1}{128}\left(c_1^2\lambda-4\phi_{\infty}\right)^2,\\
c_5=&\frac{1}{256}\left(c_1^2\lambda-4\phi_{\infty}\right)\left(48c_3-9c_1^3\lambda+20c_1\phi_{\infty}\right).
\end{split}
\end{equation}
\noindent
As can be seen, the non-linear term contributes to the coefficient $d_3$, while $c_1$ and $c_3$ are independent parameters. Note that form $c_1=({4\phi_{\infty}}/{\lambda})^{1/2}$, the logarithm contributions and higher-order coefficients in \eqref{Eq:UVSolnon-linearNegDil} vanish reducing the asymptotic solution to
\noindent
\begin{equation}
v=\sqrt{\frac{4\phi_{\infty}}{\lambda}}z+c_3z^3.
\end{equation}
\noindent
This asymptotic solution suggests that the parameter $c_1$ is fixed once we know the parameters $\phi_\infty$ and $\lambda$. Nevertheless, this result is strange since $c_1$ is related to the quark mass in the dual field theory, $c_1\propto m_q$.

In turn, the asymptotic solution in the IR may be calculated changing the variable $y= 1/z$. Then, Eq.~\eqref{Eq:NonLinTachyon5DNegDil} becomes
\noindent
\begin{equation}\label{Eq:TachyonNegDilIR}
\left[\left(y\partial_y\right)^2+2\left(2-\phi_{\infty}y^{-2}\right)\left(y\partial_y\right)+3\right]v-\frac{\lambda}{2}v^3=0.
\end{equation}
\noindent
Considering the power series ansatz, the asymptotic solution is given by
\begin{equation}
v=C_0+\frac{C_2}{z^2}+\frac{C_4}{z^4}+\cdots
\end{equation}
\noindent
where 
\begin{equation}
C_2=\frac{C_0}{8 \phi_\infty}(6-C_0^2\lambda),\quad 
C_4=\frac{3C_0}{128\phi_{\infty}^2}(C_0^2\lambda-10)(C_0^2\lambda-6).
\end{equation}
\noindent
It is interesting to realize that when $C_0=\sqrt{6/\lambda}$ the coefficients $C_2,\,C_4,\cdots$ reduce to zero, remaining $v=\sqrt{6/\lambda}$ as the exact solution. This solution is also an exact solution of the differential equation \eqref{Eq:TachyonNegDilIR}. The complete expansion in $v$ should be considered for arbitrary $C_0$ . After implementing the numerical solution, using as ``initial condition'' the asymptotic solution in the IR, then, integrating to the UV we obtained unstable solutions, highly sensitive to the change of the domain of integration. As we are looking for stable solutions we do not present those results in this paper.

\section{Holographic derivation of the GOR relation: Toy model}
\label{Sec:GORToyModel}

In this section we propose a toy model to get the Gell-Mann-Oakes-Renner relation. We start by building a simple function for the tachyon which takes into account the asymptotic solutions in the UV and IR. To simplify the analysis we consider $b=\phi_{\infty}\,z^2$ with $\phi_{\infty}=1$. Then, we consider the most simple smooth function interpolating between the asymptotic solution in the UV, $v=c_1\,z+c_3\,z^3$, and IR, $v=C_0$, which is given by
\noindent
\begin{equation}\label{Eq:ToyModelTachyon}
v=\frac{c_1\,z+c_3\,z^3}{1+\frac{c_3}{C_0}z^3}.
\end{equation}
\noindent
As can be seen, this simple function contains the relevant parameters arising in the UV and IR. We can check the UV expansion, which is given by
\noindent
\begin{equation}
v=c_1\,z+c_3\,z^3+\mathcal{O}(z^4),\qquad  z\to 0
\end{equation}
\noindent
while its IR expansion given by
\noindent
\begin{equation}
v=C_0+\mathcal{O}(z^{-2}).\qquad z\to \infty
\end{equation}
\noindent

Then, plugging \eqref{Eq:ToyModelTachyon} in Eq.~\eqref{Eq:NorCondition3} one can integrate
\noindent
\begin{equation}\label{Eq:IntegralToyModel}
\begin{split}
&\int_{0}^{\infty}\frac{u^3e^{-u^2}}{v(u)^2}du=\frac{1}{2c_1\,c_3}+\left(\frac{c_1^3}{2C_0^2\,c_3^2}+\frac{1}{2c_3^2}-\frac{3c_1^2}{2C_0^2\,c_3^2}\right)e^{c_1/c_3}\text{Ei}\left(-\frac{c_1}{c_3}\right)-\frac{c_1^2}{2C_0\,c_3^2}\\
&+\frac{1}{2C_0^2}-\frac{c_1}{C_0^2\,c_3}+\frac{c_1\sqrt{\pi}}{C_0\,c_3^2}+\frac{\sqrt{\pi}}{C_0\,c_3}-\left(\frac{3c_1^{3/2}c_3^{3/2}}{2C_0c_1c_3^3}+\frac{2c_1^{5/2}c_3^{1/2}}{2C_0c_1c_3^3}\right)\pi e^{c_1/c_3}\text{Erfc}\left(\sqrt{\frac{c_1}{c_3}}\right),
\end{split}
\end{equation}
\noindent
where $\text{Ei}[x]$ is the exponential integral function and $\text{Erfc}[x]$ the complementary error function. 
So far we did not say anything about the parameters $c_1$, $c_3$, and $C_0$, this means that the last results is valid for arbitrary values of these parameters. However, as we want to investigate the GOR relation we must consider the regime of small $c_1$ parameter. Thus, the leading term of Eq.~\eqref{Eq:IntegralToyModel} is
\noindent
\begin{equation}\label{Eq:GORAnalytic1}
\begin{split}
&\int_{0}^{\infty}\frac{u^3e^{-u^2}}{v(u)^2}du=\frac{1}{2c_1\,c_3}.
\end{split}
\end{equation}
\noindent
which is equal to Eq.~\eqref{Eq:NorCondition3}. Plugging this result in \eqref{Eq:NorCondition2}
\noindent
\begin{equation}\label{Eq:GORToyModel}
\frac{f_{\pi}^2m_{\pi}^2}{2c_1\,c_3}=1,
\end{equation}
\noindent
yields to the GOR relation.

\begin{figure}[ht!]
\centering
\includegraphics[width=7cm]{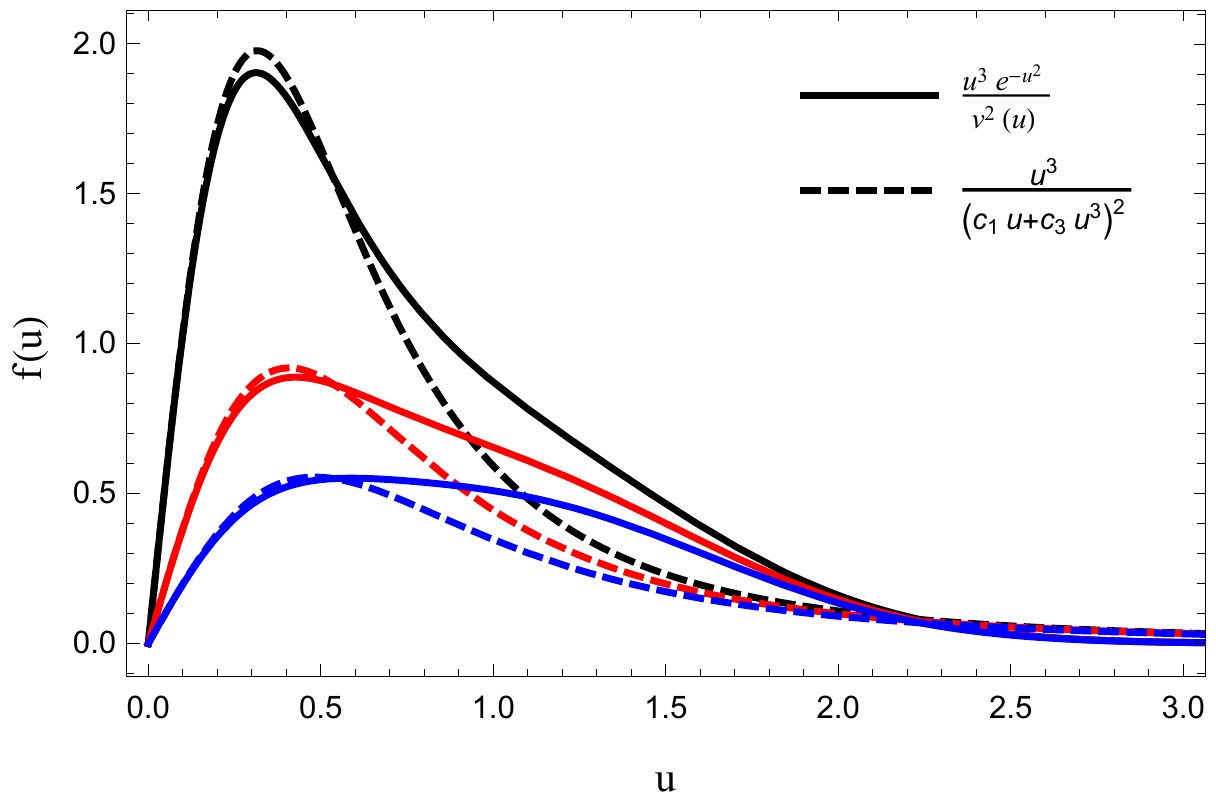}
\hfill
\includegraphics[width=7cm]{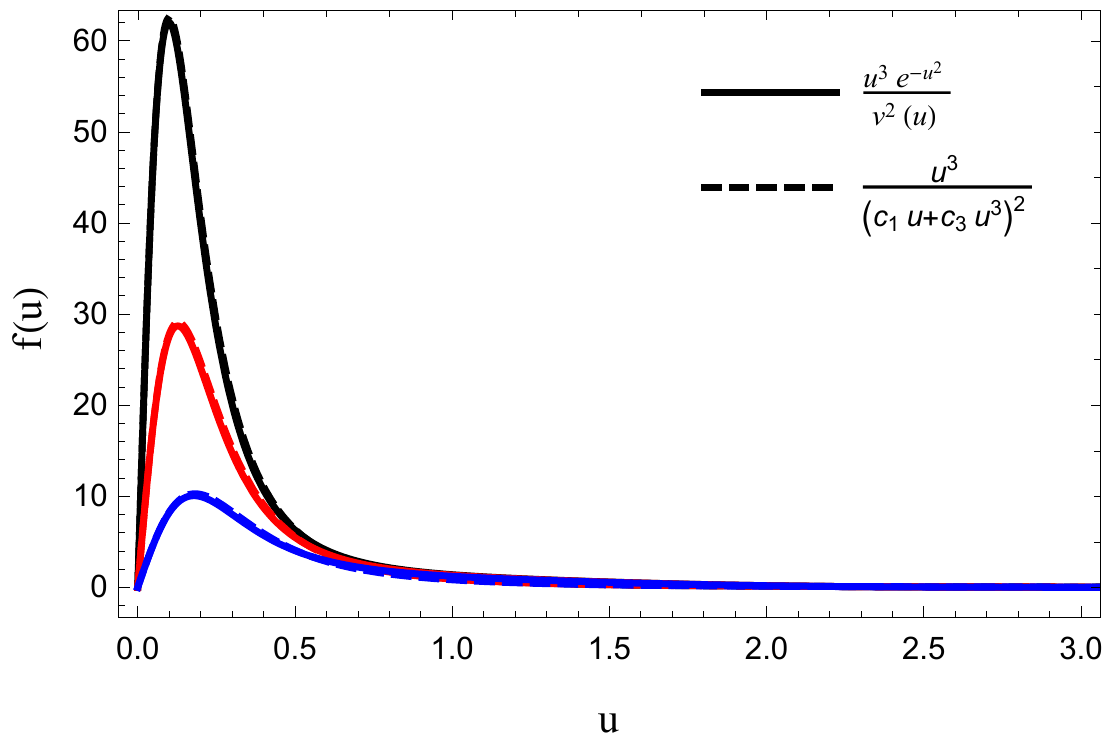}
\caption{
Left: The kernel of the integral (solid lines) against the Gaussian function (dashed lines) for selected values of $c_1$, $c_1=0.7$ (blue), $0.5$ (red), $0.3$ (black). Right: The kernel of the integral (solid lines) against the Gaussian function (dashed lines) for selected values of $c_1=0.1$ (blue), $0.05$ (red), $0.03$ (black).
}
\label{Fig:GROAnalitic}
\end{figure}

An alternative way to show that the result \eqref{Eq:GORAnalytic1} is consistent is plotting the kernel function of the integral, left side of  Eq.~\eqref{Eq:GORAnalytic1}, and investigating the behavior of this function in the region of small $c_1$. Thus, we define 
\noindent
\begin{equation}
f(u)=\frac{u^3e^{-u^2}}{v(u)^2}.
\end{equation}
\noindent
The plot of this function compared against the Gaussian function $\frac{u^3}{(c_1\,u+c_3\,u^3)^2}$ is displayed in Fig.~\ref{Fig:GROAnalitic}. As can be seen, in the left panel we compare the behavior of these functions changing the value of $c_1$, while the right panel shows the same functions for small values of $c_1$.

From Fig.~\ref{Fig:GROAnalitic} we may conclude that the kernel of the integral is equivalent to the Gaussian function in the regime of small $c_1$
\noindent
\begin{equation}
\frac{u^3e^{-u^2}}{\left(\frac{c_1u+c_3u^3}{1+\frac{c_3}{C_0}u^3}\right)^2}\equiv\frac{u^3}{\left(c_1u+c_3u^3\right)^2}.
\end{equation}
\noindent
This equivalence gets better decreasing $c_1$. Therefore, we can replace the Gaussian function in the integral getting
\noindent
\begin{equation}
\int_{0}^{\infty}\frac{u^3\,du}{\left(c_1u+c_3u^3\right)^2}=\frac{1}{2c_1c_3},
\end{equation}
\noindent
which is the result we get previously in \eqref{Eq:GORAnalytic1} and \eqref{Eq:NorCondition3}.

\bibliographystyle{utphys}

\bibliography{CSB}

\end{document}